# Lattice distortion inducing exciton splitting and coherent quantum beating in $CsPbI_3$ perovskite quantum dots

Yaoyao Han[1,2], Wenfei Liang[1], Xuyang Lin[1,2], Yulu Li[1], Fengke Sun[1,2], Fan Zhang[2,3], Peter C. Sercel[4]*, Kaifeng Wu[1]*.

[1] State Key Laboratory of Molecular Reaction Dynamics, Dalian Institute of Chemical Physics, Chinese Academy of Sciences, Dalian, Liaoning 116023, China.

[2] University of Chinese Academy of Sciences, Beijing 100049, China.

[3] Dalian National Laboratory for Clean Energy, Dalian Institute of Chemical Physics, Chinese Academy of Sciences, Dalian, Liaoning 116023, China

[4] Center for Hybrid Organic Inorganic Semiconductors for Energy, Golden, Colorado 80401, United States

* Correspondence to: pcsercel@gmail.com

* Correspondence to: kwu@dicp.ac.cn

**Anisotropic exchange-splitting in semiconductor quantum dots (QDs) results in bright-exciton fine-structure-splitting (FSS) important for quantum information processing. Direct measurement of FSS usually requires single/few QDs at liquid-helium temperatures, because of its sensitivity to QD size and shape, whereas measuring and controlling FSS at an ensemble-level seem to be impossible unless all the dots are made to be nearly the same. Here we report strong bright-exciton FSS up to 1.6 meV in solution-processed $CsPbI_3$ perovskite QDs, manifested as quantum beats in ensemble-level transient absorption at liquid-nitrogen to room temperatures. The splitting is robust to QD size and shape heterogeneity, and increases with decreasing temperature, pointing towards a mechanism associated with orthorhombic distortion of perovskite lattice. Effective-mass-approximation calculations reveal an intrinsic "fine-structure gap" that agrees well with the observed FSS. This gap stems from an avoided crossing of bright-excitons confined in orthorhombically-distorted QDs that are bounded by the pseudocubic $\{100\}$ family of planes.**

## Introduction

Semiconductor quantum dots (QDs) constitute an important material platform for quantum information science (QIS).[1] While earlier efforts on QD-related QIS focus on epitaxial QDs fabricated using complex high-temperature processes such as molecular beam epitaxy or metal-organic chemical vapor deposition, more recent attention has been paid to low-cost, solution-processable counterparts called colloidal QDs or nanocrystals.[2] Colloidal QDs have achieved great success in applications such as to light-emitting diodes, lasers, and photodetectors and in solar energy conversion.[3,4] Their implementation into QIS, however, has remained challenging due to the intrinsic “lossy” nature of these materials (defects, dangling bonds, surface ligands, etc.). For example, photon emission from colloidal QDs is often unstable (due to “blinking” phenomena)[5,6] and incoherent. Consequently they have not previously been comparable with epitaxial QDs for application as single-photon sources.[7,8] Nevertheless, recent studies suggest that a unique combination of fast radiative emission and long optical coherence in solution-grown lead halide perovskite QDs did give rise to highly coherent single-photon emission,[9,10] heralding perovskite colloidal QDs as candidates for quantum emitters.

In addition to single-photon emission, another property closely connected to QIS applications is the exciton fine-structure-splitting (FSS) induced by the electron-hole exchange interaction. In perovskite QDs, the major effect of this interaction is to split the exciton into a dark (optically inactive) singlet level and a bright (optically active) triplet.[11-13] In addition, anisotropy due to QD shape or orthorhombic lattice symmetry causes a further splitting of the bright triplet into three levels whose transition dipoles are oriented along the symmetry axes of the QD (Fig. 1a), with splitting typically ranging from a few to 100s of μeV.[11-16] In this case, optical excitation of QDs using circularly-polarized light, for example, can prepare a coherent superposition of the new eigenstates, a phenomenon that can be exploited for coherent control of quantum states for quantum computing.[17,18] Alternatively, if the QDs are excited into biexciton states,[11] the two photons emitted during biexciton cascade recombination are quantum-entangled with regard to their polarizations. This property can be exploited for entangled photon-pairs in quantum optics, although for this application it is important to suppress the magnitude of FSS in order to increase photon indistinguishability in energy.[19,20] Thus, exchange-split bright excitons are of close relevance to quantum coherence and entanglement that are essential for quantum information technologies.

Bright exciton FSS in epitaxial QDs can be observed using polarization-resolved micro-photoluminescence (μ-PL) of single/few QDs at a temperature of a few Kelvin,[21] but there have only been limited reports of such experiments for colloidal CdSe QDs due to the slow emission in these systems.[22] By contrast, the fast, efficient emission in perovskite QDs has enabled multiple reports of bright exciton FSS using single-QD μ-PL, which typically exhibits substantial dispersity due to the sensitivity of FSS to size and shape non-uniformity.[11-14] Therefore, ensemble-level measurements of FSS in perovskite QDs seem to be impractical. Here we report strong bright exciton FSS up to 1.6 meV in $CsPbI_3$ QDs measured with ensemble-level circularly-polarized transient absorption (TA) performed in the regime of liquid-nitrogen temperature to room temperature. The FSS is manifested as time-domain coherent exciton quantum beats that are surprisingly robust to QD size and shape non-uniformity, and can be readily modulated through temperature. Effective-mass model calculations, in conjunction with temperature-dependent X-ray diffraction (XRD) measurements, indicate that the observed FSS originates from an avoided crossing "fine-structure gap" between bright excitons. The excitons are confined in orthorhombic-phase QDs, but these QDs are bounded by the pseudocubic $\{100\}$ family of planes instead of the low index orthorhombic crystal planes, as revealed here by transmission electron microscope (TEM) imaging. This salient feature results in the gap that can be detected in spite of QD size and shape heterogeneity across an ensemble sample.

## Results and Discussion

**Quantum beats in ensemble QDs.** $CsPbI_3$ QDs of varying sizes were synthesized using a hot-injection method developed in prior studies;[23-25] details can be found in the Methods. Their TEM images are shown in Supplementary Fig. 1, exhibiting cuboid nanocrystals with average edge lengths ranging from 4.9 to 17.3 nm. A typical TEM image for the 7.9 nm QDs is shown in Fig. 1b. The crystal phase of these QDs will be elaborated below.

The Bohr exciton radius of $CsPbI_3$ is ~4.64 nm,[26] indicative of intermediate quantum confinement in the 4.9 to 7.9 nm QDs (See Supplementary Text 3). The optical absorption spectra of these $CsPbI_3$ QDs dispersed in hexane exhibits well-resolved exciton peaks with the first peak blueshifting from 660 to 615 nm as the QD size decreases from 7.9 to 4.9 nm (Fig. 1c). In contrast, the larger 17.3 nm QDs approach the weak confinement limit, exhibiting a continuous absorption profile with an onset at ~710 nm. The PL spectra of these QDs under 365 nm excitation are displayed in Fig. 1d and the PL quantum yields are generally >40%.[23]

For TA measurements, $CsPbI_3$ QDs were spin-coated as closely-packed films on glass substrates (Methods). A typical film has a thickness of ~400 nm and an optical density of 0.17 at its exciton peak (Supplementary Fig. 2). We assume that QDs are completely randomly oriented in the film (see below). The experimental setup for our femtosecond circularly-polarized TA are described in Methods. Briefly, QD films were excited by a wavelength-tunable pump beam (pulse duration ~230 fs) in resonance with the first exciton peak of the QDs (see Fig. 1c), and the time-dependent absorption changes were recorded by a white-light-continuum probe beam. The circular beam polarization was controlled using quarter-waveplates, and the pump beam power was minimized to avoid multiexciton excitation.[24]

Fig. 2a and 2b are the two-dimensional pseudo-color TA spectra of the 4.9 nm QDs measured with co- ($\sigma^+/\sigma^+$) and counter-polarized ($\sigma^+/\sigma^-$) pump/probe beams, respectively, at 80 K. The bleach feature at 615 nm results from the state-filling effect of photogenerated excitons,[27] whereas the induced blue- and red-side absorption features at 570 and 630 nm, respectively, can be attributed to Coulombic effects.[23] The dynamics of these spectral features were used to track the spin relaxation dynamics of the bright triplet |1, ±1> excitons generated by circularly-polarized photons, as reported in previous studies.[23,28-30] However, herein we find all the TA features display periodic oscillations on the ps time scale in both Fig. 2a and 2b. Importantly, the phases of the oscillations measured with co- and counter-polarized pump/probe beams are almost exactly anti-correlated (Fig. 2c). After the decay of oscillations, the co- and counter-polarized signals converge into a long-lived plateau that corresponds to QD excitons with ns lifetimes.[24] A subtraction between co- and counter-polarized signals results in a damped oscillatory decay, with beat frequencies corresponding to energies on the order of 1 meV. The oscillatory decays obtained at different probing wavelengths are phase-correlated and are essentially the same after normalization (Supplementary Fig. 3).

One source of TA signal oscillations is the generation of coherent optical phonons associated with distortions of the lead halide octahedra, triggered by photoexcitation and electron-phonon coupling.[31-34] However, such coherent phonon oscillations typically correspond to energies in the range of ~2-10 meV,[31-34] much larger than the beat frequencies we observed, and critically, are sensitive not to the excitation helicity but to the probing wavelengths (see Supplementary Fig. 4), contrary to the observations here. Moreover, the oscillation period at 80 K increases with increasing QD size and eventually vanishes in the weakly-confined 17.3 nm sample (Fig. 2d); see Supplementary Figs 5-9 for the corresponding TA spectra. This strong size dependence is inconsistent with coherent generation of optical phonons; for example, prior studies reported coherent phonon oscillations in weakly-confined perovskite QDs.[34] Finally, the beat frequencies of all QD samples are observed to decrease with increasing temperature and eventually vanish near room temperature (see Fig. 3 below). This strong temperature dependence is inconsistent with coherent generation of confined acoustic phonon modes either, as previous calculations have indicated much weaker variation of acoustic phonon energies with temperature[35,36].

The helicity-sensitivity and QD-size-dependence of the oscillations observed here are, however, consistent with bright exciton FSS induced by the anisotropic exchange interaction introduced above. As depicted in Fig. 1a, excitation using $\sigma^+$ or $\sigma^-$ photons with spectral width (~20-30 meV) covering the linearly-polarized bright exciton transitions creates a coherent superposition of these states and results in the TA quantum beats, as has been reported for epitaxial QDs.[37,38] This interpretation also completely explains the anti-correlated phases observed in TA oscillations measured with photons of counter helicities (See Fig. 2c and Supplementary Text 6). The oscillation/procession frequency ($\omega$) is determined by the magnitude of bright state FSS ($\Delta_{FSS}$): $\omega = \Delta_{FSS}/\hbar$. This is in contrast to carrier spin procession about an external magnetic field,[39-41] which results in long-lived quantum beats whose frequencies increase linearly with the field strength, vanishing at zero field. Because our measurements are performed without an external magnetic field, we do not detect any long-lived spin precession signatures on the longer ns timescales (Supplementary Fig. 10).

We extract $\Delta_{FSS}$ using two methods, fitting the time-domain kinetics in Fig. 2d using damped sinusoidal functions (Supplementary Table 1) and performing fast Fourier transformation on these kinetics (Fig. 2e), and obtained consistent results; see Methods. As plotted in Fig. 2f and tabulated in Supplementary Table 2, $\Delta_{FSS}$ at 80 K increases from 0.70 to 1.64 meV as the QD size is reduced from 7.9 to 4.9 nm, a manifestation of confinement-enhanced anisotropic exchange. The phenomenological decay time of the beating component ranges from 3-6 ps, without an obvious QD-size dependence (Supplementary Table 1), suggesting that the damping is dominated by inhomogeneity of QD size and shape in the ensemble. Thus, in the absence of inhomogeneity we should expect even longer coherence times. This type of exciton fine-structure coherence lifetime is considerably longer than electronic coherence lifetimes of femtoseconds found in many natural or artificial systems.[42,43]

Notably, however, there are two details not explainable by this simple two-state quantum beating model: i) to fit the kinetics in Fig. 2d, an extra pure exponential decay component is required in addition to a damped sinusoid (Methods); ii) the FFT spectra in Fig. 2e are highly-asymmetric and contain a broad tail in the high-frequency range. Related to this feature, we note that QDs in the ensemble film have a broad shape distribution (i.e., QDs are cuboids with various aspect ratios in their TEM images; see the statistics in Supplementary Figure 11). Indeed, the bright exciton $\Delta_{FSS}$ depends sensitively upon the aspect ratios due to the long-range exchange interaction.[11-14] A broad shape distribution should thus translate into a broad distribution of $\Delta_{FSS}$ and, in a simple analysis, should therefore erase any distinguishable quantum beats. Therefore, there likely exists an additional intrinsic mechanism for exciton splitting that is robust to the size and shape heterogeneity.

**Temperature-dependent FSS and lattice distortion.** In order to clarify the above issues and to reveal the origin of the observed FSS, we performed the TA measurements under varying temperatures from 80 to 300 K; see spectra in Supplementary Figs 12-14. We find the beat frequency to be temperature-sensitive, as presented in Fig. 3a-d for 4.9, 5.4, 7.4 and 7.9 nm QDs. The extracted $\Delta_{FSS}$ is plotted in Fig. 3e as a function of temperature; see also Supplementary Fig. 15 and Supplementary Tables 3-7. Smaller QDs seem to exhibit a steeper dependency of $\Delta_{FSS}$ on temperature. Specifically, for the 4.9 nm QDs, $\Delta_{FSS}$ decreases from 1.64 to 0.32 meV as the temperature increases from 80 to 300 K, whereas it ranges from 0.69 to 0.08 meV for the 7.9 nm QDs.

Because the QD shape anisotropy varies negligibly with temperature, as confirmed by our cryo-TEM measurements in Supplementary Fig. 16, we posit the temperature-dependent $\Delta_{FSS}$ is related to the lattice structure of $CsPbI_3$ crystal, which has been intensively investigated recently within the context of photovoltaics.[44-46] The room-temperature phase of bulk $CsPbI_3$ is a non-perovskite yellow-phase (δ-phase) not suitable for photovoltaics; the desired "black-phase" is stabilized only at temperatures above 320 ℃. The black-phase is further classified into cubic, tetragonal and orthorhombic phases. The symmetry-lowering is induced by distortion of the Pb-centered octahedral framework, which can be activated by decreasing temperature, as depicted in Fig. 4a.[45] For nanocrystal structures, the black-phase $CsPbI_3$ can be obtained at room temperature thanks to a contribution from the surface energy.[46,47]

The XRD patterns of our $CsPbI_3$ QDs are assigned to the orthorhombic-phase (Fig. 4b), consistent with recent studies.[47,48] With decreasing temperature, subtle changes in the diffraction peaks can be identified, as indicated for 7.9 and 17.3 nm QDs in Fig. 4b. In order to correlate these changes to lattice distortion, we performed Rietveld refinement of the XRD patterns (Methods). The refined lattice parameters are plotted in Fig. 4c. A prominent trend is that the orthorhombic lattice constants *b* and *a* increase and decrease with decreasing temperature, respectively, and hence the distinction between *b* and *a* (i.e., the lattice anisotropy) grows with decreasing temperature. The temperature dependency is steeper for 7.9 nm QDs than for 17.3 nm QDs, seemingly correlated with the steeper temperature dependency of FSS observed in smaller QDs (Fig. 3e). Note that the strong peak broadening and background in the XRD patterns of even smaller-size QDs prohibit quantitative refinement (Supplementary Fig. 17).

**"Fine-structure gap" revealed by calculation.** To reveal the physics underpinning the correlation between lattice distortion and FSS, we performed an effective-mass-approximation (EMA) calculation of the band-edge exciton fine structure of the $CsPbI_3$ QDs on the basis of a recently-developed quasi-cubic model for perovskite QDs;[16,49] full details for this model are provided in Supplementary Texts 1-5, Supplementary Figs 18-31 and Supplementary Tables 8-10. The model considers a cuboidal-shaped QD with three edge lengths of $L_x$, $L_y$ and $L_z$ (Fig. 5a, top). Importantly, according to the high-resolution TEM (Fig. 1c inset and Supplementary Figs 18, 19), these three edges are aligned to the quasi-cubic $\langle 100 \rangle_c$ directions rather than the orthorhombic symmetry axes. This interpretation is consistent with recent findings that the bounding facets of $CsPbBr_3$ nanoplatelets comprise the pseudocubic $\{100\}_c$ family of planes rather than the orthorhombic $\{100\}_o, \{010\}_o$ and $\{001\}_o$ crystal planes,[50] and also consistent with the high-resolution TEM images of cuboidal $CsPbBr_3$ nanocrystals displayed in refs[13,51] and consistent with the surface energy calculations for cuboidal $CsPbBr_3$ nanocrystals reported in ref[52]. Therefore, while the orthorhombic primitive lattice vector $\boldsymbol{c}$ is aligned with $L_z$, $\boldsymbol{a}$ and $\boldsymbol{b}$ point towards the corners rather than the edges in the $L_x$-$L_y$ plane (Fig. 5a, bottom). The temperature-induced change of lattice constants of 7.9 nm QDs revealed in Fig. 4c is described by tetragonal and orthorhombic strain components ($\delta$ and $\zeta$, respectively) relative to the cubic phase (see Supplementary Text 4.2). The temperature-dependences of $\delta$ and $\zeta$ are plotted in Fig. 5b, with the latter being the dominant component. These strain components should split the bright triplet excitons into states whose transition dipoles are aligned to the orthorhombic symmetry axes.

The energy levels of the bright excitons can be described by diagonalization of the exchange Hamiltonian taking into consideration short- and long-range exchange interaction. For cuboidal QDs whose bounding facets are orthogonal to the orthorhombic lattice vectors (***a, b, c***), the resulting bright exciton energy levels are given by:[16,49]

$$E_i = \frac{3}{2}\left(\hbar\omega_{ST} + \frac{\hbar\omega_{LT}}{3}(1 + g_\kappa)\mathcal{A}_i\right) f_i\left(\frac{\theta(L)}{\theta_{bulk}}\right), \tag{1}$$

where subscripts $i$ denote excitons labeled as $A$, $B$ and $C$, which are linearly polarized along the ***a***, ***b*** and ***c*** orthorhombic axes, respectively. The terms $\hbar\omega_{ST}$ and $\hbar\omega_{LT}$ respectively account for the short- and long-range parts of the exchange interaction, $g_\kappa$ is a dielectric-contrast factor which enhances the long-range exchange for QDs surrounded by a low-dielectric-constant medium, $\mathcal{A}_i$ are shape anisotropy functions which equate to unity for a perfect cube, $f_i$ are Bloch factors that depends on the strain via a deformation potential interaction, and $\theta(L)$ and $\theta_{bulk}$ are the exchange overlap integrals for QDs and bulk, respectively, which determine the size dependence of the FSS (Supplementary Text 4). The strain modifies the Bloch factors $f_i$ and hence affects both short- and long-range exchange. From eq 1, both lattice distortion and shape anisotropy contribute to the FSS in QDs.

By contrast, in cuboidal QDs with pseudocubic $\{100\}_c$ family of bounding facets (the situation here), while the ***c***-polarized $C$ exciton energy is still described by eq 1, the $A$, $B$ excitons are coupled (Supplementary Texts 4.3 and 5), with energies given by diagonalization of the coupling Hamiltonian,

$$\widetilde{H}^{ex}_{A,B} = \frac{3}{2}\left\{ \hbar\omega_{ST}\begin{pmatrix} f_A & 0 \\ 0 & f_B \end{pmatrix} + \frac{\hbar\omega_{LT}}{3}(1 + g_\kappa)\begin{pmatrix} f_A\mathcal{A}_{A,A} & \sqrt{f_A f_B}\mathcal{A}_{A,B} \\ \sqrt{f_A f_B}\,\mathcal{A}_{BA} & f_B\mathcal{A}_{BB} \end{pmatrix}\right\}\left(\frac{\Theta(L_e)}{\Theta_{bulk}}\right), \qquad (2)$$

where the off-diagonal terms $\mathcal{A}_{A,B} = \mathcal{A}_{B,A}$ reflect coupling of the *A*, *B* excitons that are not orthogonal to the bounding facets of the QD. Note that for $L_x = L_y$ these off-diagonal terms vanish and eq 2 reduces to eq 1. We label the resulting coupled states $\alpha, \beta$ to distinguish them from the uncoupled *A, B*. Fig. 5c shows $E_\alpha$, $E_\beta$, and $E_C$ as a function of $L_y/L_x$ with $\zeta$ = -0.03 (the lattice strain at ~80 K). For comparison, we also show in the same plot the energies of the uncoupled *A*, *B* excitons, calculated from eq 1. Significantly, there is an avoided crossing gap of $\Delta_{\alpha,\beta}$ ~0.75 meV between $E_\alpha$ and $E_\beta$ at $L_y/L_x$ = 1. By contrast, the gap is absent between $E_A$ and $E_B$ which intersect at $L_y/L_x$ ~1.12. Thus, it is the rotation of the ***a, b*** lattice vectors with respect to the pseudo-cubic $\langle 100\rangle_c$ axes that creates this avoided crossing gap. We also evaluated the effect of elongation or shortening of $L_z$, in the ***c*** direction, on exciton splitting by assuming $L_y/L_x$ =1 and varying $L_z/L_x$. Fig. 5d shows the result for $\zeta$ = -0.03. In this case the ordering of $E_C$ depends sensitively on the aspect ratio so that variation in $L_z/L_x$ results in a broad spread of the energy differences $\Delta_{\alpha,C}$ and $\Delta_{\beta,C}$.

**Agreement between experiment and theory.** Because the QDs in the film are randomly oriented, a circularly-polarized pump beam should excite linear combinations of $\alpha, \beta$ excitons for QDs whose ***c***-axis is aligned parallel to the light wave vector, and linear combinations of $\alpha, \beta$ and $C$ excitons for QDs whose ***c***-axis is aligned with a component perpendicular to it. Therefore, in principle there should be three beat frequencies if the QD morphological heterogeneity is not considered. Our calculation based on the realistic morphological statistics in Supplementary Figs 11 and 20 indicates that, however, the broad distributions of the edge length ratios $L_y/L_x$, $L_z/L_x$ and $L_z/L_y$ wash out all resolved beating frequencies associated with these shape anisotropies; see Supplementary Text 6 for details. These broadly-distributed frequencies associated with shape anisotropy are manifested as the pure-exponential-like decay in the fitting of TA kinetics in Fig. 2d and the broad high-frequency tail in the FFT spectra in Fig. 2e. The only distinguished beat frequency is the one corresponding to the minimum avoided crossing gap $\Delta_{\alpha,\beta}$ found in the subset of QDs with $L_y/L_x$ ~1, which correspond to a peak in the distribution of beat frequencies taken across the distribution of QD shapes. Note the distinction between this ensemble-level quantum beating derived FSS and the FSS reported in previous single-dot studies[11-14]. In the latter, FSS can be detected for any QDs with shape anisotropy and/or lattice distortion, whereas in our ensemble measurement here only those that survive size shape and orientational averaging are manifested as quantum beats.

By taking into account the experimental size and shape distributions of 7.9 nm QDs and assuming that QDs are completely randomly oriented in the film, we can reproduce the damping behavior of the TA kinetics and the asymmetric and broad shape of the FFT spectrum (Fig. 5e and 5f); See Supplementary Text 6 and Supplementary Figs 32-39 for modeling details. However, the simulated TA kinetics is more strongly damped, and the FFT spectrum is broader, than the experimental ones, suggesting that the QDs might be preferentially orientated with $L_z$ perpendicular to the substrate and/or that the shape distribution is overestimated in our TEM statistics.

Calculating the gap $\Delta_{\alpha,\beta}$ for strains corresponding to the measured lattice constants at different temperatures, we obtain a temperature-dependent $\Delta_{\alpha,\beta}$ curve that fits well to the experimental $\Delta_{FSS}$ of 7.9 nm QDs (Fig. 5g), using only literature values for the exciton and material parameters (Supplementary Table 10), and a single fit parameter, the deformation potential $U_d$ relating the measured strains to the lattice symmetry breaking through the Bloch factors $f_i$. If we assume $\zeta$ = -0.03 for QDs of other sizes at 80 K, we can also generate the size-dependent $\Delta_{FSS}$ with this model (Fig. 5h), although this is an oversimplified assumption because strains at a given temperature should be size-dependent. As a result, the modeled size-dependent curve deviates from experimental $\Delta_{FSS}$, especially for small QD sizes. In principle, we should perform XRD refinement for all QD sizes and extract their respective lattice distortion as input of the model, which, however, is hampered the strong XRD broadening in small-size QDs.

The quasi-cubic model clarifies why we can observe one (and only one) resolved quantum beat frequency in an ensemble measurement with broad distributions of QD shapes and orientations. As emphasized above, this beat frequency corresponds to an avoided-crossing "fine-structure gap" $\Delta_{\alpha,\beta}$ arising from the "misalignment" of the ***a, b*** axes with respect to the $L_x$, $L_y$ edge directions (Fig. 5a). This previously overlooked feature of the orthorhombic $CsPbI_3$ perovskite QDs turns out to have important consequences on their bright-exciton FSS. Without this feature, quantum beats will be completely washed out at an ensemble-level due to the broad distribution of the FSS continuously from zero to a few meV induced by QD shape and size heterogeneity.

Moreover, because it is associated with an intrinsic lattice symmetry breaking, the fine-structure gap can be quantitatively controlled through temperature.

**Reconciliation with previous studies.** The temperature-dependent TA quantum beats are not unique to $CsPbI_3$ QDs but rather can be extended to $CsPbBr_3$ and organic-inorganic hybrid $FAPbBr_3$ (FA: formamidinium) perovskite QDs (Supplementary Figs 40, 41). At 80 K, quantum beats are observed for both $CsPbBr_3$ and $FAPbBr_3$ QDs, but the oscillations are less than one-cycle, likely because the lattice distortion (and hence FSS) is weaker in $CsPbBr_3$ and $FAPbBr_3$, and/or the dephasing is faster, than $CsPbI_3$. These quantum beats disappear at 300 K, a behavior qualitatively similar to $CsPbI_3$ QDs.

It is also important to reconcile the current results with previous reports.[23,28,30] As mentioned above, in these studies the circularly-polarized TA kinetics were interpreted as spin relaxation dynamics of the bright $|1,\pm1>$ excitons. This interpretation, however, is valid only for QDs with ideal cubic lattice and cubic shape (i.e., no lattice distortion or shape anisotropy). Our current study indicates that, even in the absence of shape anisotropy, because of lattice distortion, circularly-polarized photons should excite coherent superpositions of bright excitons rather than populations of $|1,\pm1>$ excitons. Therefore, the circularly-polarized TA kinetics mostly reflects a dephasing process. Quantum beats are not observed in refs 23 and 30 because all the experiments therein were performed at room temperature, for which the FSS is too weak to induce observable beating. Ref 28 performed low-temperature measurements, but the sample is bulk-like $CsPbI_3$ nanocrystals for which exchange splitting is also too weak. Collectively, these observations are fully consistent with corresponding results here, but their interpretations need to be revised.

**Implications for quantum information science.** As introduced at the beginning, bright-exciton FSS is of close relevance to QIS. Therefore, the lattice-distortion dependent FSS in $CsPbI_3$ QDs has many important consequences for their applications in QIS. It strongly complicates their application for entangled photon-pairs. Nevertheless, in principle, along certain directions it is still possible to minimize FSS for energy-indistinguishable photon-pairs (see $\alpha$ and $C$ excitons, or $\beta$ and $C$ excitons, in Fig. 5c). On the other hand, however, the polarization-entangled FSS excitons allow performing a two-qubit conditional rotation gate (CROT),[53] which explicitly relies on the energy difference between the ground-to-exciton and exciton-to-biexciton transitions. Further, we envision that a sizable FSS can be harnessed for coherence control. In principle, by introducing another circularly-polarized off-resonance pulse ("tipping" pulse) between the pump and probe pulses, one can coherently manipulate the exciton coherence, through the optical Stark effect,[54] on a Bloch sphere with $\alpha$ and $\beta$ excitons as north and south poles. This magnetic-field-free, all-optical method of coherence control could be useful for high-speed quantum information processing.

## Conclusion

Our study identified strong, temperature-dependent bright exciton FSS in solution-processed $CsPbI_3$ QDs through ensemble quantum beating spectroscopy and rationalized it using a sophisticated exciton fine-structure model. It opens a new opportunity of facilely controlling FSS by leveraging the temperature-programmable lattice distortion of $CsPbI_3$ perovskite materials. The highly dynamic nature of lead halide lattice has been proposed to account for the "defect-tolerance" and long-lived hot carriers in these materials that are exciting for photovoltaics.[55] This study, on the other hand, reveals the potential of this property for quantum information technologies.

## Methods

**Chemicals and Synthesis of $CsPbI_3$ QDs.** Cesium carbonate ($Cs_2CO_3$, 99.9%), oleic acid (OA, 90%), oleylamine (OAm, 70%), Bis(2,4,4-trimethylpentyl) phosphinic acid (TMPPA, 90% ) 1-octadecene (ODE, 90%) were purchased from Sigma-Aldrich. Lead(II) iodide ($PbI_2$, 99.99%) and zinc iodide ($ZnI_2$, 99.99%) were purchased from Alfa Aesar. Hexane and methyl acetate were purchased from Energy Chemical. All chemicals were used directly without any further purification.

$CsPbI_3$ QDs were synthesized by using a hot injection approach.[23,24] In a typical synthesis, the Cs-oleate precursors solution was prepared by mixing 0.25 g $Cs_2CO_3$, 0.98 mL oleic acid (OA) and 9 mL 1-octadecene (ODE) in a 25 mL 3-neck flask and vacuum-dried for 1 hour at 120 ℃ using a Schlenk line. The mixture was heated under an nitrogen atmosphere to 150 ℃ for 10 minutes to dissolve all the $Cs_2CO_3$. The Cs-oleate precursor solution was kept at 100 ℃ to avoid precipitation. The precursor solution of Pb and I was prepared by dissolving $PbI_2$ (120 mg) and $ZnI_2$ (250 mg) in a mixture of 5 mL ODE, 2 mL oleylamine (OAm) and 2 mL Bis (2,4,4-trimethylpentyl) phosphinic acid (TMPPA) at 120 ℃ under vacuum for 1 hour in another 25 mL 3-neck flask. The mixture was further set to the reaction temperature, which was varied between 145-190 ℃ depending on the desired QD sizes, under nitrogen atmosphere. 0.4 mL of Cs precursor solution was swiftly injected. After 20 seconds, the reaction was quenched by cooling the flask in an ice bath. The product was centrifuged at 1290 rcf for 30 minutes to remove the unreacted salts, and the QDs dispersed in the supernatant were collected. Methyl acetate was used as the antisolvent. The precipitated QDs were used for further characterization after resuspending in hexane. Synthesis of $CsPbBr_3$ and $FAPbBr_3$ perovskite QDs also followed previous work.[24]

**Preparation of $CsPbI_3$ QD-films.** $CsPbI_3$ QD-films were prepared by a one-step spin-coating method in a glovebox. All QD samples underwent multiple rounds of purification by methyl acetate. Subsequently, $CsPbI_3$ QDs in hexane solutions were directly spin-coated (2,000 rpm., 40 seconds) onto pre-cleaned, oxygen-plasma-treated glass slides.

**Femtosecond TA experiments.** Femtosecond TA experiments were based on a Pharos laser (1030 nm, 100 kHz, 230 fs pulse-duration; Light Conversion) and Orpheus-HP optical parameter amplifier (OPA; Light Conversion). The 1030-nm output laser was split into two beams with 80/20 ratio. The 80% part was used to pump the OPA to generate a wavelength tunable pump beam. The 20% part was further split into two parts with 75/25 ratio. The 75% part was attenuated with a neutral density filter and focused into a 13-mm thick YAG crystal to generate a white light continuum used as the probe beam. The probe beam was focused with an Al parabolic reflector onto the sample. After the sample, the probe beam was collimated and then focused into a fiber-coupled spectrometer with CMOS cameras and detected at a frequency of 10 kHz. The pump pulses were chopped by a synchronized chopper at 5 kHz and the absorbance change was calculated with twenty adjacent probe pulses (ten pump-blocked and ten pump-unblocked). The intensity of the pump pulse used in the experiment was controlled by a variable neutral-density filter wheel. The delay between the pump and probe pulses was controlled by a motorized delay stage. Circular polarizations of the pump and probe beams were controlled by quarter waveplates. The samples were $CsPbI_3$ QDs films on glass substrates mounted into a liquid-nitrogen-cooled cryostat. The samples were stabilized for >20 minutes at each preset temperature in order to reach this temperature.

**Structure Characterization.** X-Ray Diffractometer (Empyrean) with a low-temperature reaction cell (TTK 450, Anton Paar GmbH) was used to perform the in situ XRD measurements using Cu Kα radiation operated at 40 kV and 40 mA, with a step width of 0.026° in the 2θ range from 10° to 50°. QDs were placed on glass substrates and then loaded into the reaction cell for each measurement. The samples were stabilized for ~10 minutes at each preset temperature in order to reach this temperature.

The TEM characterizations were performed on two machines at electron microscopy center of Dalian Institute of Chemical Physics. The QD solution samples were dropped onto ultrathin carbon TEM grids for analysis. For the regular size and morphology measurements in Fig. 1b and Supplementary Figure 1, we used JEOL JEM-2100 transmission electron microscopy performed at 200 kV accelerating voltage. For the varying-temperature measurements, we using Titan Themis G3 environmental transmission electron microscope (ETEM, Thermo Scientific Company) with a spherical-aberration (Cs) corrector for parallel imaging (CEOS GmbH) performed at 300 kV accelerating voltage. Note that the cryo-TEM images were obtained at desired temperatures under low dose conditions (~100 $e \cdot Å^{-2}\ s^{-1}$). At higher dose conditions, we observed amorphization of the perovskite QDs induced by the electron beams, especially at low temperatures. This observation is consistent with recent TEM studies on lead halide perovskite materials.[56]

**TA kinetics analysis.** The damped quantum beating signals were fitted phenomenologically with the following equation:

$$S(t) = A_0 \cdot \left[ F \cdot e^{-(t/T_\delta)^2} \cdot \cos(\frac{2\pi}{T_{FSS}} t + \varphi) + (1-F) \cdot e^{-t/T_{dec}} \right] \quad (3),$$

where $A_0$ is the total amplitude, $F$ is the fraction of QDs that exhibit a beating component, $T_\delta$ accounts for an inhomogeneous broadening that shortens the apparent beating lifetime, $T_{FSS}$ is the beating period (= $2\pi/\omega$), $\varphi$ is the initial phase, and $T_{dec}$ is a phenomenological decay time for the non-beating component. See Supplementary Text 7 and Supplementary Fig. 42 for detailed explanations of the choice of the empirical fitting equation.

The fitting parameters for quantum beats kinetics of varying-size QDs at 80 K are summarized in Supplementary Table 1, and the corresponding values of the fine structure splitting ($\Delta_{FSS} = \hbar\omega$) are calculated in Supplementary Table 2. The fitting parameters for the quantum beats kinetics of three QD samples at varying temperatures are summarized in Supplementary Tables 3-6, and the values of the fine structure splitting ($\Delta_{FSS} = \hbar\omega$) are calculated in Supplementary Table 7.

In order to further corroborate our fitting results, we also performed fast Fourier transformation (FFT) on the time-domain kinetics (using OrginPro 2018) to directly obtain the oscillation frequencies. The results from kinetics fitting and FFT are compared in Supplementary Tables 2 and 7.

**XRD refinement.** Rietveld refinement of the XRD patterns was performed using the general structure analysis system (GSAS) package with EXPGUI interface and following the Rietveld refinement guidelines formulated by the International Union of Crystallography Commission on powder diffraction.[57] The backgrounds were first fitted using the Chebishev (type #1) function. The peak profiles were then fitted using a convolution of pseudo-Voigt function (type #3) and asymmetry function, together with the microstrain broadening model.[58-60] All variables were refined stepwise until the refinement converged to chi-squared ($\chi^2$) minimum.

**Effective-mass-approximation calculation.** Theoretical modeling was performed using the Mathematica, V12 programming environment, using custom written software for numerical integration, Hamiltonian diagonalization, and simulation of quantum beating results; see details in the Supplementary Text.

## Acknowledgments

We thank Dr. Yang Zhao, Dr. Wei Liu, Dr. Qike Jiang for TEM measurements, Dr. Chang Wang for XRD measurements, and Dr. Peng Guo for discussions on XRD refinement. K.W. acknowledged financial support from the the Ministry of Science and Technology of China (2018YFA0208703), the Chinese Academy of Sciences (YSBR-007), and Dalian Institute of Chemical Physics (DICP I201914). Theoretical calculations were supported by the Center for Hybrid Organic Inorganic Semiconductors for Energy (CHOISE) an Energy Frontier Research Center funded by the Office of Basic Energy Sciences, Office of Science within the US Department of Energy.

## Author contributions

K.W. and Y.H. conceived the idea and initiated the study. K.W. supervised the project. Y.H. synthesized the samples and measured their spectroscopy. W.L. made the XRD Rietveld refinement. X.L., Y.L., F.S. and F.Z. helped with experiments or data analysis. P.C.S. developed the theoretical model for exciton fine structure and transient absorption. K.W., Y.H. and P.C.S. wrote the manuscript with contributions from all authors.

**Competing interests:** Authors declare no competing interests.

**Data and materials availability:** All data is available in the main text or the supplementary materials and can be obtained upon reasonable request from K.W. (kwu@dicp.ac.cn).

**Code availability:** Custom software developed for theoretical modeling associated with this study is available for verification purposes upon reasonable request from P.C.S. (pcsercel@gmail.com).

**Supplementary Materials:** Supplementary Texts 1-7; Supplementary Figures 1-42**;** Supplementary Tables 1-10**;** SM References

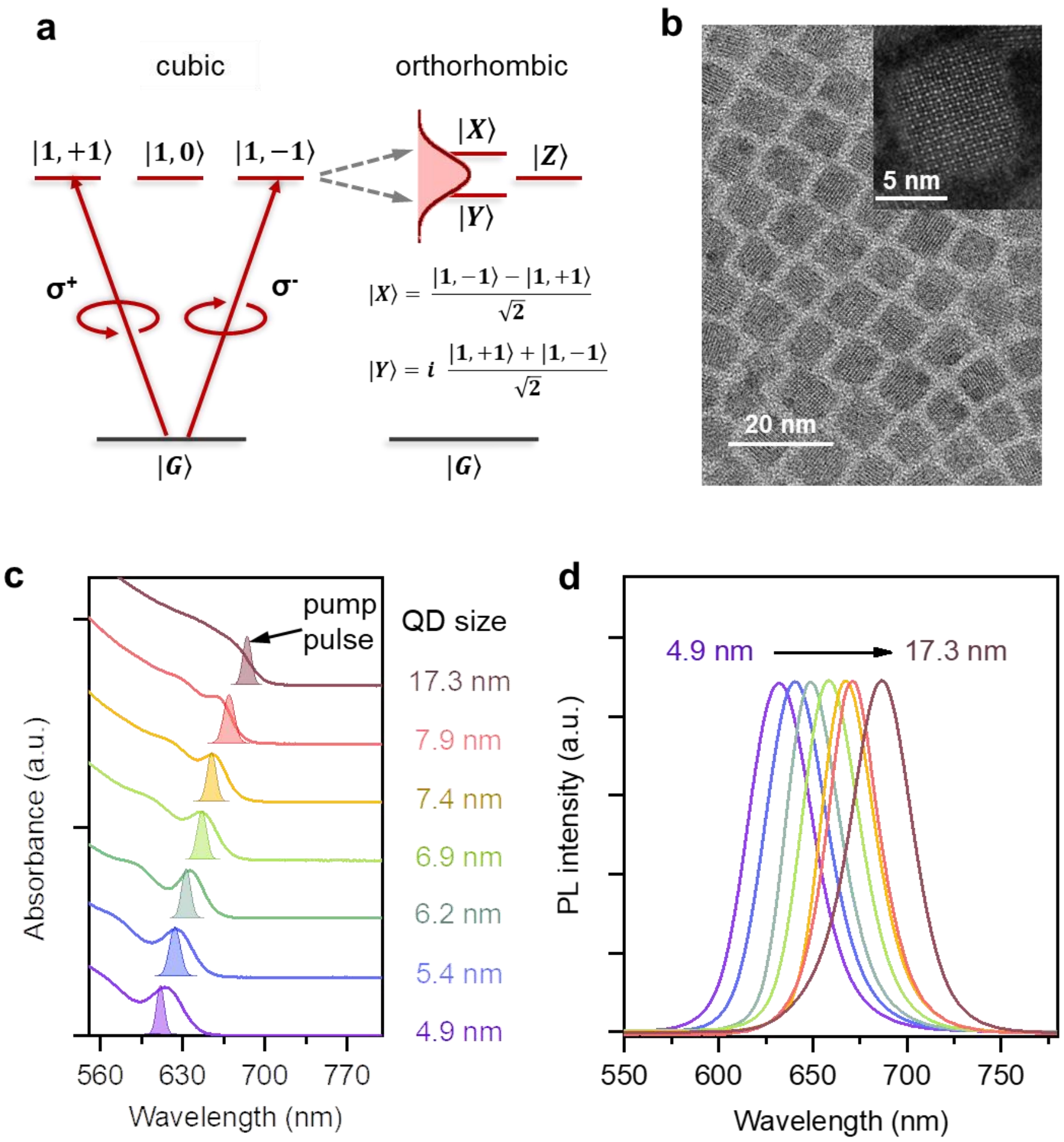

**Figure 1. Principle of fine structure splitting (FSS) and sample information.** (a) The bright triplet |1, ±1> exciton states in cubic symmetry transform into two new eigenstates |*X*> and |*Y*> in orthorhombic QDs. The splitting between |*X*> and |*Y*> is $\Delta_{FSS}$. For illustration we consider excitation using a circularly-polarized pulse directed along the *Z* axis with bandwidth larger than $\Delta_{FSS}$, which can create a coherent superposition of |*X*> and |*Y*>. (b) A representative TEM image of 7.9 nm QDs. Inset is a high-resolution dark-field image of a typical QD. (c, d) Absorption (c) and photoluminescence (d) spectra of varying-size QDs dispersed in hexane.

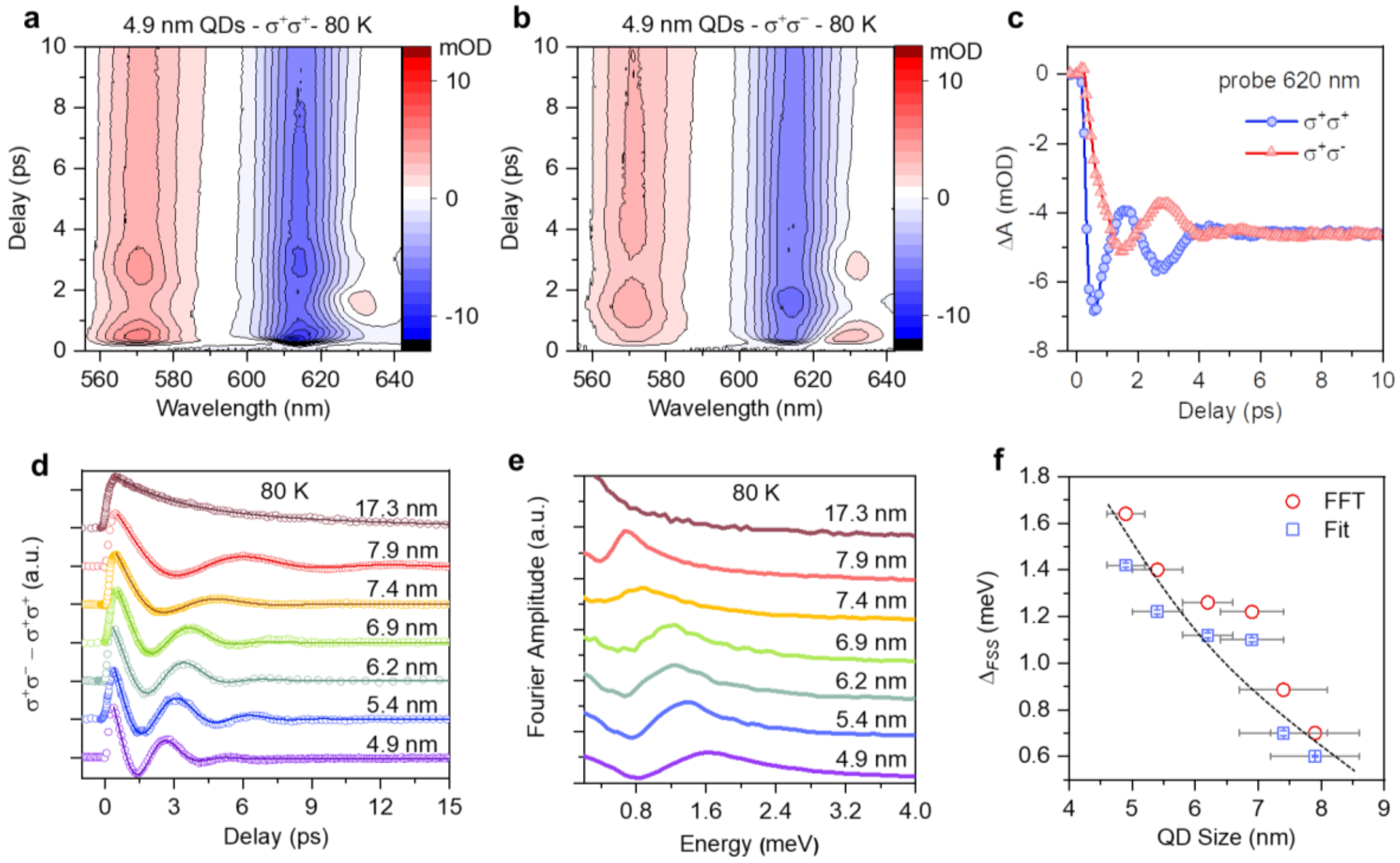

**Figure 2. Quantum beats and FSS in ensemble $CsPbI_3$ QD-films.** (a, b) 2D pseudo-color TA spectra of 4.9 nm QDs measured with (a) co- ($\sigma^+/\sigma^+$) and (b) counter-polarized ($\sigma^+/\sigma^-$) pump/probe beams at 80 K. (c) TA kinetics probed at 620 nm revealing opposite phases measured with co- (blue circles) and counter- (red triangles) polarized pump/probe beams. (d) Quantum beating kinetics measured for varying-size QDs at 80 K (colored circles) and their damped sinusoidal fits (colored lines). Signal sizes are in general a few mOD but are scaled for clarity. The signal after ~15 ps is at the zero base-line for each sample. (e) FFT of the kinetics of varying-size CsPbI3 QDs measured at 80 K. The amplitude at larger than ca. 5 meV is at the zero base-line for each sample. (f) Size-dependent FSS ($\Delta_{FSS}$) obtained from damped sinusoidal fits (blue squares) and fast Fourier transformation (red circles). Horizontal error bars are the standard deviations of the QD sizes. Vertical error bars are the fitting errors. The dashed line is a guide to the eye.

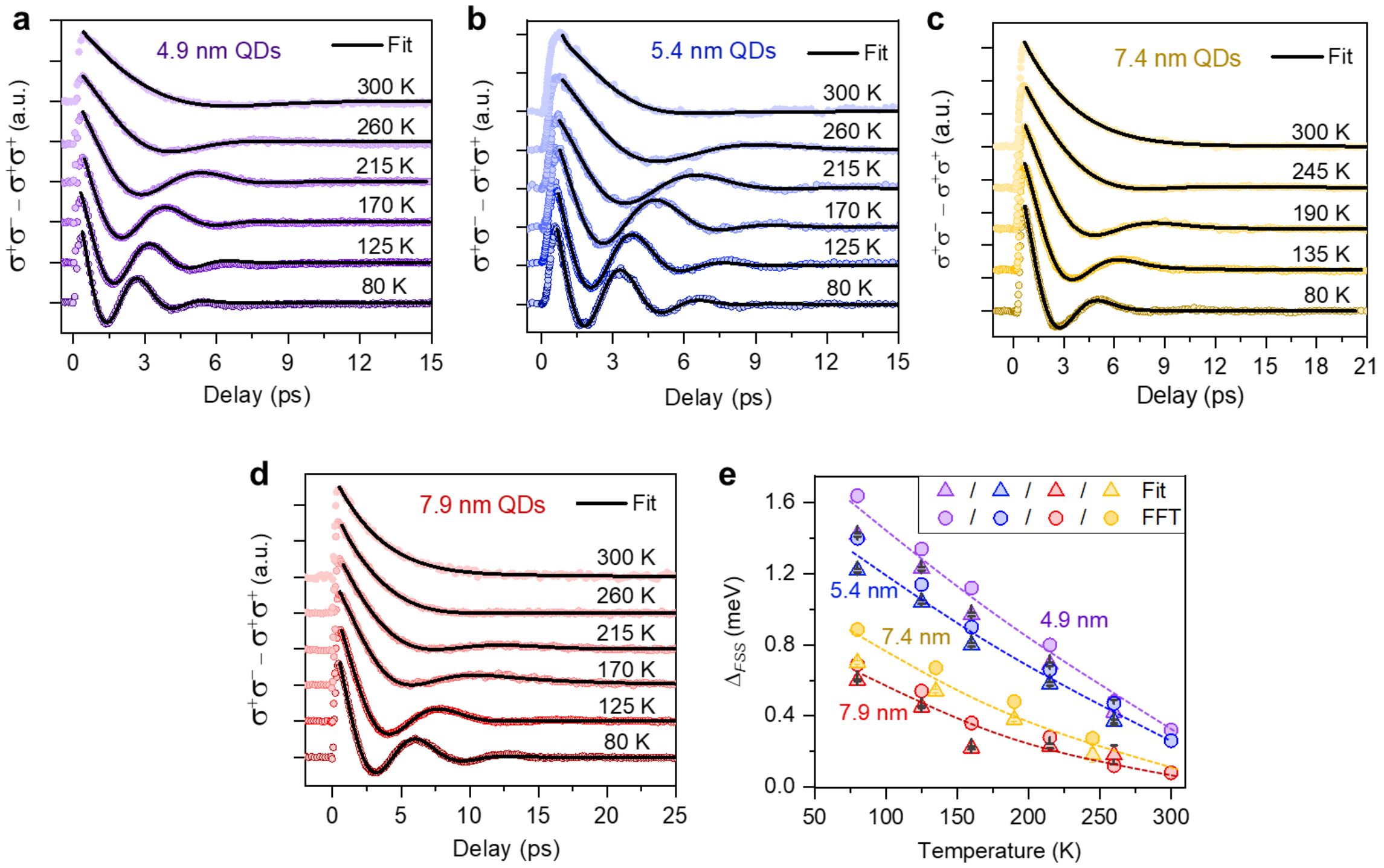

**Figure 3. Temperature-dependent FSS in $CsPbI_3$ QDs.** (a-d) Quantum beats observed for (a) 4.9 nm, (b) 5.4 nm, (c) 7.4 nm, and (d) 7.9 nm QDs at varying temperatures. Data points are show as circles while the lines represent damped sinusoidal fits. Signal sizes are in general a few mOD but are scaled for clarity. The signal after ~15 ps is at the zero base-line for each sample. (e) Temperature-dependent $\Delta_{FSS}$ obtained from damped sinusoidal fits (triangles) and from FFT (circles) for 4.9 nm (purple), 5.4 nm (blue), 7.4 nm (yellow), and 7.9 nm (red) QDs. The dashed lines are guides to the eye. Vertical error bars are the fitting errors.

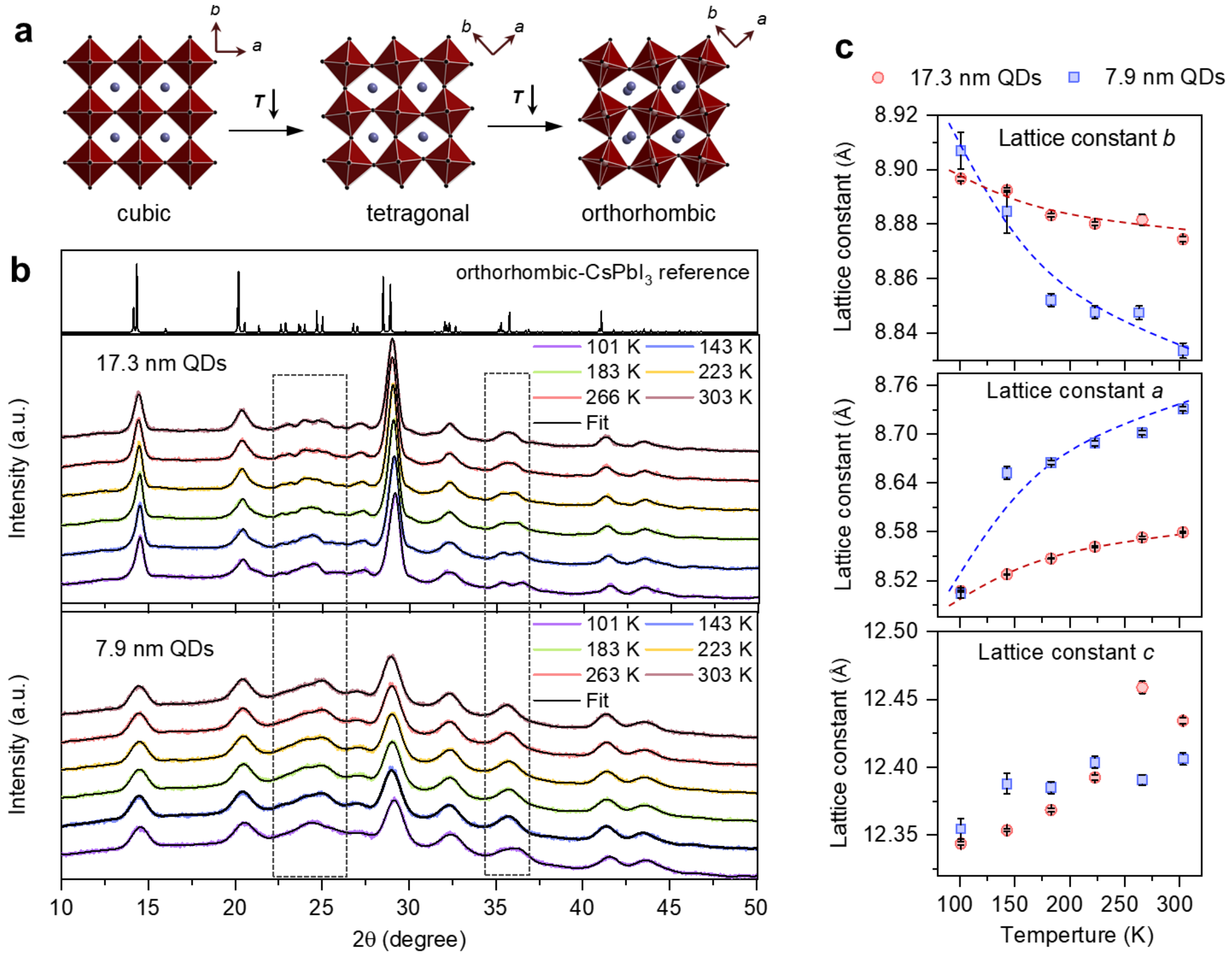

**Figure 4. Temperature-dependent lattice distortion in $CsPbI_3$ QDs.** (a) Temperature-induced phase changes in $CsPbI_3$ perovskites. The $PbI_6$ octahedras are shown in red while Cs atoms are shown with gray spheres. (b) XRD patterns of 17.3 nm (middle) and 7.9 nm (bottom) QDs measured at varying temperatures, in comparison to the reference of bulk orthorhombic-$CsPbI_3$ (top). The dashed squares highlight subtle changes in XRD patterns. The black solid lines are Rietveld-refined patterns. (c) Temperature-dependent lattice parameters *b* (top), *a* (middle) and *c* (bottom) obtained from refinement of 17.3 nm (red circles) and 7.9 nm (blue squares) QDs. Vertical error bars are the refinement errors. The dashed lines are guides to the eye highlighting the increase of *b* and decrease of *a* with decreasing temperature. These lattice constant changes correspond to a primarily orthorhombic strain relative to the cubic phase.

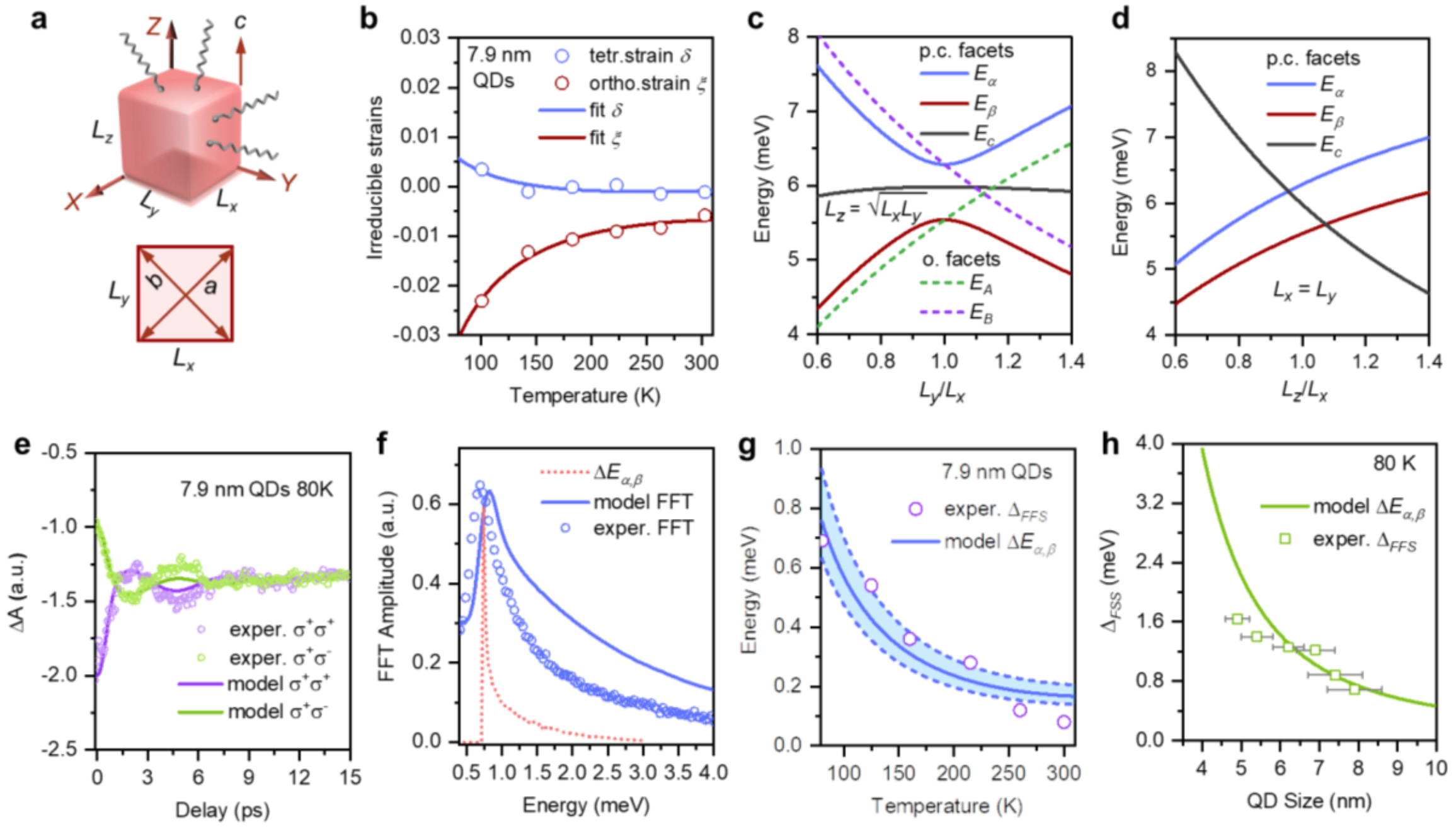

**Figure 5. Quasi-cubic model calculation and "fine-structure gap".** (a) A cuboid-shaped QD with edge lengths $L_x$, $L_y$ and $L_z$ considered in the calculation (top). While the orthorhombic primitive lattice vector $\boldsymbol{c}$ is aligned with $L_z$, $\boldsymbol{a}$ and $\boldsymbol{b}$ point towards the corners rather than the edges in the $L_x$-$L_y$ plane (bottom). (b) Tetragonal and orthorhombic strain components ($\delta$ and $\zeta$, respectively) relative to the cubic phase in 7.9 nm QDs as a function of temperature, calculated from lattice constants in Fig. 4c. (c) Bright exciton energy levels versus the edge ratio $L_y/L_x$ at fixed $L_z = L_e$, where the effective edge length $L_e = (L_xL_yL_z)^{1/3} = 7.9$ nm. Energies calculated in the pseudocubic (p.c.) and orthorhombic (o.) facet models are shown in solid and dashed lines, respectively. The $C$ exciton energy is the same in both models. (d) Bright exciton energies versus the edge ratio $L_z/L_x$ ( $L_x = L_y$) in the pseudocubic (p.c.) facet model. (e) Simulated QD TA kinetics (solid lines) in comparison to experimental data (circles) for 7.9 nm QDs at 80 K. In the simulation, QDs are assumed to be completely randomly oriented in the film and their shape and size inhomogeneities have been explicitly accounted for. (f) Corresponding FFT spectrum (blue solid line) in comparison to experimental data (blue circles). The red dashed line represents the distribution of the energy difference between $\alpha$ and $\beta$ excitons calculated in panels (c) and (d). (g) Single-parameter fit (blue solid line) to measured temperature-dependent $\alpha$, $\beta$ exciton FSS (violet circles) for 7.9 nm QDs. The shaded area encompassed by the dashed line corresponds to the calculation results within one standard deviation of the $L_e$ distributions. (h) QD size-dependent FSS at 80 K (green squares) and its fit (green solid line) using the model assuming the same lattice strain for all QD sizes. Horizontal error bars reflect the standard deviations of the QD sizes. See main text and SM for details of the models.

Supplementary Materials for:

# Lattice distortion inducing exciton splitting and coherent quantum beating in $CsPbI_3$ perovskite quantum dots

Yaoyao Han[1,2], Wenfei Liang[1], Xuyang Lin[1,2], Yulu Li[1], Fengke Sun[1,2], Fan Zhang[2,3], Peter C. Sercel[4]*, Kaifeng Wu[1]*

[1] State Key Laboratory of Molecular Reaction Dynamics, Dalian Institute of Chemical Physics, Chinese Academy of Sciences, Dalian, Liaoning 116023, China.

[2] University of Chinese Academy of Sciences, Beijing 100049, China.

[3] Dalian National Laboratory for Clean Energy, Dalian Institute of Chemical Physics, Chinese Academy of Sciences, Dalian, Liaoning 116023, China

[4] Center for Hybrid Organic Inorganic Semiconductors for Energy, Golden, Colorado 80401, United States

* Correspondence to: pcsercel@gmail.com

* Correspondence to: kwu@dicp.ac.cn

# Contents

**List of Figures.**

**List of Tables.**

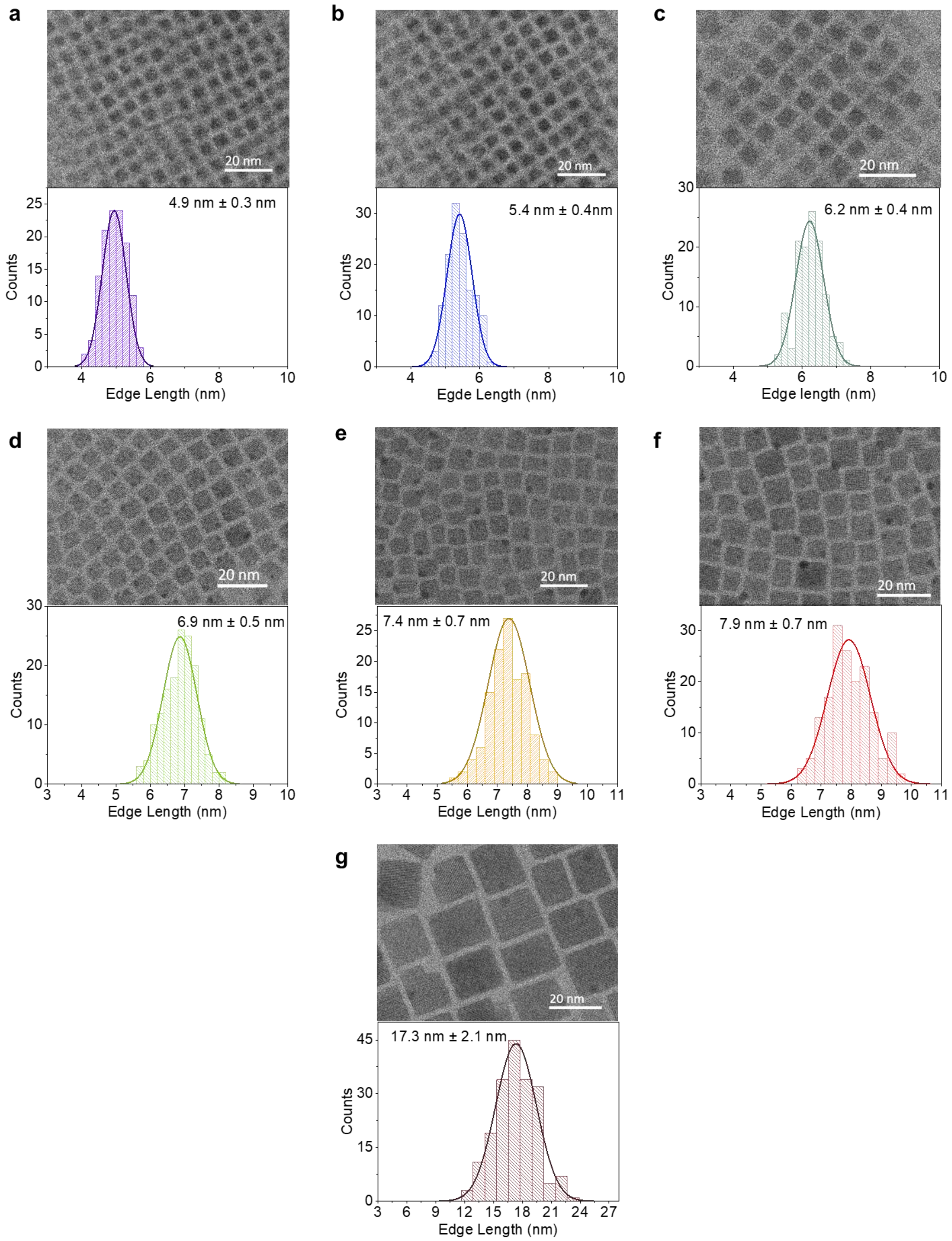

Supplementary Figure 1. TEM images of varying-size $CsPbI_3$ QDs.

The statistic histograms of the edge lengths of these cube-shaped QDs are shown at the bottom.

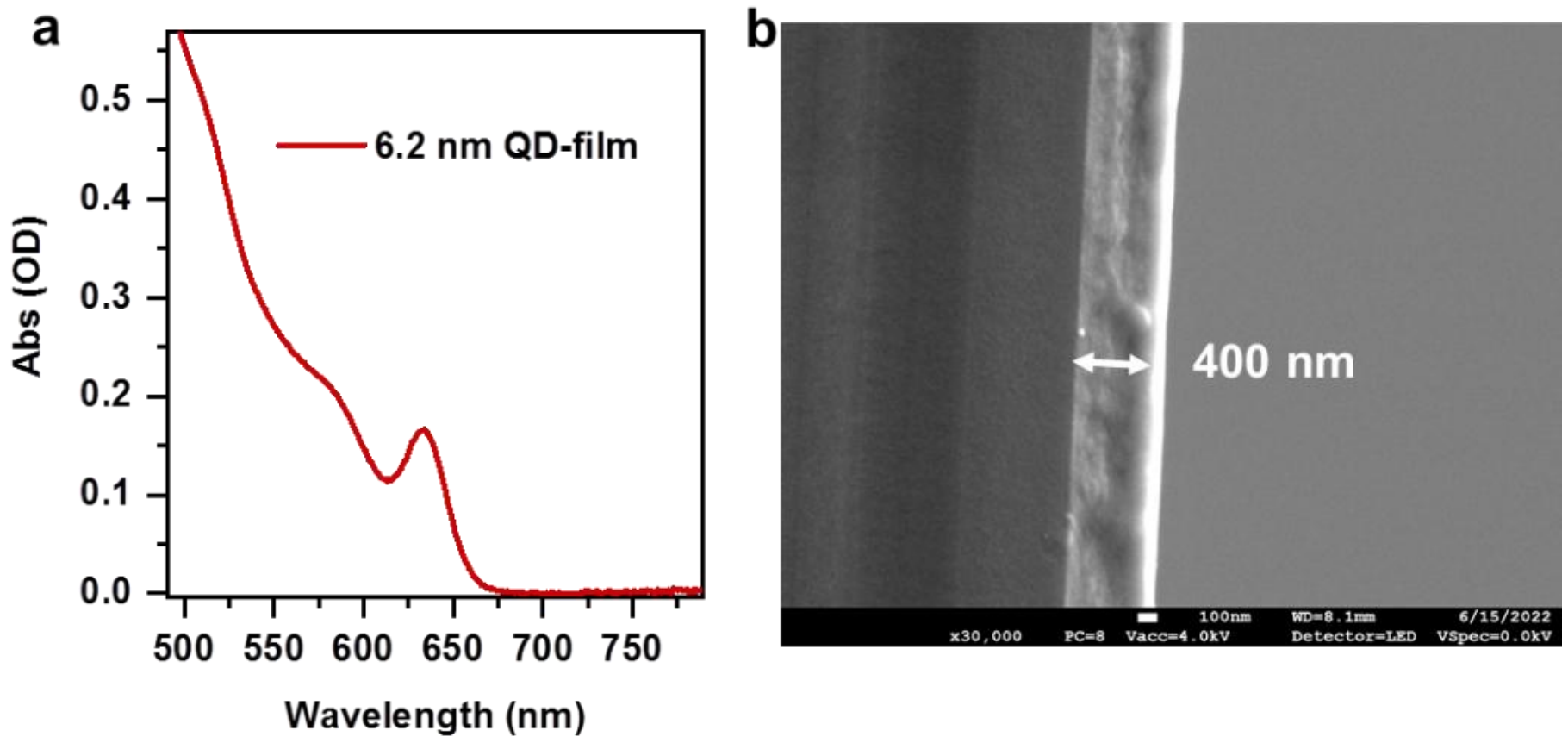

Supplementary Figure 2. Characterization of a typical film of $CsPbI_3$ QDs.

(a) Absorption spectrum of a typical film spin-coated from 6.2 nm QDs. The film has an optical density of ~0.17 at the exciton peak. (b) Scanning electron microscope (SEM) image of the film, with a thickness of ca. 400 nm.

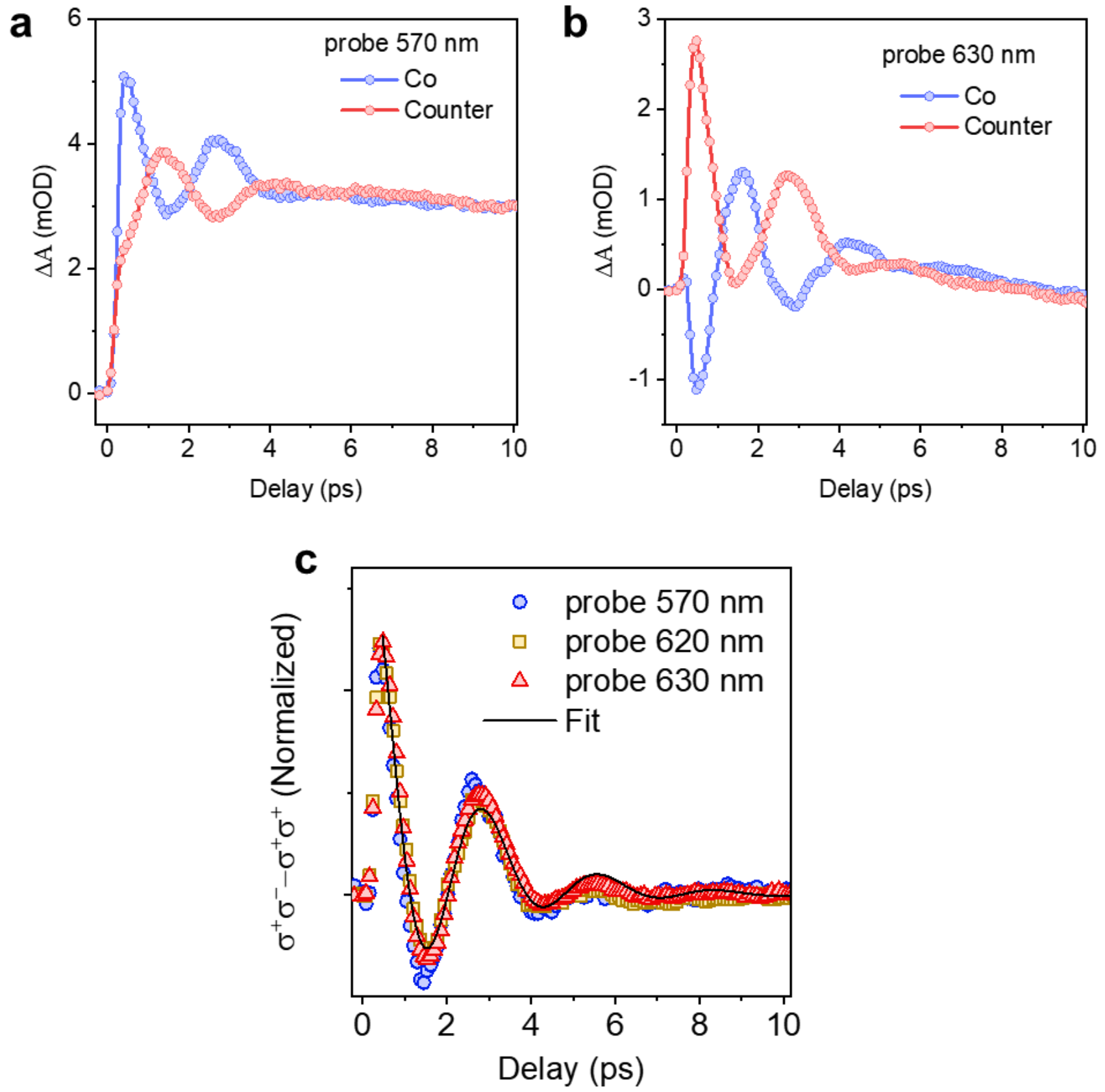

Supplementary Figure 3. TA kinetics at different wavelengths for 4.9 nm $CsPbI_3$ QDs at 80 K.

TA kinetics probed at (a) 570 nm and (b) 630 nm of 4.9 nm $CsPbI_3$ QDs revealing opposite phases measured with co- (blue) and counter-polarized (red) pump/probe beams. (c) Extracted quantum beating kinetics (difference between co- and counter-TA signals) at 570 nm (blue), 620 nm (yellow) and 630 nm (red) and their damped sinusoidal fit (black line). The signal after 10 ps is at the zero base-line.

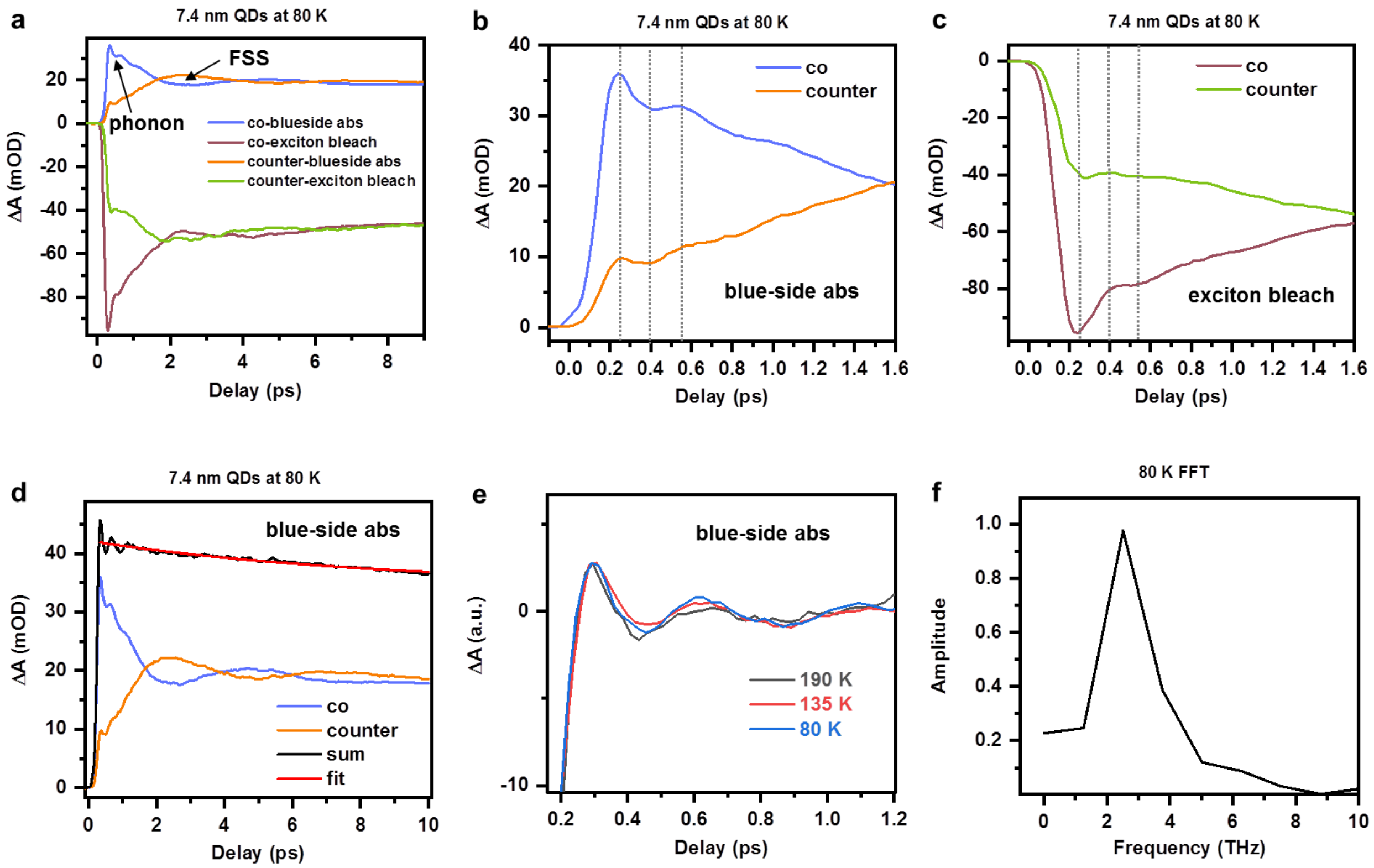

Supplementary Figure 4. Coherent phonon oscillations .

(a) Co and counter circular TA kinetics measured at a large TA signal size for 7.4 nm QDs at 80 K. Kinetics are probed at both the exciton bleach and the blue-side induced absorption positions. Oscillations corresponding to FSS and coherent phonons are distinguishable by their different frequencies. (b,c) Zoom-in of the coherent phonons oscillations probed at (b) blue-side absorption and (c) exciton bleach. The phases are the same for co and counter kinetics at a certain wavelength (as indicated by dashed lines), but are nearly opposite between (b) and (c). (d) Taking the sum of co and counter kinetics at the blue-side absorption, to extract pure coherent phonon oscillations. (e) Coherent phonon oscillations probed at three different temperatures, which are barely temperature-dependent in this range. (f) FFT of kinetics at 80 K, resulting in a coherent phonon frequency of ca. 3 THz.

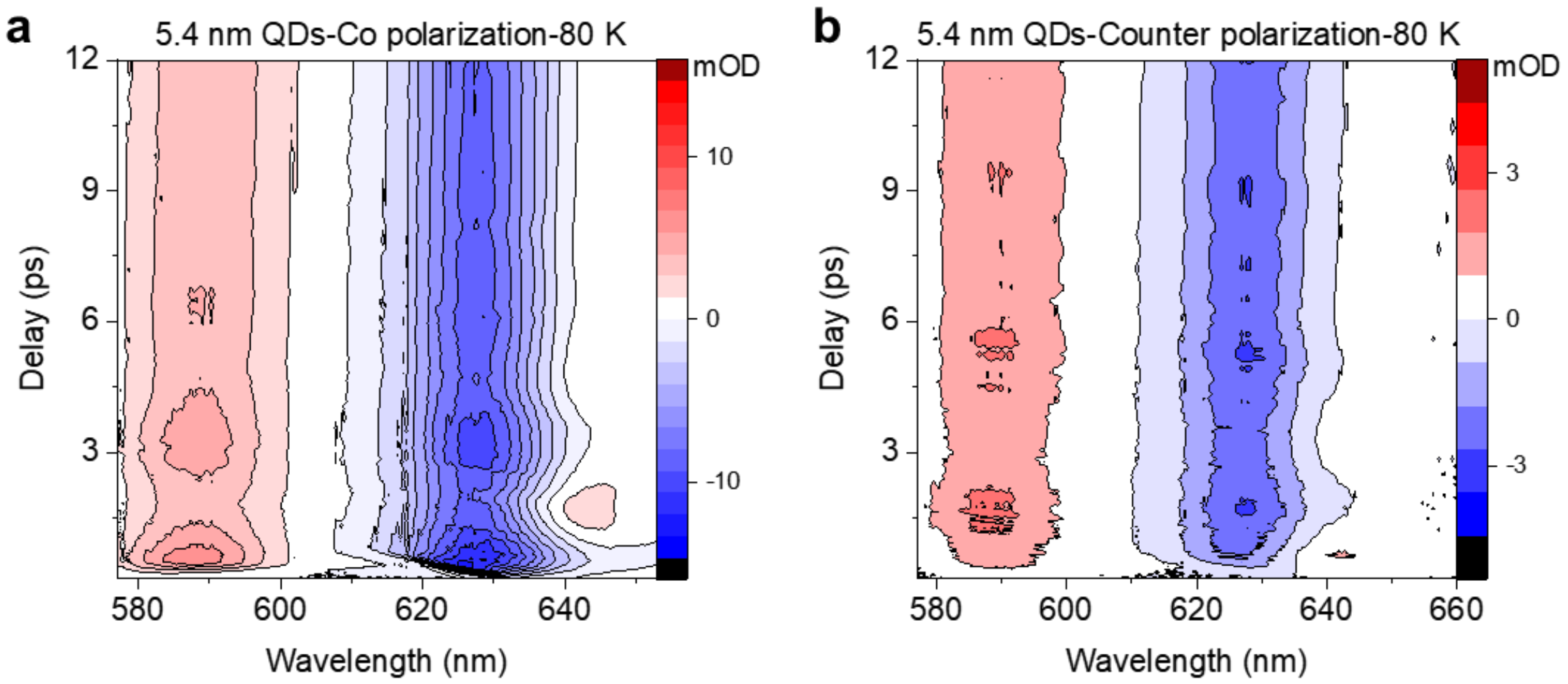

Supplementary Figure 5. 2D pseudo-color TA spectra of 5.4 nm $CsPbI_3$ QDs at 80 K.

2D pseudo-color TA spectra of 5.4 nm $CsPbI_3$ QDs measured with (a) co- ($\sigma^+/\sigma^+$) and (b) counter-polarized ($\sigma^-/\sigma^+$) pump/probe beams at 80 K. Note the TA signal for co-polarization is stronger than for the counter-polarization. Because the QD film could be damaged under long-term excitation, after finishing measurement at co- or counter-polarization, we moved the excitation to another spot on the film after we changed polarization. The variation in the optical densities at different spots on the film could result in TA signal changes. Since this is not associated with any additional physics, we chose to normalize the co- and counter-polarization kinetics to their long-lived tail.

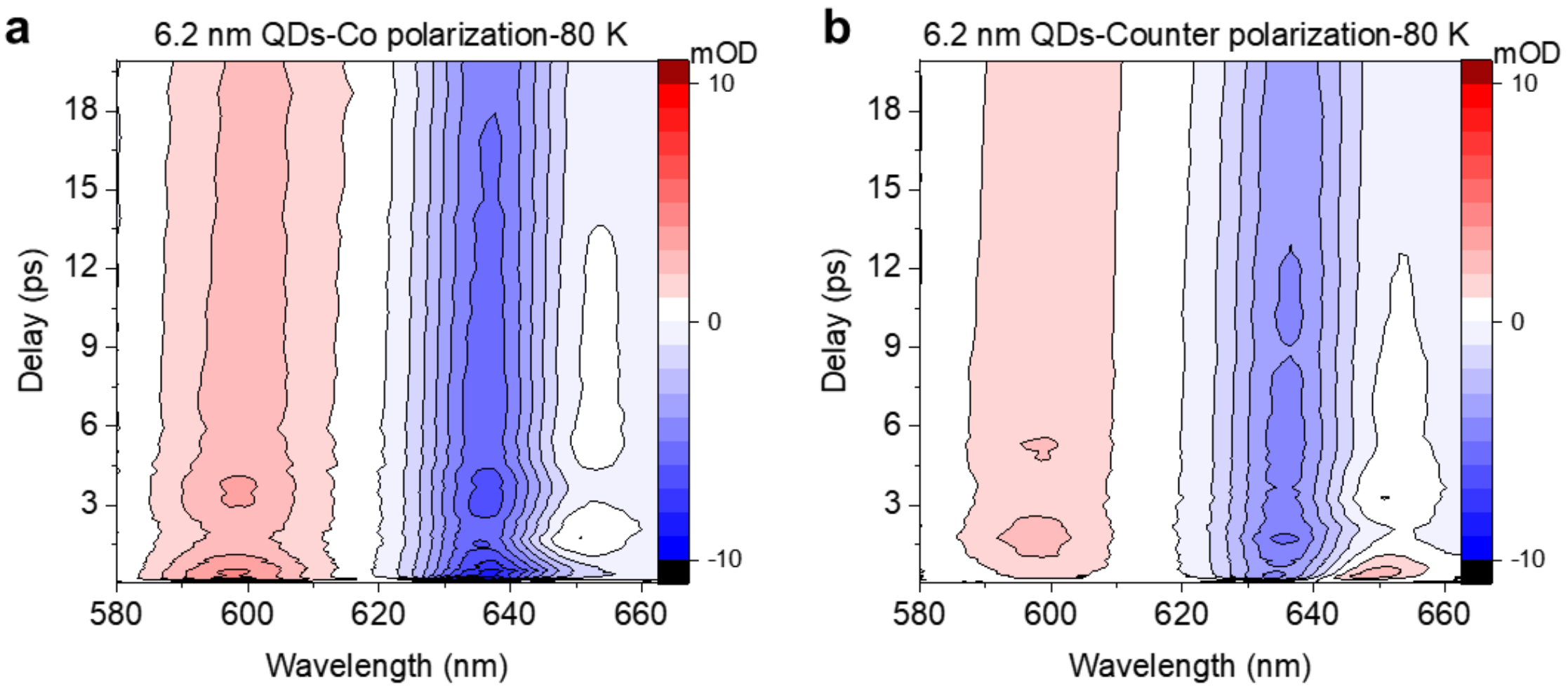

Supplementary Figure 6. 2D pseudo-color TA spectra of 6.2 nm $CsPbI_3$ QDs at 80 K.

2D pseudo-color TA spectra of 6.2 nm $CsPbI_3$ QDs measured with (a) co- ($\sigma^+/\sigma^+$) and (b) counter-polarized ($\sigma^-/\sigma^+$) pump/probe beams at 80 K.

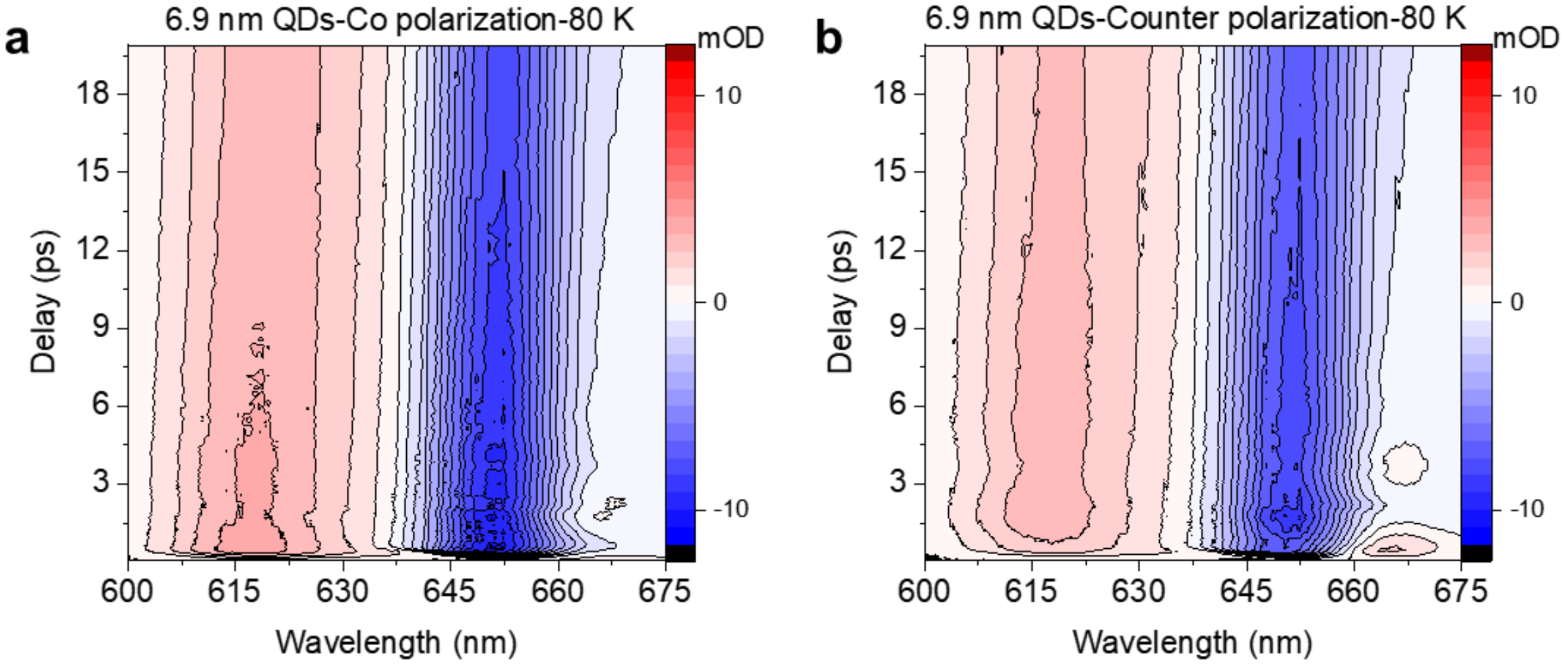

Supplementary Figure 7. 2D pseudo-color TA spectra of 6.9 nm CsPbI3 QDs at 80 K.

2D pseudo-color TA spectra of 6.9 nm CsPbI3 QDs measured with (a) co- (σ+/σ+) and (b) counter-polarized (σ-/σ+) pump/probe beams at 80 K.

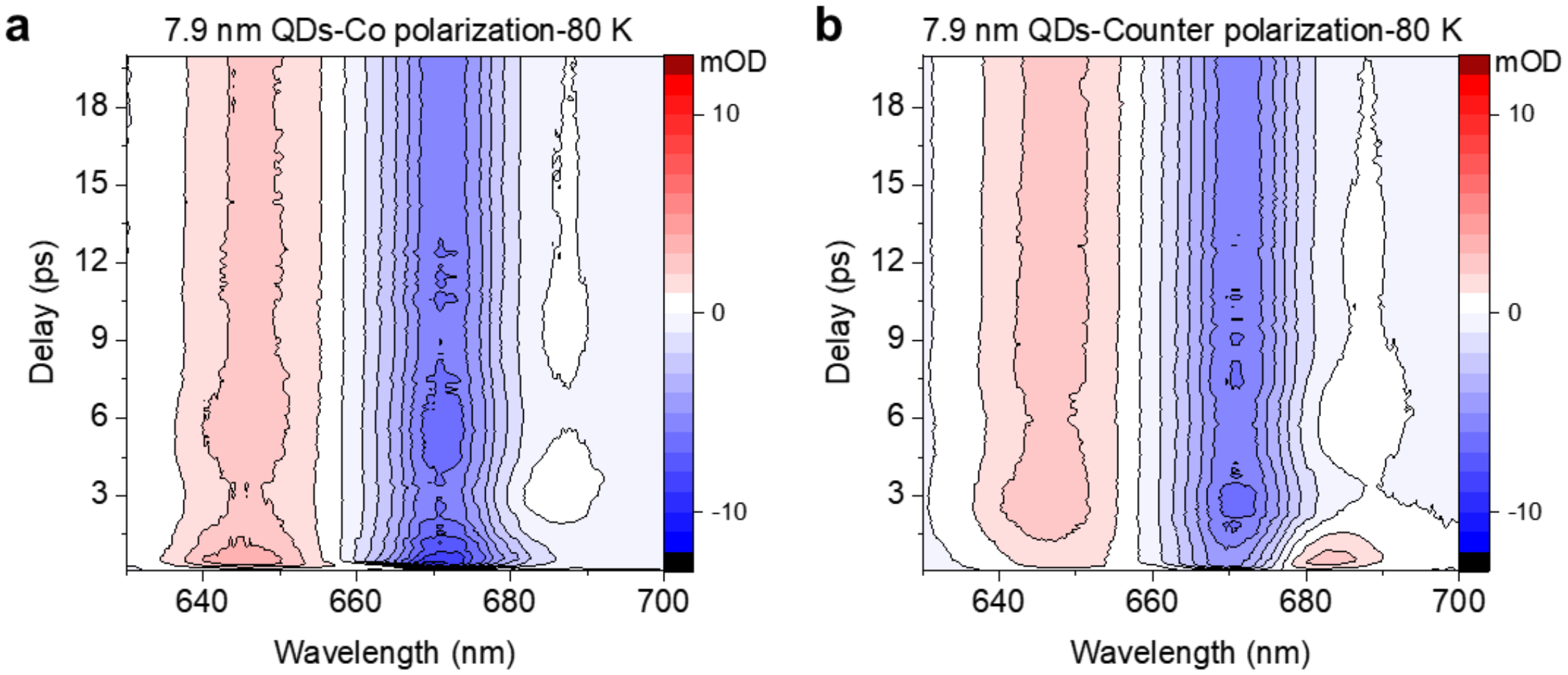

Supplementary Figure 8. 2D pseudo-color TA spectra of 7.9 nm $CsPbI_3$ QDs at 80 K.

2D pseudo-color TA spectra of 7.9 nm $CsPbI_3$ QDs measured with (a) co- ($\sigma^+/\sigma^+$) and (b) counter-polarized ($\sigma^-/\sigma^+$) pump/probe beams at 80 K.

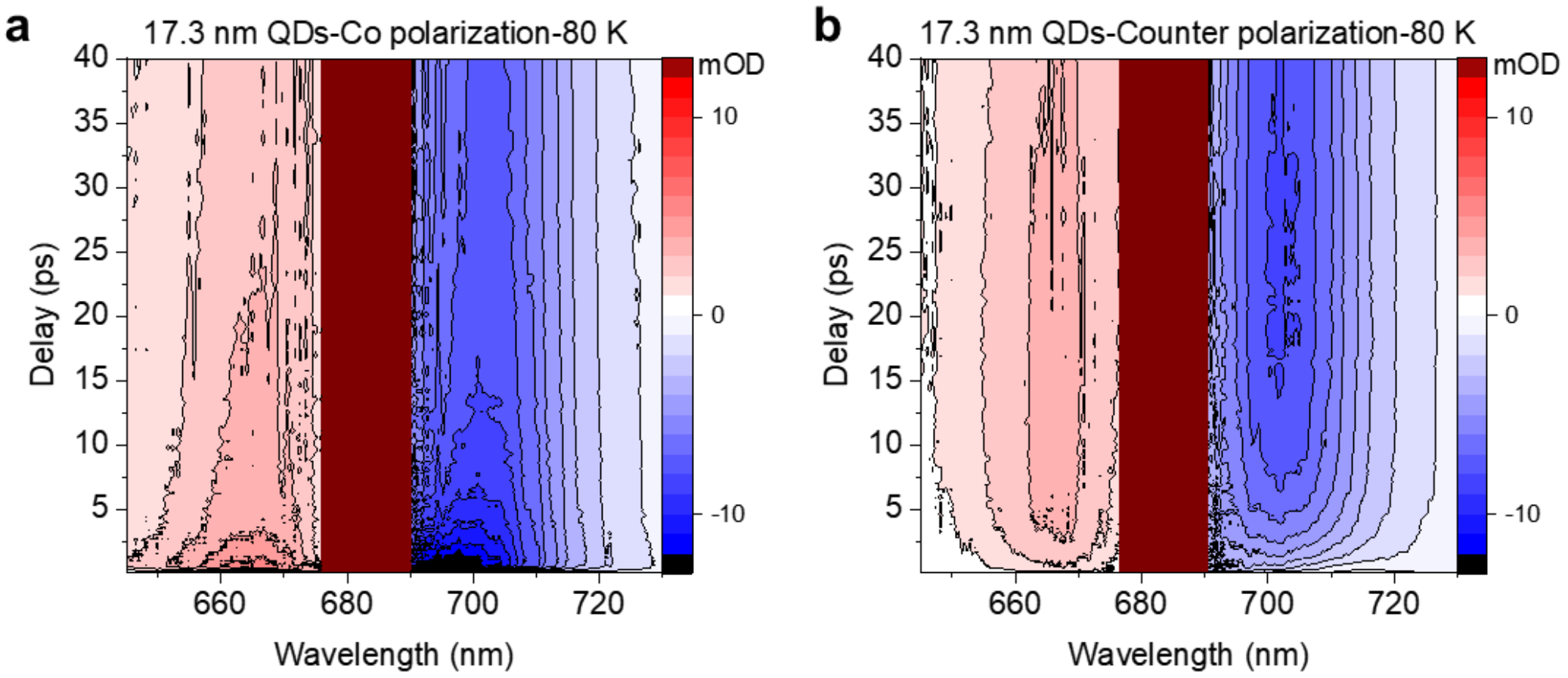

Supplementary Figure 9. 2D pseudo-color TA spectra of 17.3 nm $CsPbI_3$ QDs at 80 K.

2D pseudo-color TA spectra of 17.3 nm $CsPbI_3$ QDs measured with (a) co- ($\sigma^+/\sigma^+$) and (b) counter-polarized ($\sigma^-/\sigma^+$) pump/probe beams at 80 K.

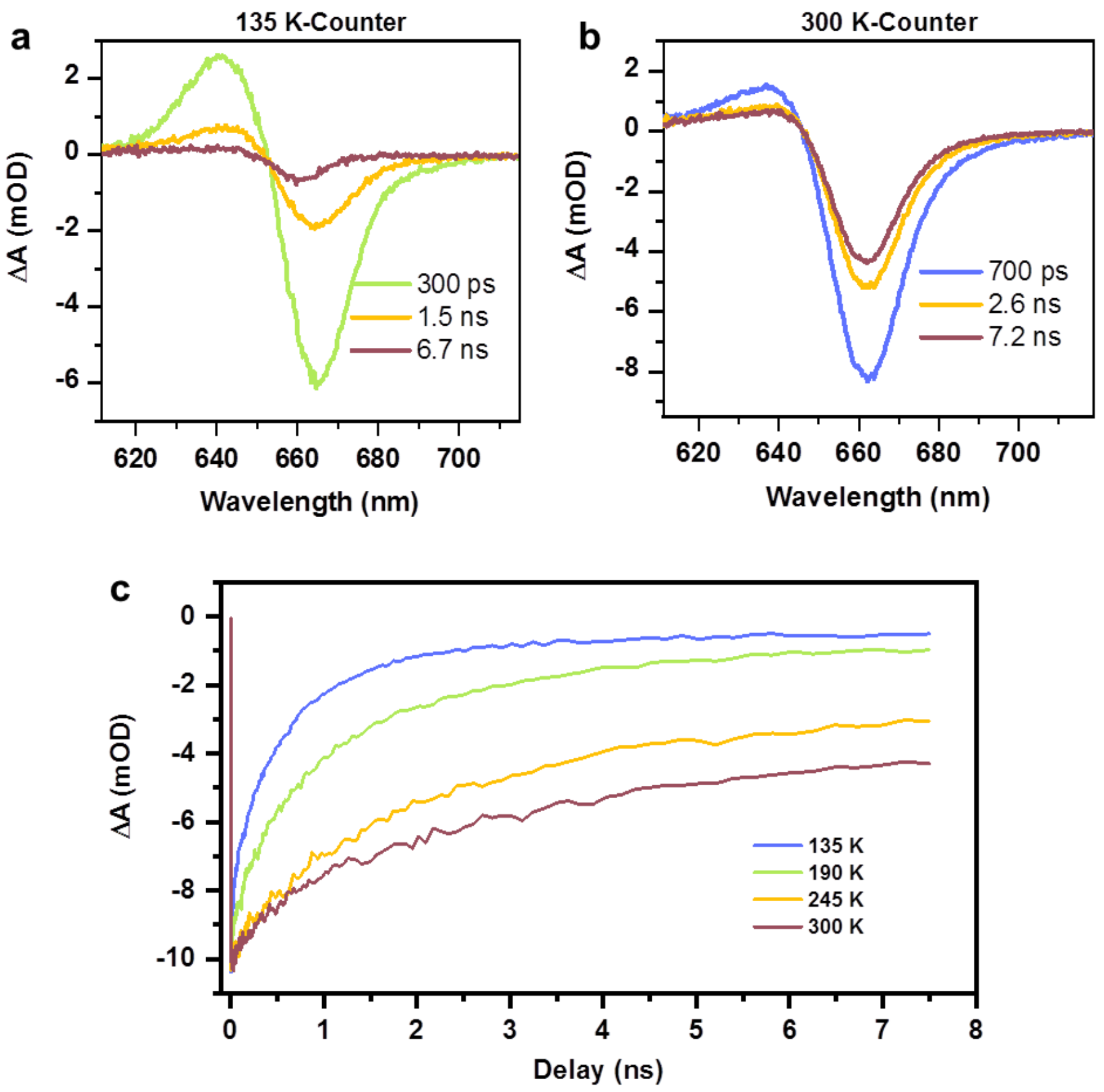

Supplementary Figure 10. TA kinetics on the ns timescale.

(a,b) Long-time scale TA spectra of 7.4 nm $CsPbI_3$ QDs measured at (a) 135 K and (b) 300 K, with a counter-polarized pump/probe configuration. The co-configuration gives essentially identical spectral at this long-time scale. (c) Temperature-dependent exciton bleach lifetime on the long-time scale.

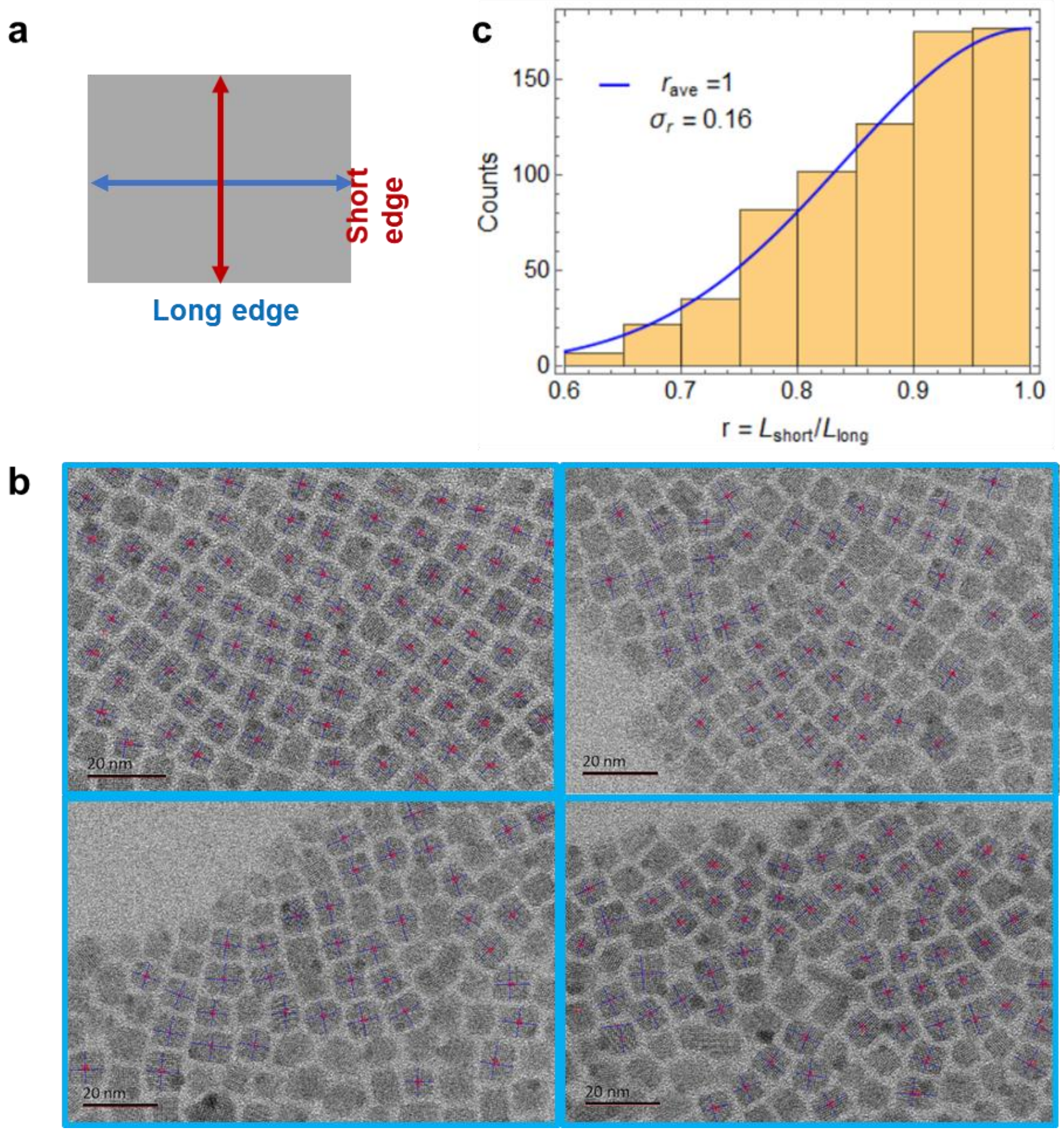

Supplementary Figure 11. Statistics on the aspect ratio of 7.9 nm QDs.

(a) A scheme shows the long and short edges of a cuboid lying on a substrate. (b) Some representative TEM images of 7.9 nm QDs for the long and short edge measurements. (c) A histogram of the ratio, $r$, of the short and long edge lengths obtained from the TEM images of 703 individual QDs. The average ratio $r$ is 1 with a standard deviation of 0.16. A prefect cube shape would be characterized by $r = 1$.

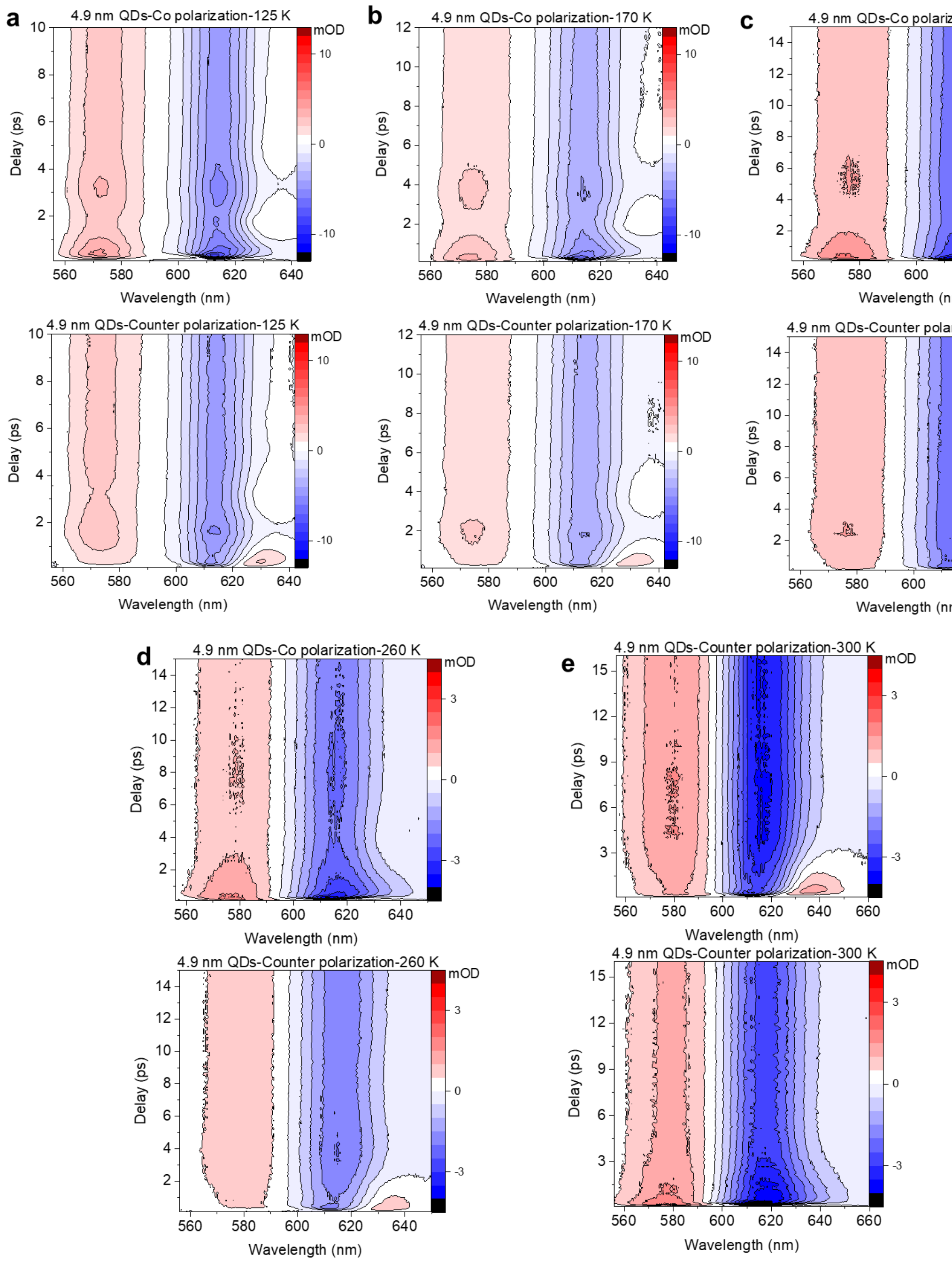

Supplementary Figure 12. 2D pseudo-color TA spectra of 4.9 nm $CsPbI_3$ QDs measured at varying temperatures.

2D pseudo-color TA spectra of 4.9 nm $CsPbI_3$ QDs measured with co- ($\sigma^+/\sigma^+$;top) and counter-polarized ($\sigma^-/\sigma^+$; bottom) pump/probe beams at (a) 125 K, (b) 170 K, (c) 215 K, (d) 260 K and (e) 300 K.

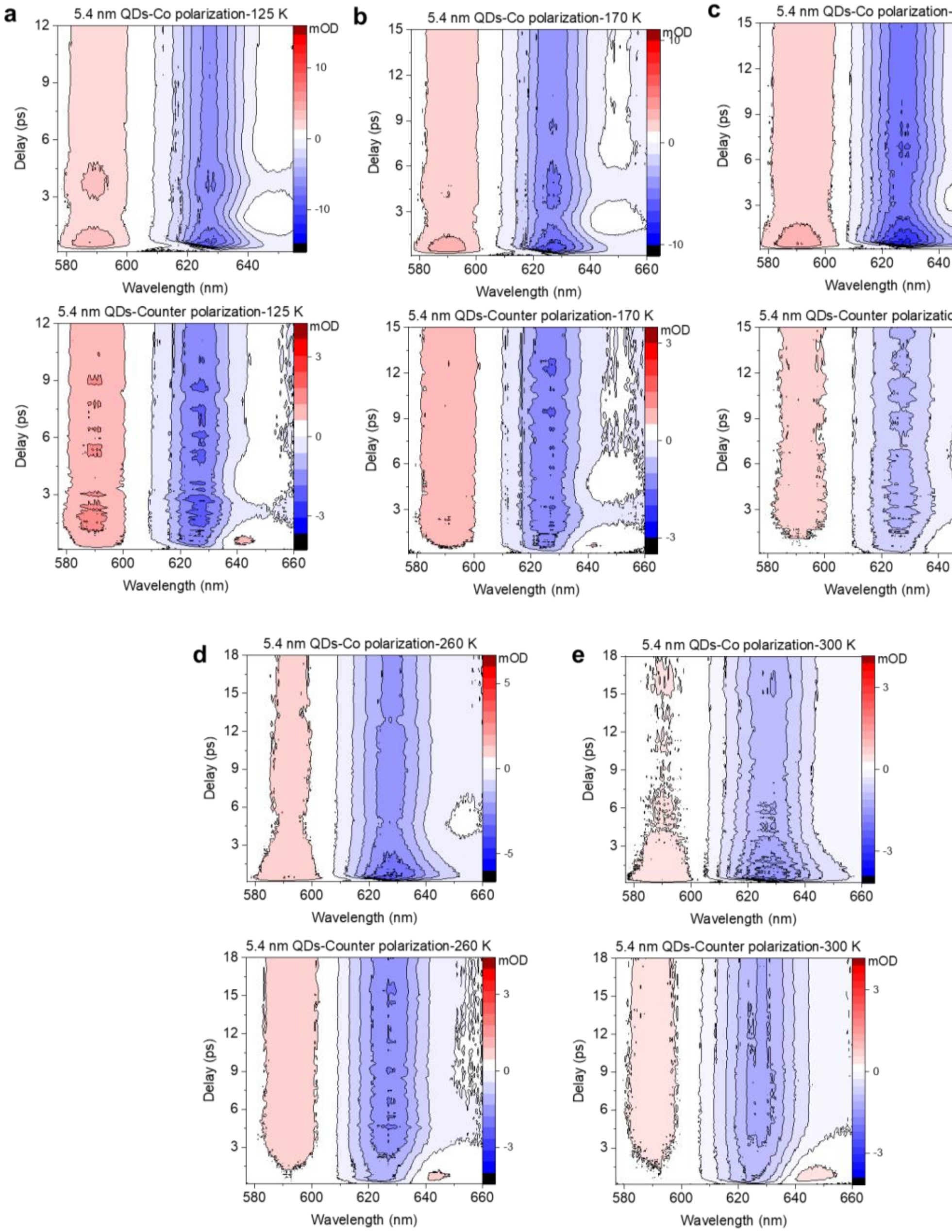

Supplementary Figure 13. 2D pseudo-color TA spectra of 5.4 nm $CsPbI_3$ QDs measured at varying temperatures.

2D pseudo-color TA spectra of 5.4 nm $CsPbI_3$ QDs measured with co- ($\sigma^+/\sigma^+$; top) and counter-polarized ($\sigma^-/\sigma^+$; bottom) pump/probe beams at (a) 125 K, (b) 170 K, (c) 215 K, (d) 260 K and (e) 300 K. Note at some temperatures the TA signal of co-polarization is stronger than the counter-polarization. Because the QD film could be damaged under long-term excitation, after finishing measurement at co- or counter-polarization, we moved the excitation to another spot on the film after we changed polarization. The variation in the optical densities at different spots on the film could result in TA signal changes. Since this is not associated with any additional physics, we chose to normalize the co- and counter-polarization kinetics to their long-lived tail.

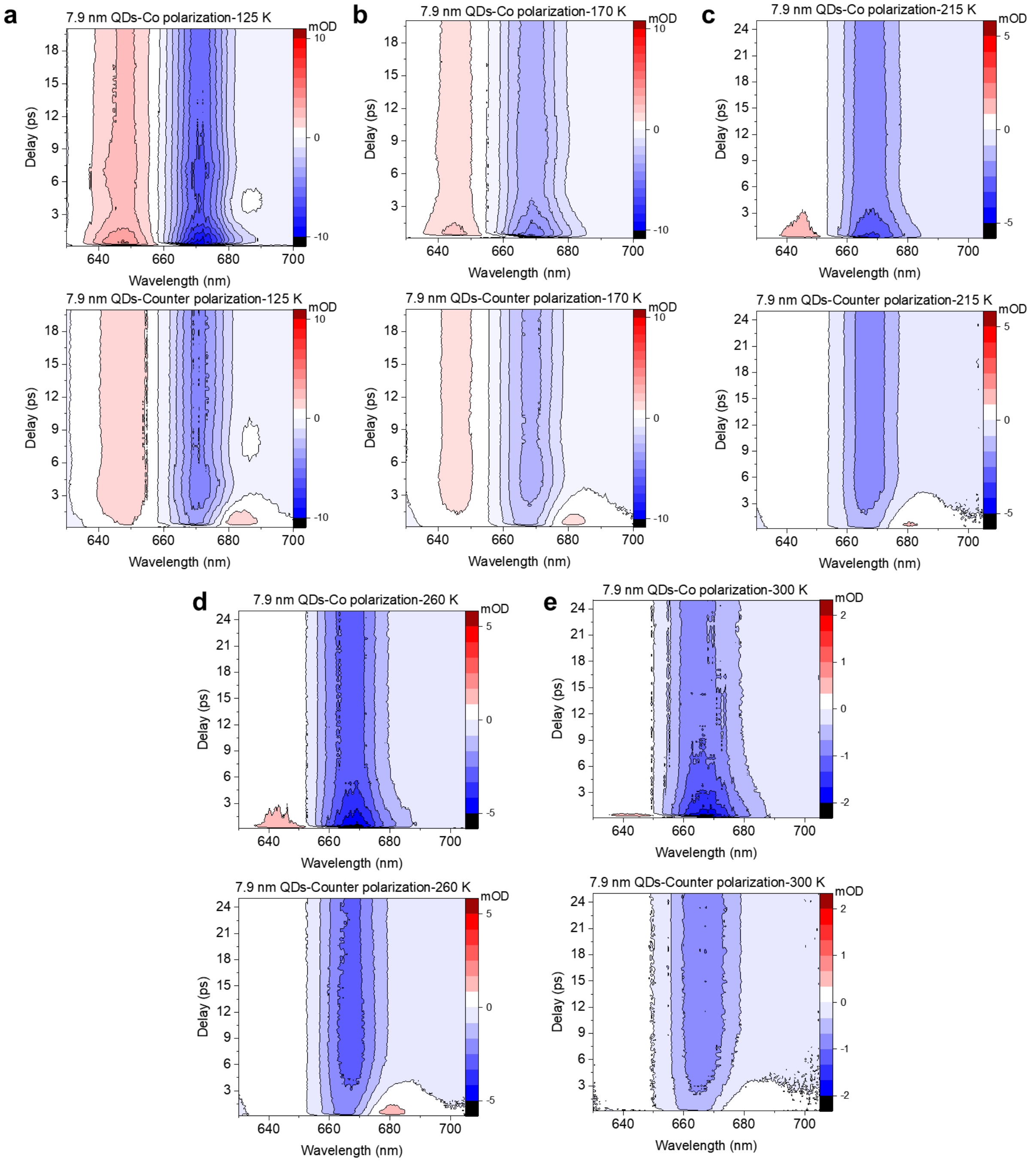

Supplementary Figure 14. 2D pseudo-color TA spectra of 7.9 nm $CsPbI_3$ QDs measured at varying temperatures.

2D pseudo-color TA spectra of 7.9 nm $CsPbI_3$ QDs measured with co- ($\sigma^+/\sigma^+$; top) and counter-polarized ($\sigma^-/\sigma^+$; bottom) pump/probe beams at (a) 125 K, (b) 170 K, (c) 215 K, (d) 260 K and (e) 300 K.

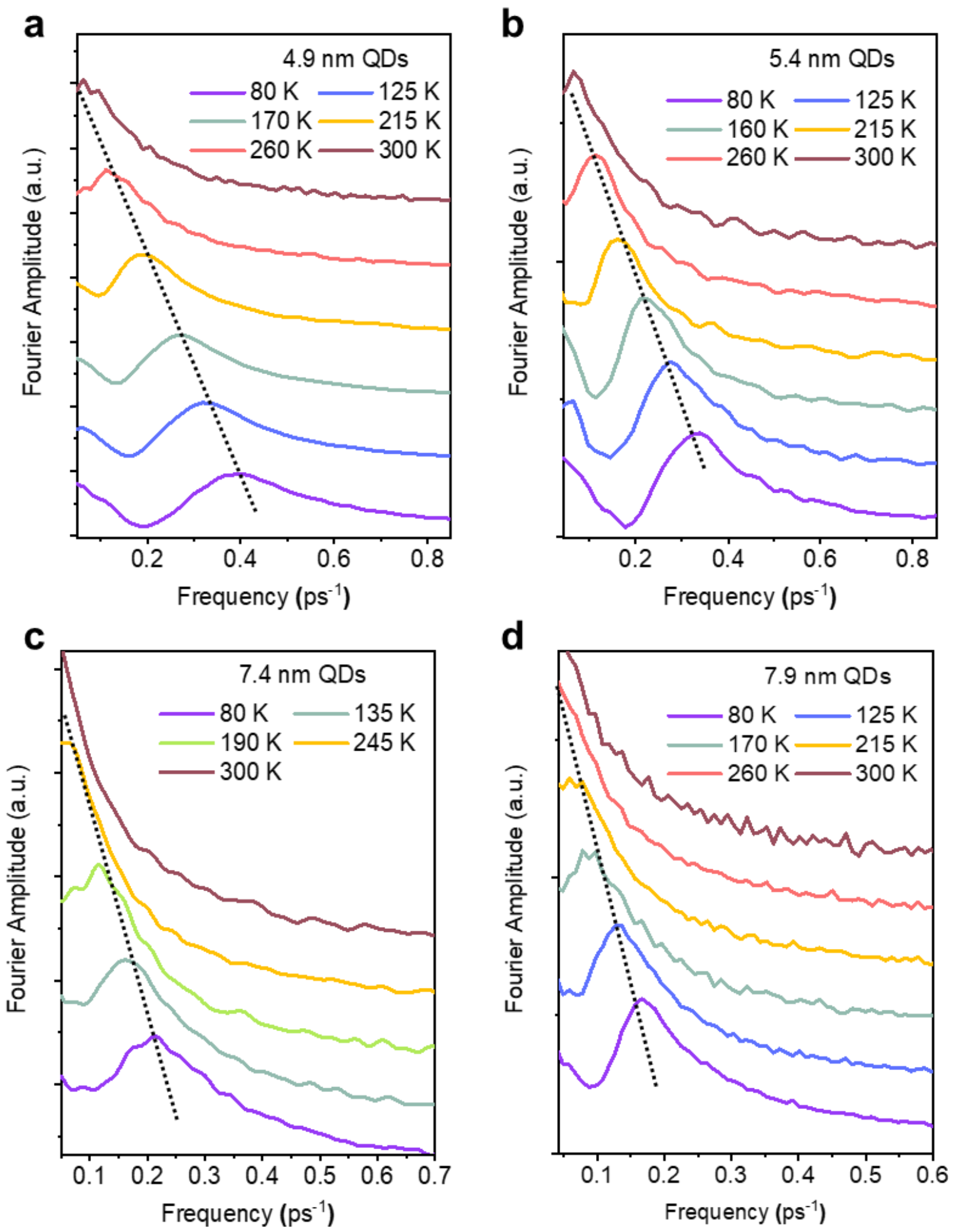

Supplementary Figure 15. FFT of quantum beats in $CsPbI_3$ QDs measured at varying temperatures.

FFT of the quantum beats kinetics of (a) 4.9 nm, (b) 5.4 nm, (c) 7.4 nm and (d) 7.9 nm $CsPbI_3$ QDs measured at varying temperatures from 80 to 300 K.

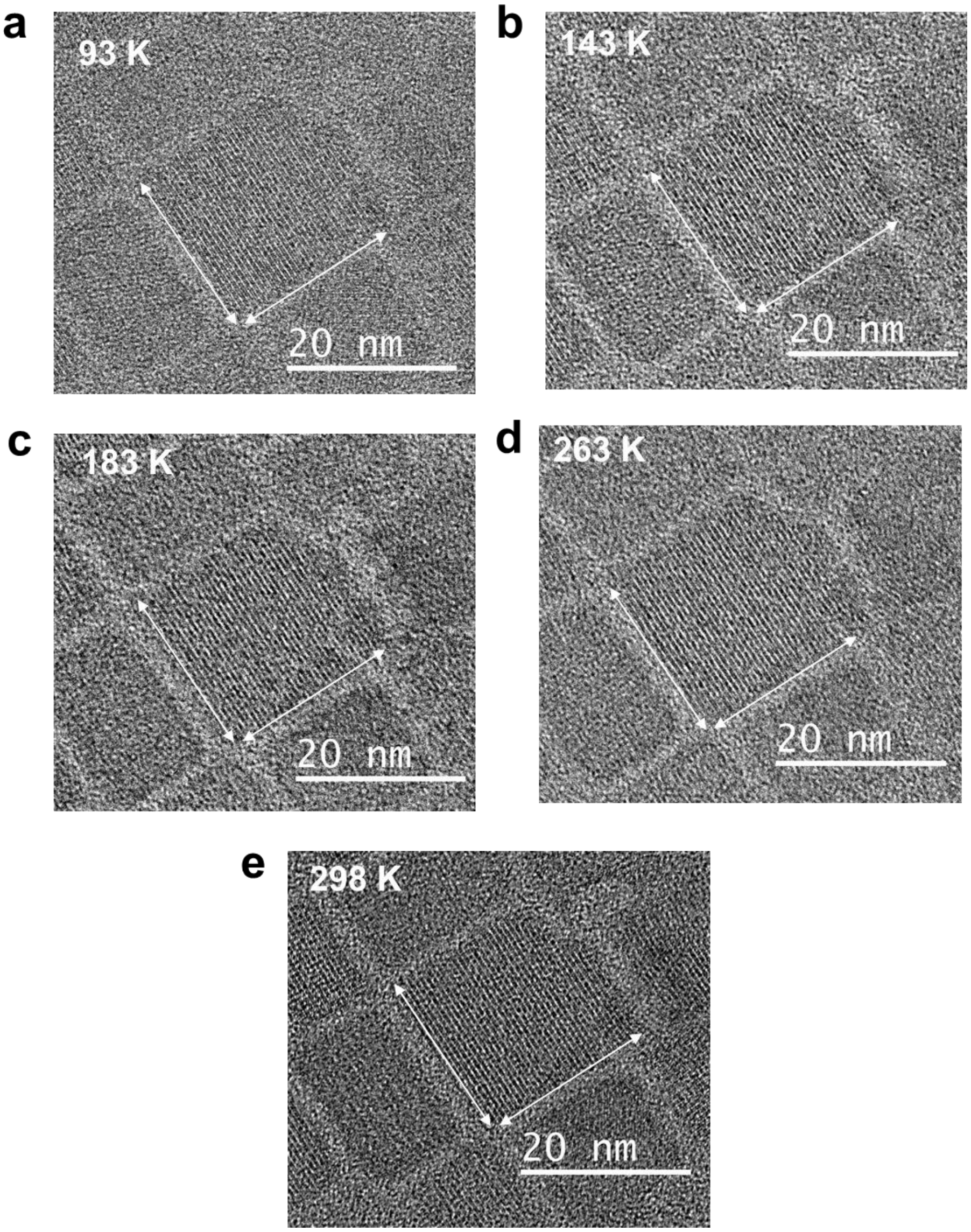

Supplementary Figure 16. Varying-temperature TEM images of 17.3 nm $CsPbI_3$ QDs.

High-resolution TEM images of 17.3 nm $CsPbI_3$ QDs at varying temperatures. The white arrows have the same length, indicative of a negligible morphologic change of the QDs at varying temperatures. The images contrast was relatively low because they were obtained under low electron dose conditions (~100 e ·$Å^{-2}$ $s^{-1}$). Higher-dose conditions induced amorphization of the perovskite QDs, especially at low temperatures, as explained in Materials and Methods. Smaller-size QDs were even more sensitive to the electron beams, prohibiting varying-temperature image acquisition under our experimental conditions.

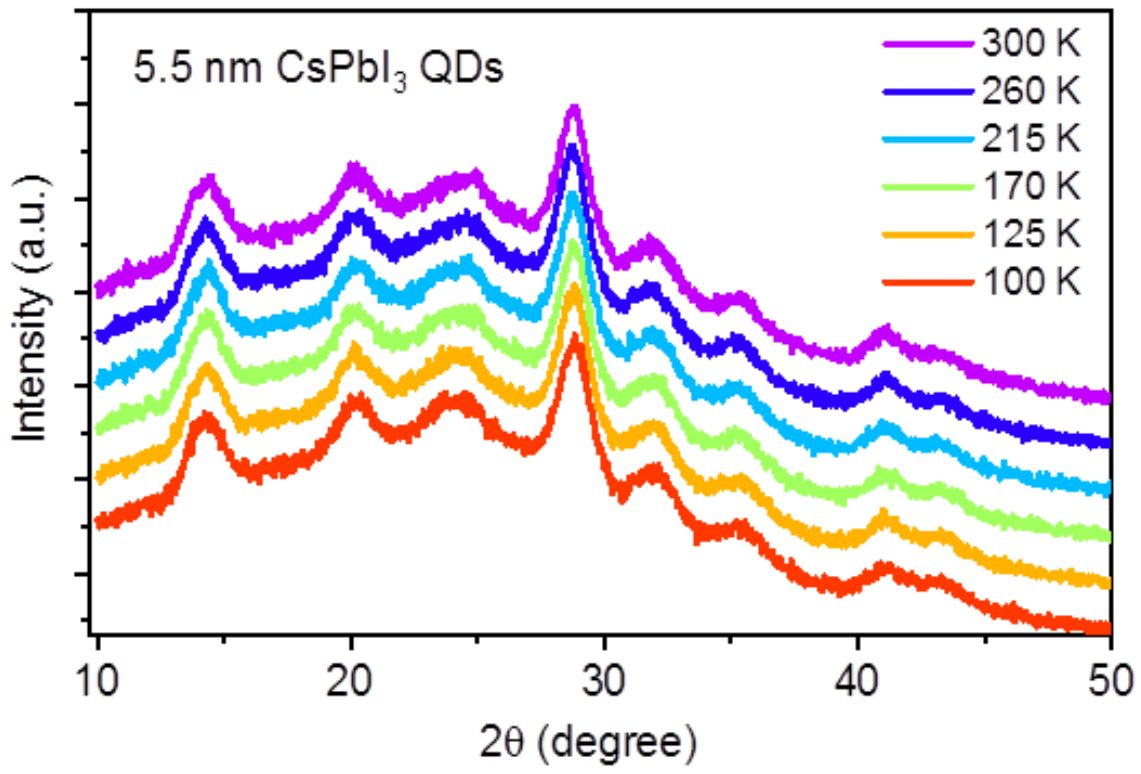

Supplementary Figure 17. Varying-temperature XRD of 5.5 nm $CsPbI_3$ QDs.

XRD patterns of 5.5 nm $CsPbI_3$ QDs measured at varying temperatures. The peaks are strongly broadened because of the small QD size. Moreover, the surface-to-volume ratio of small-size QDs is large, and hence the contributions from surface-bound organic ligands become important, resulting in the broad diffraction background. For these reasons, quantitative refinement of the XRD patterns is not performed.

Supplementary Table 1. Fitting parameters for varying-size QDs at 80 K using eq. 3.

| QD size | $F$ | $T_{dec}$ (ps) | $T_{\delta}$ (ps) | $T_{FSS}$ (ps) | φ |
|---|---|---|---|---|---|
| 4.9 nm | 0.39 ±0.02 | 0.97 ±0.04 | 3.06 ±0.05 | 2.92 ±0.03 | 0.19 ±0.05 |
| 5.4 nm | 0.41 ±0.01 | 1.44 ±0.06 | 4.00 ±0.06 | 3.40 ±0.03 | -0.13 ±0.04 |
| 6.2 nm | 0.50 ±0.02 | 2.04 ±0.11 | 3.62 ±0.09 | 3.70 ±0.04 | 0.04 ±0.05 |
| 6.9 nm | 0.43 ±0.01 | 1.71 ±0.06 | 3.66 ±0.06 | 3.78 ±0.03 | -0.25 ±0.04 |
| 7.4 nm | 0.43 ±0.02 | 1.79 ±0.08 | 3.48 ±0.07 | 5.93 ±0.08 | 0.08 ±0.04 |
| 7.9 nm | 0.29 ±0.01 | 1.72 ±0.03 | 6.46 ±0.08 | 6.88 ±0.04 | 0.38 ±0.02 |

Supplementary Table 2. Fine structure splitting ($\Delta_{FSS}$) of varying-size QDs at 80 K obtained from kinetics fitting and FFT.

| QD size | $\Delta_{FSS,Fit}$ (meV) | $\Delta_{FSS,FFT}$ (meV) |
|---|---|---|
| 4.9 nm | 1.42 ±0.01 | 1.64 |
| 5.4 nm | 1.22 ±0.01 | 1.40 |
| 6.2 nm | 1.12 ±0.01 | 1.26 |
| 6.9 nm | 1.10 ±0.01 | 1.22 |
| 7.4 nm | 0.70 ±0.01 | 0.89 |
| 7.9 nm | 0.60 ±0.003 | 0.70 |

Supplementary Table 3. Fitting parameters for 4.9 nm QDs at varying temperatures using eq. 3.

| Temperature | $F$ | $T_{dec}$ (ps) | $T_{\delta}$ (ps) | $T_{FSS}$ (ps) | φ |
|---|---|---|---|---|---|
| 80 K | 0.39 ± 0.02 | 0.97 ± 0.04 | 3.06 ± 0.05 | 2.92 ± 0.03 | 0.19 ± 0.05 |
| 125 K | 0.40 ± 0.02 | 0.92 ± 0.06 | 3.64 ± 0.09 | 3.36 ± 0.04 | $5.08*10^{-5}$ ± 0.05 |
| 170 K | 0.33 ± 0.01 | 1.09 ± 0.03 | 4.06 ± 0.06 | 4.27 ± 0.03 | 0.23 ± 0.03 |
| 215 K | 0.33 ± 0.01 | 1.41 ± 0.05 | 4.84 ± 0.09 | 5.98 ± 0.06 | 0.17 ± 0.03 |
| 260 K | 0.51 ± 0.12 | 1.32 ± 0.78 | 4.00 ± 0.28 | 10.13 ± 1.59 | 0.05 ± 0.42 |

Supplementary Table 4. Fitting parameters for 5.4 nm QDs at varying temperatures using eq. 3.

| Temperature | $F$ | $T_{dec}$ (ps) | $T_{\delta}$ (ps) | $T_{FSS}$ (ps) | φ |
|---|---|---|---|---|---|
| 80 K | 0.41 ± 0.01 | 1.44 ± 0.06 | 4.00 ± 0.06 | 3.40 ± 0.03 | -0.13 ± 0.04 |
| 125 K | 0.46 ± 0.02 | 1.86 ± 0.08 | 4.24 ± 0.07 | 3.97 ± 0.04 | -0.11 ± 0.04 |
| 170 K | 0.43 ± 0.01 | 2.33 ± 0.09 | 5.05 ± 0.09 | 5.15 ± 0.05 | 0.02 ± 0.04 |
| 215 K | 0.47 ± 0.02 | 2.76 ± 0.18 | 5.56 ± 0.17 | 7.18 ± 0.11 | 0.02 ± 0.05 |
| 260 K | 0.46 ± 0.05 | 3.12 ± 0.32 | 6.00 ± 0.26 | 11.29 ± 0.41 | 0.24 ± 0.06 |

Supplementary Table 5. Fitting parameters for 7.4 nm QDs at varying temperatures using eq. 3.

| **Temperature** | ***F*** | **$T_{dec}$ (ps)** | **$T_{\delta}$ (ps)** | **$T_{FSS}$ (ps)** | **φ** |
|---|---|---|---|---|---|
| 80 K | 0.43 ±0.02 | 1.79 ±0.08 | 3.48 ±0.07 | 5.93 ±0.08 | 0.08 ±0.04 |
| 135 K | 0.44 ±0.02 | 2.63 ±0.09 | 4.18 ±0.08 | 7.72 ±0.11 | 0.11 ±0.03 |
| 190 K | 0.33 ±0.03 | 2.68 ±0.13 | 5.59 ±0.16 | 10.78 ±0.32 | 0.24 ±0.05 |
| 245 K | 0.31 ±0.09 | 3.14 ±0.22 | 6.47 ±0.36 | 24.43 ±5.88 | 0.71 ±0.22 |

Supplementary Table 6. Fitting parameters for 7.9 nm QDs at varying temperatures using eq. 3.

| **Temperature** | ***F*** | **$T_{dec}$ (ps)** | **$T_{\delta}$ (ps)** | **$T_{FSS}$ (ps)** | **φ** |
|---|---|---|---|---|---|
| 80 K | 0.29 ±0.01 | 1.72 ±0.03 | 6.46 ±0.08 | 6.88 ±0.04 | 0.38 ±0.02 |
| 125 K | 0.36 ±0.01 | 2.61 ±0.06 | 6.45 ±0.08 | 9.23 ±0.07 | 0.34 ±0.02 |
| 170 K | 0.53 ±0.02 | 6.32 ±0.19 | 5.99 ±0.15 | 18.48 ±0.66 | 0.74 ±0.05 |
| 215 K | 0.21 ±0.02 | 3.57 ±0.18 | 8.76 ±0.40 | 17.69 ±0.91 | 0.68 ±0.09 |
| 260 K | 0.16 ±0.06 | 2.77 ±0.29 | 7.24 ±0.76 | 24.69 ±7.30 | 0.20 ±0.55 |

Supplementary Table 7. Fine structure splitting ($\Delta_{FSS}$) of varying-size QDs at varying temperatures obtained from kinetics fitting and FFT.

| | 4.9 nm QDs | | 5.4 nm QDs | | 7.9 nm QDs | |
|---|---|---|---|---|---|---|
| | $\Delta_{FSS,Fit}$ (meV) | $\Delta_{FSS,FFT}$ (meV) | $\Delta_{FSS,Fit}$ (meV) | $\Delta_{FSS,FFT}$ (meV) | $\Delta_{FSS,Fit}$ (meV) | $\Delta_{FSS,FFT}$ (meV) |
| 80 K | 1.42 ±0.01 | 1.64 | 1.22 ±0.01 | 1.4 | 0.60 ±0.003 | 0.69 |
| 125 K | 1.23 ±0.01 | 1.34 | 1.04 ±0.01 | 1.14 | 0.45 ±0.003 | 0.54 |
| 170 K | 0.97 ±0.01 | 1.12 | 0.80 ±0.01 | 0.9 | 0.22 ±0.008 | 0.36 |
| 215 K | 0.69 ±0.01 | 0.8 | 0.58 ±0.01 | 0.66 | 0.23 ±0.012 | 0.28 |
| 260 K | 0.42 ±0.07 | 0.48 | 0.37 ±0.01 | 0.47 | 0.18 ±0.054 | 0.12 |
| 300 K | - | 0.32 | - | 0.26 | - | 0.08 |

| | 7.4 nm QDs | |
|---|---|---|
| | $\Delta_{FSS,Fit}$ (meV) | $\Delta_{FSS,FFT}$ (meV) |
| 80 K | 0.70 ±0.01 | 0.886 |
| 135 K | 0.54 ±0.01 | 0.671 |
| 190 K | 0.38 ±0.01 | 0.480 |
| 245 K | 0.18 ±0.04 | 0.273 |

# QD fine structure model

## Supplementary Text 1 . Introduction and overview

In this section we describe a quasi-cubic model for the electronic structure of $CsPbI_3$ QDs, including the exciton fine structure and optical properties of the lowest energy confined exciton, how these are affected by the nanocrystal lattice structure as well as the QD size and shape. We then apply the model developed to simulate the quantum beating observed in polarized transient absorption (TA) measurements described in the main text. We find that the temperature-dependent bright-exciton fine structure splitting (FSS) revealed by the quantum beating measurements can be modelled in terms of the measured temperature-dependent lattice constants with a single fitting parameter, the strain deformation potential constant, with all other model parameters taken from measurements in the literature.

The electronic structure and fine structure model are based on the quasi-cubic model of Ref. [1], which treats the symmetry breaking of the orthorhombic structure as a strain perturbation on the high-temperature cubic crystal structure. As shown in the main text, the $CsPbI_3$ QDs studied here have orthorhombic, not cubic, crystal structure; moreover the orthorhombic lattice constants measured by XRD change with temperature, showing an increasing departure from the pseudocubic lattice constants ($\sqrt{2}a \times \sqrt{2}a \times 2a$) as the temperature is reduced. At the same time, measurement of quantum beating in circularly-polarized degenerate transient absorption reveals temperature-dependent FSS (see Fig 3 in the main text), which is roughly correlated with the temperature-dependence of the measured lattice constants (Fig 4 in the main text). Within the context of the quasi-cubic model, we expect the temperature dependence of the lattice constants to be correlated to the FSS via the effect of symmetry breaking which breaks the degeneracy of the triplet bright exciton states.

In addition to the lattice symmetry breaking, the QD shape effects the bright exciton FSS via the long-range exchange, which entails an interaction between the exciton polarization and the NC bounding facets. In Ref.[1], as in a number of other similar recent studies, it has been assumed that the QD bounding facets are comprised of the orthorhombic $\{100\}_o$ , $\{010\}_o$, and $\{001\}_o$ crystal planes, which are orthogonal to the bright exciton transition dipoles in orthorhombic QDs. Note that the subscript "*o*" on the Miller indices denotes specifically that the Miller indices are referenced to the primitive vectors of the orthorhombic crystal system. Here, in contrast to Ref.[1], we consider the alternate structural model suggested by recent studies on $CsPbBr_3$ nanoplatelets [2,3] that the nanocrystals bounding facets are formed, not by the low index orthorhombic crystal planes, but by the *pseudocubic* $\{100\}_c$ crystal planes. Here the subscript "*c*" denotes reference to the pseudo-cubic crystal system. Directions in the two systems are related as follows:

$$[100]_o \leftrightarrow [110]_c; \quad [010]_o \leftrightarrow [1\bar{1}0]_c\,; \quad [001]_o \leftrightarrow [002]_c\,, \qquad \text{(S1)}$$

with parallel relations between the lattice planes in the two systems. The distinction between the two structure models is shown in Supplementary Fig. 18. Our TEM image in Supplementary Fig. 18 tends to support the *pseudocubic facet model*.

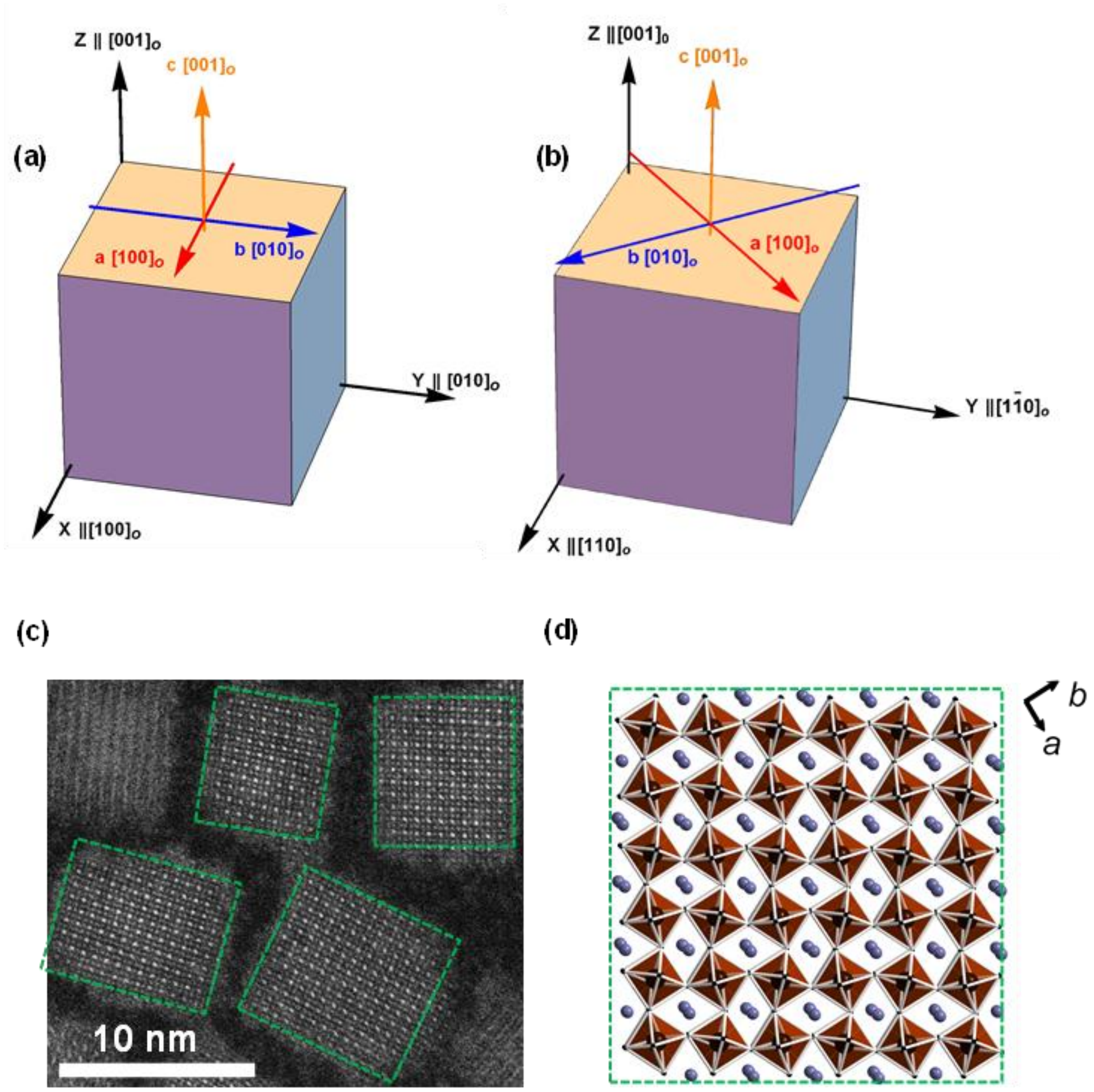

Supplementary Figure 18. Orthorhombic facet model and pseudo-cubic facet model for $CsPbI_3$ QDs.

Panel (a) shows the orthorhombic facet model, in which the bounding facets are $\{100\}_o$, $\{010\}_o$, and $\{001\}_o$ crystal planes, where the subscript "$o$" refers to the orthorhombic crystal system. Panel (b) shows the pseudocubic facet model, where the basal planes are comprised of the orthorhombic $\{110\}_o$ and $\{1\bar{1}0\}_o$ crystal planes. These are alternately described as the $\{100\}_c$ and $\{010\}_c$ planes within the pseudocubic crystal system, where the subscript "$c$" refers to the cubic system. We follow the convention that the $[001]_o$ direction, namely the orthorhombic primitive ***c*** axis, is always taken as the Z direction, while the X direction is either the $[100]_o$ or $[110]_o$ direction in the case of the orthorhombic or pseudocubic facet models, respectively. In both cases the directions of the primitive ***a, b, c*** orthorhombic lattice vectors are shown for reference. Panel (c) shows the dark-field high-resolution TEM images taken for 7.9 nm $CsPbI_3$ QDs. The bright spots are the Pb atoms, i.e., the centers of the $PbI_6$ octahedra. Clearly, the square-shaped networks of the $PbI_6$ octahedra have their edge lines aligned to the bounding facets of the QDs, as illustrated by the cartoon in (d). Thus, albeit the lattice of the $CsPbI_3$ QDs is distorted to the orthorhombic phase, their bounding facets are still the pseudocubic $\{100\}_c$ and $\{010\}_c$ families of planes.

The difference between the *orthorhombic facet model* and the *pseudocubic facet model* is critically important: Since the bright excitons whose transition dipoles lie in the plane spanned by the orthorhombic ***a, b*** primitive vectors (in a plane parallel to the $\{001\}_c$ family of planes) are not orthogonal to the QD bounding facets in the pseudocubic model (see Supplementary Fig. 18), the corresponding excitons *couple* via long-range exchange. These exciton states are therefore *always* separated by an avoided crossing energy gap, even in the presence of variations of the edge length ratios from 1:1:1 characteristic of a perfect cube shape, as is the case based on our TEM imaging. This will be shown below. Such an energy gap does *not* occur in the orthorhombic facet model: That is, with a distribution of the edge length ratios, the energy difference between any two exciton states considered across the QD distribution will always be continuously distributed from zero.

Applying the model developed to simulations of the transient absorption measurements described in the main text, we find consequently that with any significant shape dispersity, *quantum beating signatures in the orthorhombic facet model are completely obscured by shape inhomogeneity*. By contrast, in the pseudocubic facet model, quantum beating between the bright ***a, b*** polarized excitons survive averaging by virtue of the aforementioned coupling gap.

Consequently, the observation of quantum beating in highly inhomogeneously broadened $CsPbI_3$ QDs is a hallmark feature arising from pseudocubic bounding facets.

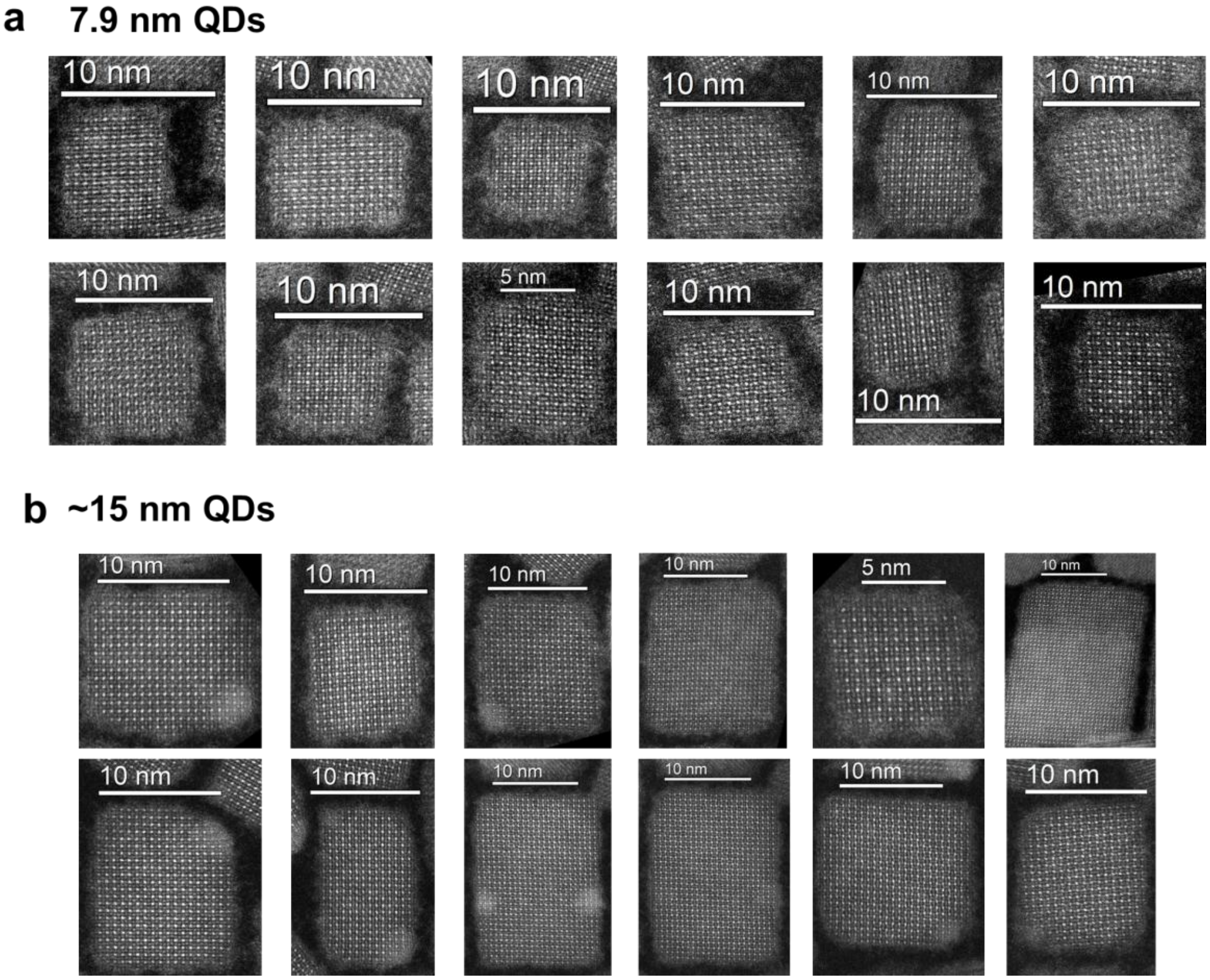

Supplementary Figure 19. More TEM images supporting the pseudocubic facet model.

TEM images of 7.9 nm (upper) and 15 nm (lower) $CsPbI_3$ QDs, showing that they are mostly bounded by pseudocubic $\{100\}_c$ family of planes. The slight imperfection in the QD shapes is likely induced by sample damage under electron beams.

The pseudocubic facet model can be well supported by prior literature. First, in refs [4,5], it has been shown both experimentally and computationally that cuboidal $CsPbX_3$ nanocrystals are bounded by the pseudocubic facets, not the lowest index orthorhombic facets. According to these studies, these cuboidal shaped nanocrystals adopt the pseudocubic facets in order to minimize the surface energy. Closely related experimental results were reported for $CsPbBr_3$ nanoplatelets [6,7].

We also provide further experimental evidence by taking extensive HR-TEM images of our $CsPbI_3$ QDs. In addition to Supplementary Fig. 18, we show in Supplementary Fig. 19 more TEM images of two QD samples with average sizes of ~7.9 nm and ~15 nm. In almost all QDs for which we can see the TEM lattices, the square-shaped networks of the $PbI_6$ octahedra align well with the QD facets.

## Supplementary Text 2. Size and shape of $CsPbI_3$ QDs

From the discussion above, it is clear that an essential starting point in our analysis of the electronic and optical properties of the $CsPbI_3$ NCs is to describe their size and shape distributions. Measurement by TEM (Supplementary Fig. 11) was employed to determine the distribution of size and to quantify the distribution of shapes. Supplementary Fig. 20 panel (a) shows a size histogram for the QDs with average edge length 7.9 nm. This QD size is chosen for particular detailed theoretical study since these QDs show both pronounced quantum beating signatures (unlike the larger 17.3 nm QDs, see Fig. 2d) while being large enough that size broadening of the XRD peaks does not prohibit Rietveld refinement (unlike the smaller sized QDs, see Supplementary Fig. 17). This refinement is necessary to determine the lattice constants and to quantify the quasi-cubic symmetry breaking as discussed later on.

The size distribution for this sample distribution is approximately normally distributed with a standard deviation of 0.7 nm. The shape distribution was also assessed, by measuring the lengths of the short and long edges of each of 703 QDs in TEM. The resulting distribution, plotted in Supplementary Fig. 20 panel (b), has an average short-to-long edge length ratio ($r$) of 1 with a standard deviation of 0.16. A perfect cube shape would be characterized by $r = 1$. Between the variation in size and the variation in shape, the more significant source of inhomogeneous spectral broadening is due to the shape distribution reflected in Supplementary Fig. 20 panel (b) as will become clear in the analysis to follow (See Supplementary Text 6). To assess this source of inhomogeneous broadening, we need to construct a model for the shape of the QDs that is consistent with the measured distribution of the side edge length ratios.

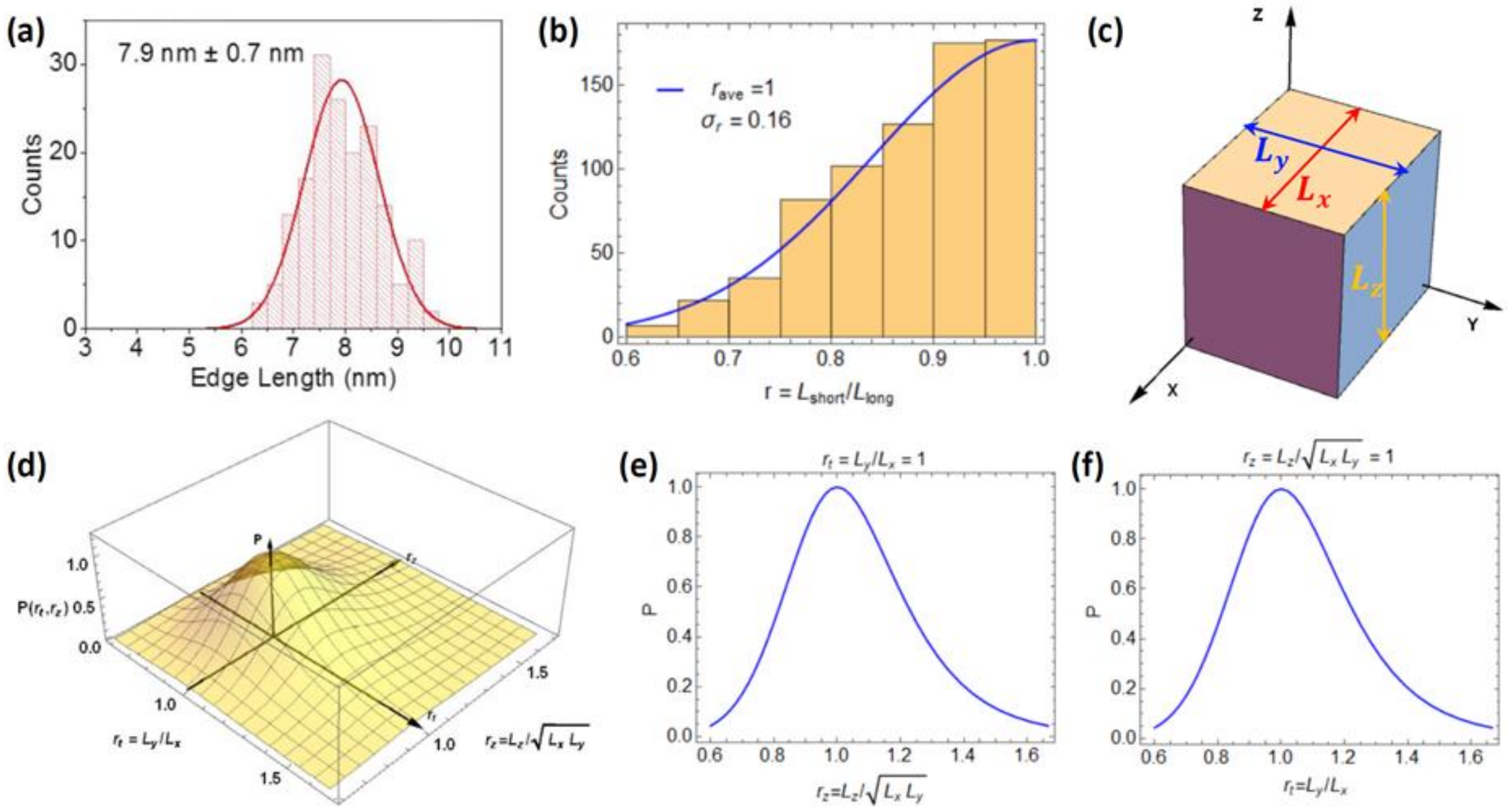

Supplementary Figure 20. Size and shape distribution for $CsPbI_3$ QDs with average edge length 7.9 nm.

Panel (a) shows the size histogram and a fit to a normal distribution with a standard deviation of 0.7 nm. Panel (b) shows a histogram of the ratio, $r$, of the short and long edge lengths for the 7.9 nm QDs obtained from the TEM images of 703 individual QDs in Supplementary Fig. 8. The average ratio $r$ is 1 with a standard deviation of 0.16. A perfect cube shape would be characterized by $r = 1$. Panel (c) shows a schematic of a cuboid with unequal edge lengths $L_x, L_y, L_z$ in the $x$ ,$y$, and $z$ directions. Considering the random orientations of the QDs standing on the TEM grid, the histograms in (b) reflects results for $L_y/L_x$ (or $L_x/L_y$), $L_z/L_x$ (or $L_x/L_z$), and $L_z/L_y$ (or $L_y/L_z$). The corresponding biaxial-model shape distribution using Eq. S2 is shown in panel (d), plotted against the length ratios, $r_t = L_y/L_x$ and $r_z = L_z/\sqrt{L_x L_y}$ . Line plots along the axis $r_z = 1$ and $r_t = 1$ are shown in panels (e) and (f), respectively. Note that the probability $P(r_t) = P(1/r_t)$ and $P(r_z) = P(1/r_z)$.

Since the QDs imaged in TEM lay randomly with one of six facets oriented with its perpendicular along the imaging direction, the distribution in Supplementary Fig. 20 panel (b) is not consistent with a distribution of uniaxially distorted shapes, since in that case, we would expect to see two separate sub-distributions. One population, expected to comprise 1/3 of the QDs, would consist of the QDs whose uniaxially-distorted axis is parallel to the TEM beam axis. This sub-population would be characterized by a sharp distribution near $r = L_{short}/L_{long}$ =1. The second population would comprise those QDs whose uniaxially-distorted axis lies perpendicular to the TEM imaging axis. These QDs would be expected to show a broader distribution corresponding to the single distorted axis. Since these features are not observed in the measured shape distribution, we assume that the ratio of any two edge lengths on the nano-cuboid has essentially the same distribution in $r = L_{short}/L_{long}$.

A model that approximately fulfills this condition is a biaxially distorted shape in which the edge lengths in the x, y and z directions are determined by two $r$ ratios, each of which has the distribution shown in Supplementary Fig. 20 panel (b), and which we denote $r_z$ and $r_t$ . These ratios are defined respectively as $r_t = L_y/L_x$ for any value of $L_z$, and $r_z = L_z/L_t$, where $L_t$ is the transverse length parameter defined as $L_t = \sqrt{L_x L_y}$ for any $L_z$ . Then for a given shape configuration $r_z$ , $r_t$, the individual QD edge lengths are given by,

$$L_x = L_e/(r_t^{1/2} r_z^{1/3}) , \qquad L_y = L_e r_t^{1/2}/r_z^{1/3}, \qquad L_z = L_e r_z^{2/3} . \tag{S2}$$

Here, the effective edge length is $L_e = \left(L_x\, L_y\, L_z\right)^{1/3}$ so the distortion described is volume preserving with respect to a cube of three equal edge lengths $L_e$. The average configuration is thus a nano-cube with equal edge lengths along the x, y, and z directions, consistent with the measured shape distribution in Supplementary Fig. 20(b). The corresponding biaxial model shape distribution using Eq. S2 is shown in Supplementary Fig. 20(d), plotted against the length ratios, $r_t = L_y/L_x$ and $r_z = L_z/\sqrt{L_x L_y}$. Line plots along the axis $r_z = 1$ and $r_t = 1$ are shown in panels (e) and (f), respectively. We also note that in the TEM measurement of the edge length ratios, two QDs with their Z axes aligned along the imaging direction with configurations described by $r_t$ and $1/r_t$ , respectively, are not distinguishable since only the long and the short axes are distinguishable. \Similarly, the ratios $r_z$ and $1/r_z$ are not distinguishable, i.e., $P(r_t) = P(1/r_t)$ and $P(r_z) = P(1/r_z)$.

## Supplementary Text 3. Quantum confinement model

In this work, $CsPbI_3$ QDs of cuboidal shape with edge lengths in the size range 5-17 nm edge length are grown and characterized. For comparison, the bulk exciton radius of $CsPbI_3$ is $a_x = 4.64$ nm [8] (see Supplementary Text 5.1), so that the QDs studied are of size comparable to or larger than the bulk exciton radius. Consequently, the assumption of strong quantum confinement, which neglects correlated electron-hole motion, provides an inadequate description. Conversely, QDs in the size range studied are sufficiently small that the weak confinement approximation, in which the exciton center-of-mass confinement is accounted for but the exciton internal relative electron/hole motion is considered un-modified relative to the exciton in bulk material, is not valid. The condition of validity is that the quantum confinement energy separations of confined levels of the non-interacting electron and hole be small relative to the exciton binding energy.

To bridge between these two limits we must describe the QDs within the intermediate confinement regime where the Coulomb interaction of the electron and hole is significant, but the quantum confinement of the underlying carriers is also accounted for [9]. For this description we must solve the effective mass equation for an exciton within the QD,

$$\widehat{H}^{eff} f(\boldsymbol{r}_e, \boldsymbol{r}_h) = E_X\, f(\boldsymbol{r}_e, \boldsymbol{r}_h)\,, \qquad \text{(S3)}$$

subject to the condition that the envelope function vanish on the QD surface. Here, $f(\boldsymbol{r}_e, \boldsymbol{r}_h)$ is the envelope function for the exciton, which is written as a function of the electron and hole coordinates, $\boldsymbol{r}_e$, $\boldsymbol{r}_h$, respectively, $E_x$ is the exciton energy relative to the band gap. The effective mass Hamiltonian for the electron/hole pair is given by the sum of their kinetic energies and their Coulomb interaction energy:

$$\widehat{H}^{eff} = -\frac{\hbar^2}{2m_e}\nabla_e^2 - \frac{\hbar^2}{2m_h}\nabla_h^2 - \frac{e^2}{\epsilon_{eff}|\boldsymbol{r}_e - \boldsymbol{r}_h|}\ . \qquad \text{(S4)}$$

The terms $m_e$, $m_h$ are the electron and hole effective masses, respectively, while $\epsilon_{eff}$ is the effective dielectric constant screening the electron/hole Coulomb interaction. Note that $\epsilon_{eff}$ is intermediate between the static and the high frequency dielectric function as it includes lattice polarization effects [10]. We apply a variational approach to solving this problem within the parabolic band approximation [9,11]. The wavefunction of the lowest energy exciton is assumed to be of the form,

$$f(\boldsymbol{r}_e, \boldsymbol{r}_h) = \frac{1}{\sqrt{N(\beta)}} e^{-\beta|r_e - r_h|}\varphi(\boldsymbol{r}_e)\varphi(\boldsymbol{r}_h). \qquad \text{(S5)}$$

In this expression, $\varphi$ represents lowest energy electron and hole quantum confined levels in a cuboid of edge lengths $L_{x,}L_{y,}L_z$, aligned to the pseudocubic lattice directions, given by,

$$\varphi(x, y, z) = \sqrt{\frac{8}{L_x L_y L_z}}\cos\left(\frac{\pi x}{L_x}\right)\cos\left(\frac{\pi y}{L_y}\right)\cos\left(\frac{\pi z}{L_z}\right), \qquad \text{(S6)}$$

while the term involving the variational parameter $\beta$ imparts correlated relative motion of the electron and hole. Using this ansatz function, the exciton energy $E_x$ is calculated as,

$$E_X(\beta) = \frac{\langle f|\hat{H}^{eff}|f\rangle}{\langle f|f\rangle} \quad \text{(S7)}$$

The optimum variational parameter $\beta_{opt} = \beta_{opt}(L_x, L_y, L_z)$ is determined as the value of $\beta$ which minimizes the energy for a QD of given mutually perpendicular edge lengths, $L_x, L_y, L_z$. For large sizes it has been shown that $\beta \to 1/a_x$, where $a_x$ is the bulk exciton Bohr radius [11]. The energy is given in the form,

$$E_X = -\frac{\hbar^2}{2\mu L_{eff}{}^2}\frac{I_K(\beta_{opt})}{N(\beta_{opt})} + \frac{e^2}{\epsilon_{eff}\, L_{eff}}\frac{I_C(\beta_{opt})}{N(\beta_{opt})} \quad , \quad \text{(S8)}$$

where $\mu = (1/m_e + 1/m_h)^{-1}$ is the reduced effective mass, $I_K$, $I_C$ and $N$ are dimensionless integrals for the kinetic and potential energies and the normalization, respectively. These integrals are written in terms of dimensionless coordinates, $\boldsymbol{u}_e = \boldsymbol{r}_e/L_{eff}$ and $\boldsymbol{u}_h = \boldsymbol{r}_h/L_{eff,}$ where the effective length, $L_{eff,}$ is chosen as the edge length of a cube with the same kinetic energy for the lowest energy exciton, i.e., $L_{eff} = \sqrt{3/(L_x^{-2} + L_y^{-2} + L_z^{-2})}$ [12]. Note the difference between $L_{eff}$ here and $L_e$ defined above. In Supplementary Fig. 21 we show the energy of the lowest exciton versus $L_{eff,}$ for QDs of cube shape as well as QDs with a uniaxial shape distortion with $L_x = L_y = L_t$ , with aspect ratio $r = L_z/L_t$ set to 1.6 and 0.625. As seen in the figure, the energy of the lowest exciton level *is independent of deviations in shape from cubic shape*. Shown in the figure for comparison are the energies of the lowest exciton in the strong and weak confinement limits calculated using the expressions in Supplementary Table 8. Plots of $\beta_{opt}$ and $N(\beta_{opt})$ vs $L_{eff}/a_X$ are given in Ref. [11].

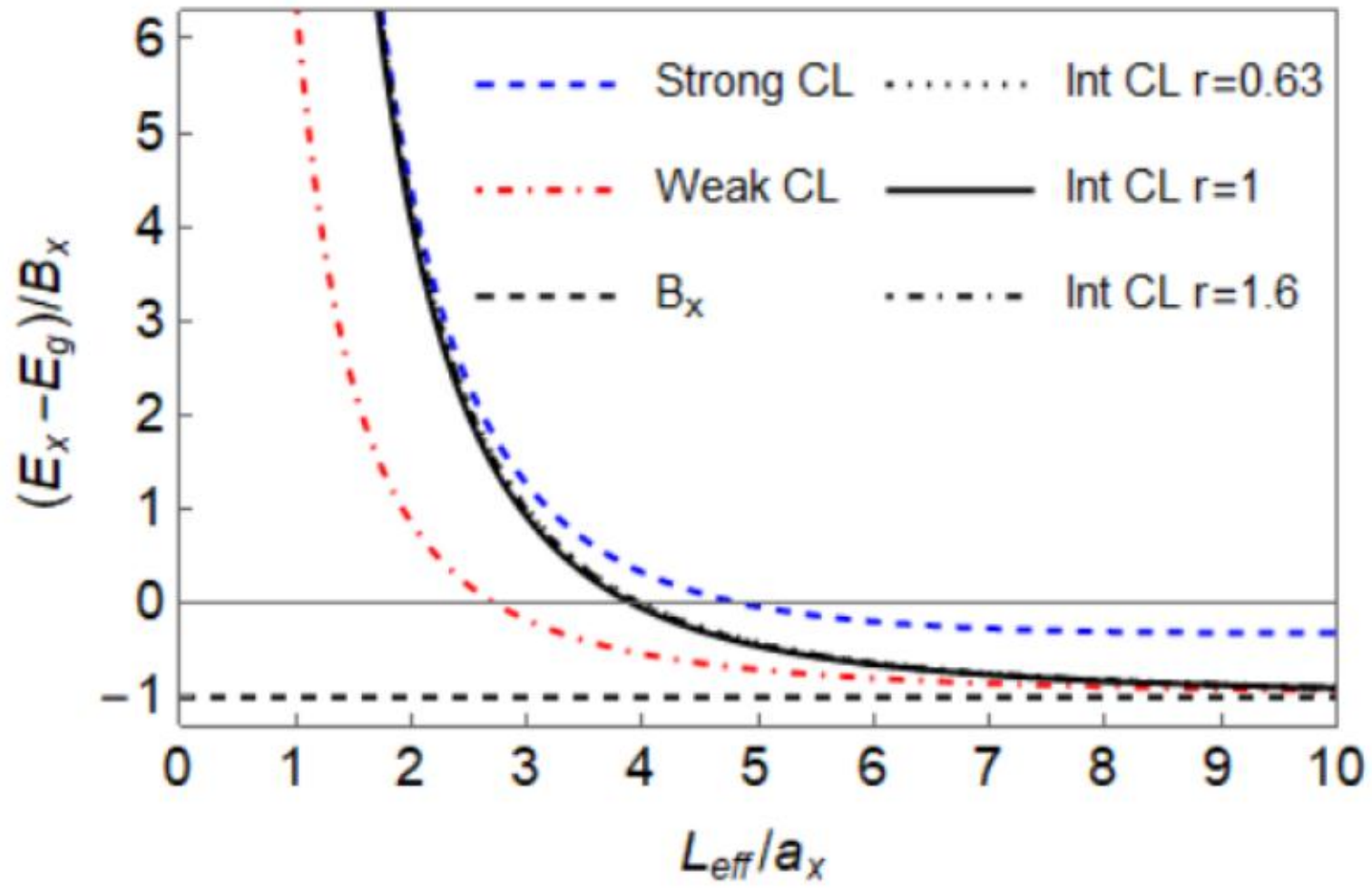

Supplementary Figure 21. Exciton energy versus QD size.

Exciton energy $E_X - E_g$ is plotted in units of the bulk exciton binding energy, $B_X$, versus the ratio of the effective edge length, $L_{eff}$, to the exciton Bohr radius $a_X$. The solid black line represents the result of a variation calculation for cube shaped QDs valid for the intermediate confinement limit (CL), $L \sim a_X$ ; dotted and dot-dashed black lines show calculations for QDs shortened or elongated along the z direction by factors $r_z$ = 0.625 and $r_z$ = 1.6 respectively, where $r_z = L_z/L_t$ and $L_x = L_y = L_t$. Shown for reference are the calculated exciton energies in the strong confinement limit (CL) ($L_{eff} \ll a_X$, blue dashed line), the weak CL ($L_{eff} \gg a_X$; red dashed line), and the bulk exciton binding energy $B_X$ (black dashed line). Values of $E_g$, $B_X$, and $a_X$ can be found in Supplementary Table 10.

### Supplementary Table 8. Exciton energy in the strong and weak confinement limits

Energy of the lowest energy exciton for a cube-shaped QD of effective edge length $L_{eff,}$, in the strong and weak confinement limits. The bulk exciton Bohr radius is $a_X = a_o\, \epsilon_{eff}/\mu$, where $a_o$ is the hydrogen Bohr radius, $\epsilon_{eff}$ is the effective dielectric constant and $\mu$ is the reduced effective mass. The bulk exciton binding energy is $B_X = \frac{\mu}{{\epsilon_{eff}}^2} Ry = \frac{\hbar^2}{2\mu a_X^2}$, where $Ry$ is the hydrogen Rydberg. The numerical factors within the Coulomb energy term for the strong confinement approximation represent the result of numerical integration of the Coulomb energy within first-order perturbation theory.

| | Lowest exciton energy |
|---|---|
| Strong confinement | $E_X = E_g + B_X \left\{ 3\left(\frac{\pi}{L_{eff}/a_X}\right)^2 - 2\left(\frac{3.047}{L_{eff}/a_X}\right) \right\}$ |
| Weak confinement | $E_X = E_g + B_X \left\{ \frac{3}{4}\left(\frac{\pi}{L_{eff}/a_X}\right)^2 - 1 \right\}$ |

## Supplementary Text 4. Exciton fine structure

Having described the quantum confined exciton levels in Supplementary Text 3, we now consider the problem of the fine structure of the lowest exciton state. We first discuss the effect of short-range (SR) electron-hole exchange interaction, and then proceed to discuss the effect of long-range exchange.

### Supplementary Text 4.1. Short-range exchange: Cubic symmetry

The short-range (SR) electron-hole exchange interaction can be written in the form of a spin-dependent contact interaction [13],

$$H_{ex}^{SR} = \frac{1}{2} C_{ex} \Omega \left[ \hat{I} - (\sigma_e \cdot \sigma_h) \right] \delta(r_e - r_h)\,. \tag{S9}$$

In this expression, $C_{ex}$ is the SR exchange constant for the material, $\Omega$ is the volume of the crystal unit cell, $\hat{I}$ is the 4×4 unit matrix, and $\boldsymbol{\sigma_e}, \boldsymbol{\sigma_h}$ are Pauli operators representing the electron and hole spin. To obtain the size-dependent SR exchange energies we calculate the exchange interaction over the total exciton wavefunction, including the underlying periodic part, which is given by the product of the band-edge Bloch functions of the conduction and valence bands labelled $i, j$ respectively:

$$\psi_{i,j}(\boldsymbol{r}_e, \boldsymbol{r}_h) = u_i^e(\boldsymbol{r_e}) u_j^h(\boldsymbol{r}_h) f(\boldsymbol{r}_e, \boldsymbol{r}_h) \quad . \tag{S10}$$

Here, $f(\mathbf{r_e}, \mathbf{r_h})$ is the envelope function for the exciton in the intermediate confinement regime given by Eq. S5. Integrating over the envelope functions we find that the short range exchange interaction can be written as an effective spin operator in the form [11],

$$H_{SR} = \frac{1}{2} C_{ex}\, \Theta \left[\hat{I} - \boldsymbol{\sigma}_e . \boldsymbol{\sigma}_h\right] . \tag{S11}$$

The term $\Theta$ is the electron-hole exchange overlap factor, representing the probability that the electron and hole reside in the same unit cell:

$$\Theta = \Theta(L_x, L_y, L_z) = \Omega \int_V d^3\boldsymbol{r}\ |f(\boldsymbol{r}_e, \boldsymbol{r}_h)|^2. \tag{S12}$$

The integration above is taken over the volume, $V$, of the QD. We later will find it convenient to express the QD size in terms of an effective length $L_e$ such that $V = L_e^3$. For the ground exciton state in the bulk, the overlap factor is [11],

$$\Theta_{bulk} = \frac{\Omega}{\pi a_{\mathrm{X}}^3} \ , \tag{S13}$$

where $a_{\mathrm{X}}$ is the bulk exciton Bohr radius. With this result we can rewrite Eq. S11 for the ground exciton in terms of the average singlet-triplet splitting in the bulk, $\hbar\omega_{ST}$,

$$H_{SR} = \frac{3}{4} \hbar\omega_{st} \left[\hat{I} - \boldsymbol{\sigma}_e . \boldsymbol{\sigma}_h\right] \left(\frac{\Theta(L_x, L_y, L_z)}{\Theta_{bulk}}\right), \tag{S14}$$

$$\hbar\omega_{ST} = 2/3\, C_{ex}\Theta_{bulk} \, . \tag{S15}$$

This equation is our effective spin operator for the electron-hole exchange interaction, where the size dependence sets in via the electron-hole exchange overlap function, $\Theta$, with is dependent on the QD size through the size-dependence of the exciton envelope function, Eq. S5.

To proceed further we need expressions for the electron and hole Bloch functions. For a QD of cubic lattice symmetry, which we address first for simplicity, the electron and hole Bloch functions are eigenstates of total angular momentum: The hole Bloch functions $u^h_{J_h, J_{h,z}}(\mathbf{r_e})$ have s-symmetry, and are the even parity states of angular momentum which we write [1]:

$$|u^h_{1/2,1/2}\rangle = |s\rangle|\uparrow\rangle$$
$$|u^h_{1/2,-1/2}\rangle = |s\rangle|\downarrow\rangle \tag{S16}$$

where the spinor functions ↑ and ↓ are the eigenfunctions of the electron spin projection along z, and the lower-case symbol s denotes an orbital function that transforms as an invariant under the operations of the crystal point symmetry group. The conduction band Bloch functions, $u^c_{J_e, J_{e,z}}(\mathbf{r_h})$, which have p-symmetry, comprise states with angular momentum $J$ = ½ with odd parity given by [1]:

$$|u^e_{1/2,1/2}\rangle = \frac{-1}{\sqrt{3}} \left[(|x\rangle + i|y\rangle)|\downarrow\rangle + |z\rangle|\uparrow\rangle\right] ,$$

$$|u^e_{1/2,-1/2}\rangle = \frac{1}{\sqrt{3}}[-(|x\rangle - i|y\rangle)|\uparrow\rangle + |z\rangle|\downarrow\rangle]. \tag{S17}$$

In this expression the lower case symbols $|x\rangle, |y\rangle, |z\rangle$ denote orbital functions that transform like $x$, $y$, $z$ under rotations. Using these expressions in Eq. S10, and diagonalizing the SR exchange Hamiltonian, we find the well-known result that the exchange interaction splits the exciton into a bright upper lying triplet level with total angular momentum $F = 1$ with projection quantum numbers $F_z = \pm 1,0$ and a lower lying "dark" singlet $F = 0$ state. The energies of the triplet and singlet states are given by,

$$E_{1;\,0,\pm 1} = \hbar\omega_{ST}\left(\frac{\Theta(L_x, L_y, L_z)}{\Theta_{bulk}}\right),$$
$$E_{0;0} = 0 \ . \tag{S18}$$

The size dependence of the singlet-triplet spacing is thus governed by the electron-hole exchange overlap factor $\Theta(L_x, L_y, L_z)$. This function was verified numerically to be independent of the shape of the QD. In Supplementary Fig. 22 we show the size dependence of the exchange overlap factor versus $L_e = (L_x L_y L_z)^{1/3}$ for QDs of cube shape, as well as for QDs with a uniaxial shape distortion with $L_x = L_y = L_t$ , with aspect ratio $r_z = L_z/L_t$ set to 1.6 and 0.625. As seen in the figure, overlap function and thus the singlet-triplet splitting is independent of deviations in shape from cubic shape. Also shown in the figure for comparison are exchange overlap functions calculated in the strong and weak confinement limits using the expressions in Supplementary Table 9 from Ref.[11]. It is clear that in the limit of small (large) size, the intermediate confinement calculation goes to the strong (weak) confinement limits.

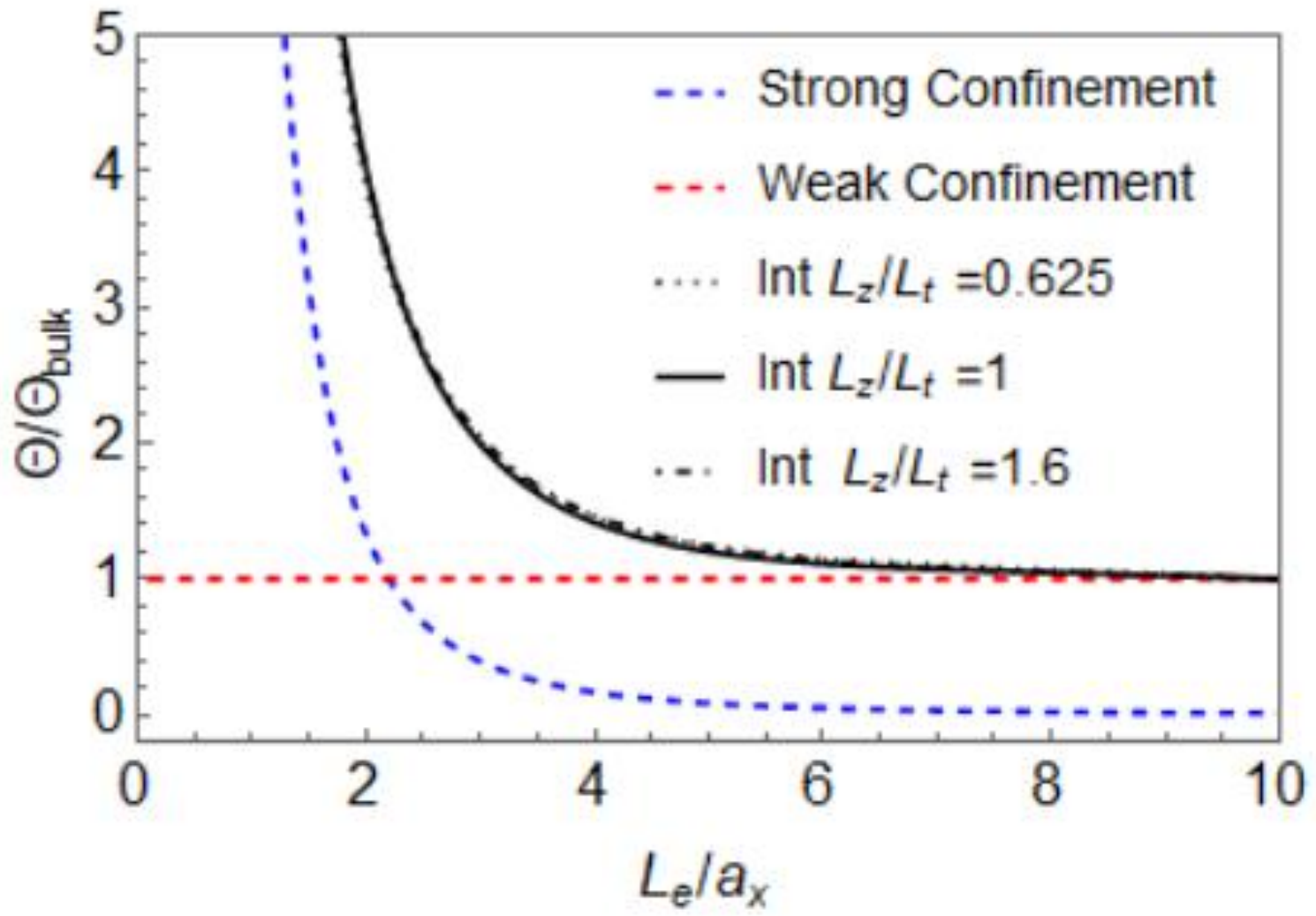

Supplementary Figure 22. Exchange overlap factor versus QD size.

Exchange overlap factor $\Theta(L_x, L_y, L_z) = \Theta(L_e)$ in a QD relative to that of the bulk exciton is plotted versus $L_e/a_X$, the ratio of the effective edge length, $L_e = (L_x L_y L_z)^{1/3}$, to the bulk exciton radius $a_X$. The solid black line represents the result of a variation calculation for cube-shaped QDs valid for the intermediate confinement limit, $L \sim a_X$ ; dotted and dot-dashed black lines show calculations for QDs shortened or elongated along the z direction by factors of $r_z$=0.625 and $r_z$=1.6 respectively, where $r_z = L_z/L_t$ . In the plots, we set $L_x = L_y = L_t$ for simplicity. Shown for reference are the calculated overlap factors in the strong confinement ($L_e \ll a_X$, blue dashed line) and the weak confinement ($L_e \gg a_X$; red dashed line) approximations. The exchange overlap factor is seen from the calculation to be independent of the shape for a given QD volume.

Supplementary Table 9. Exchange overlap versus QD size for strong, intermediate, and weak confinement.

Exchange overlap functions for the lowest energy exciton in a QD of effective edge length, $L_e$, in the strong, intermediate, and weak confinement limits and for the bulk. The exciton radius is $a_X$ while the unit cell volume is $\Omega$. For intermediate confinement the term $N(\beta_{opt}(L_e))$ is the normalization integral evaluated at the optimum variational parameter, $\beta_{opt}$, for a given size QD. Plots of $\beta_{\mathrm{opt}}$ and $N(\beta_{\mathrm{opt}})$ vs $L_{eff}/a_X$ are given in Ref. [11].

| | Exchange overlap factor |
|---|---|
| Bulk semiconductor & Weak confinement | $\Theta(L_e) = \frac{\Omega}{\pi a_X^3}$ |
| Strong confinement | $\Theta(L_e) = \frac{27}{8}\frac{\Omega}{L_e^3}$ |
| Intermediate confinement | $\Theta(L_e) = \frac{1}{\mathrm{N}(\beta_{\mathrm{opt}}(L_e))}\frac{27}{8}\frac{\Omega}{L_e^3}$ |

A deficiency of the analysis leading to Eq. S18 is that there is no splitting among the degenerate bright exciton levels. This is a consequence of the assumed cubic symmetry. We show in the next section that symmetry breaking from cubic to orthorhombic lattice symmetry, quantified via the temperature-dependent lattice constants measured by XRD and structural refinement as described in the main text, breaks the degeneracy of the bright exciton levels.

## Supplementary Text 4.2. SR exchange with orthorhombic lattice distortion

As shown in the main text, the $CsPbI_3$ QDs studied here have orthorhombic, not cubic, crystal structure; moreover the orthorhombic lattice constants measured by XRD change with temperature, showing an increasing departure from the pseudocubic lattice constants ($\sqrt{2}a \times \sqrt{2}a \times 2a$) as the temperature is reduced. At the same time, measurement of quantum beating in co-circular polarized and counter-circular polarized degenerate transient absorption reveals temperature-dependent fine structure splitting (FSS) (see Fig. 3 in the main text and Supplementary Fig. 23 below), which is roughly correlated with the temperature-dependence of the measured lattice constants. Within the context of the quasi-cubic model, we expect the temperature dependence of the lattice constants to be correlated to the FSS via the effect of symmetry breaking which breaks the degeneracy of the triplet bright exciton states.

Here we analyze the effect of the temperature-dependent lattice distortion within the context of a quasi-cubic model as elaborated in Ref. [1]. The approach is motivated by the observation that the lattice constants of the tetragonal and orthorhombic phases are typically within $\sim 2 - 3\%$ of those of the cubic $\alpha$ phase, viewed in a non-primitive supercell aligned to the orthorhombic unit cell. The differences can be quantified by defining strains along the principle axes ***a, b, c*** of the orthorhombic crystal structure relative to the pseudocubic structure viewed in a non-primitive supercell of dimension $(\sqrt{2}\bar{a}_n \times \sqrt{2}\bar{a}_n \times 2\bar{a}_n)$ and aligned to the orthorhombic primitive vectors. Here, $\bar{a}_n$ is the pseudocubic lattice constant, given by,

$$\bar{a}_n = (a\, b\, c/4)^{1/3}, \tag{S19}$$

where $a$, $b$, $c$ are the lattice parameters of the orthorhombic structure. The strain components are thus [1],

$$e_{aa} = \frac{a - \sqrt{2}\bar{a}_n}{\sqrt{2}\bar{a}_n} \;;\quad e_{bb} = \frac{b - \sqrt{2}\bar{a}_n}{\sqrt{2}\bar{a}_n} \;;\quad e_{cc} = \frac{c - 2\bar{a}_n}{2\,\bar{a}_n}. \tag{S20}$$

Using these strain components we then parameterize the effect on the conduction band edge states using a deformation potential model [1,14]. The strain deformation Hamiltonian is constructed using the theory of invariants as [1],

$$H_d = U_d\left\{e_{aa}\hat{L}_a^2 + e_{bb}\hat{L}_b^2 + e_{cc}\hat{L}_c^2 - 2/3(e_{aa} + e_{bb} + e_{cc})\hat{I}\right\}, \tag{S21}$$

where $U_d$ is a deformation potential; here $e_{ii}$ denote the principle components of the strain tensor with index $i$ running over the orthogonal directions ***a, b, c***; $\hat{L}_i$ are the matrices representing the projections of the angular momentum operator $l = 1$ along unit vectors in the ***a, b, c*** directions, and $\hat{I}$ here denotes the $3\times3$ identity matrix. Representing the strain Hamiltonian in a basis of $l = 1$ is appropriate to the orbital p-basis associated with the conduction band Bloch functions, it is straightforward to show that Eq. S21 can be re-cast in terms of irreducible tetragonal and orthorhombic strains, $\delta$ and $\xi$, given respectively by,

$$\delta = e_{cc} - \frac{e_{aa} + e_{bb}}{2}\,, \tag{S22}$$

$$\zeta = \frac{e_{aa} - e_{bb}}{2}\,. \tag{S23}$$

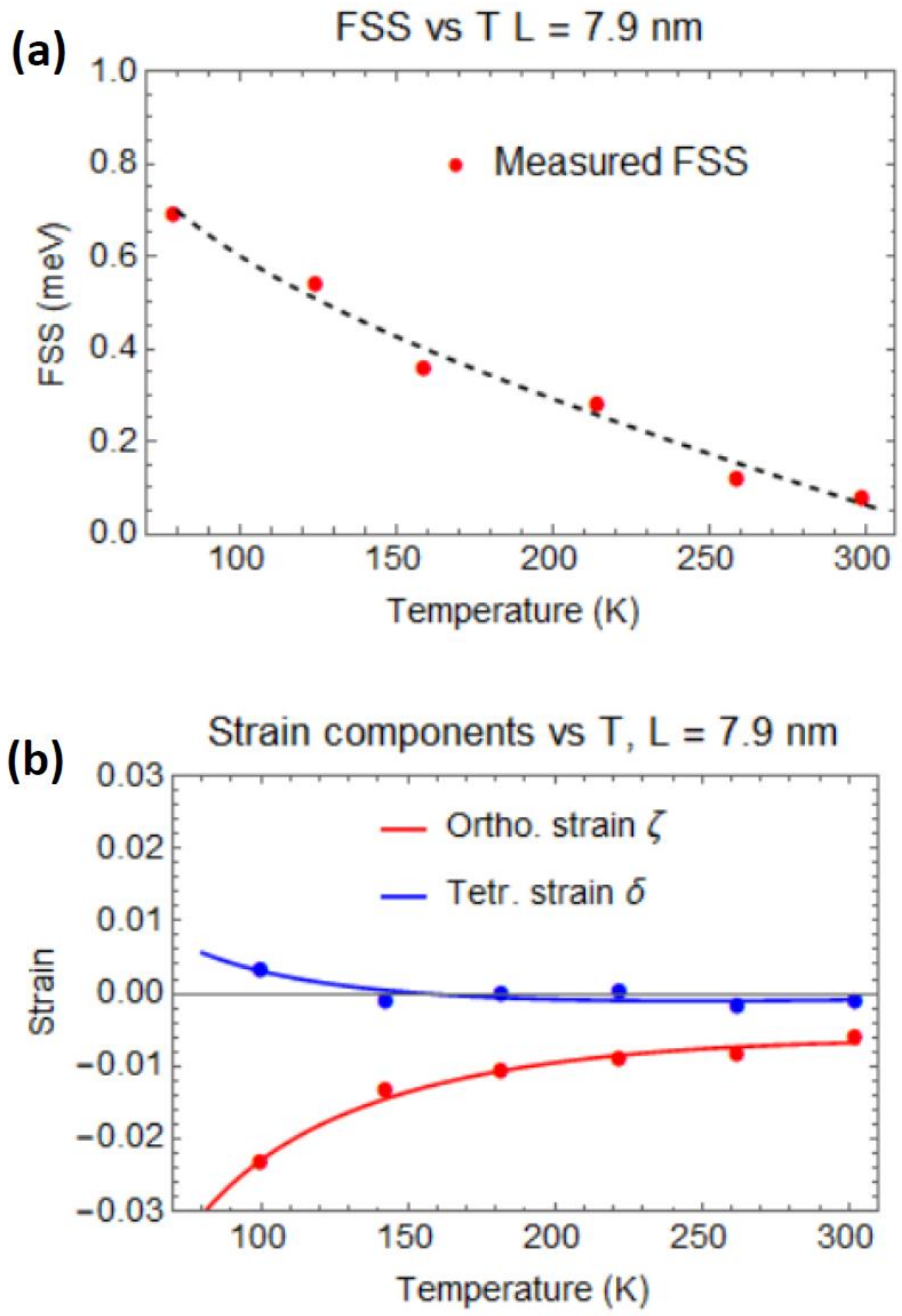

Supplementary Figure 23. Fine-structure splitting (FSS) and strain versus temperature.

Panel (a) shows the FSS determined versus temperature via quantum beating for $CsPbI_3$ QDs of average edge length $L_e$ = 7.9 nm while panel (b) shows the tetragonal and orthorhombic strain components. In panel (b), data points for the orthorhombic strain (red points) and the tetragonal strain (blue points) are calculated from the *measured* lattice constants versus temperature, using Eqs. S20, S22 and S23. The solid lines in both panels are fits to the data using the empirical function of the form $f(T) = c_1 + c_2 T + c_3/T$, where $T$ is the temperature and the $c_i$ are fitting constants.

Supplementary Fig. 23 panel (b) shows a plot of the temperature dependence of the tetragonal and orthorhombic strains, $\delta$ and $\zeta$, calculated from the measured XRD data for the $L_e$= 7.9 nm QDs. The figure shows that the dominant strain is the orthorhombic component. The non-zero strains break the degeneracy of the ***a***, ***b***, and ***c*** directions and the associated orbital p states:

$$H_d = U_d \left\{ \left( \zeta - \frac{\delta}{3} \right) L_a^2 + \left( -\zeta - \frac{\delta}{3} \right) L_b^2 + \left( \frac{2}{3}\delta \right) L_c^2 \right\}. \quad \text{(S24)}$$

In Ref. [1] the effect of the strain deformation on the conduction band edge Bloch functions was considered; adding the strain deformation Hamiltonian to the spin orbit coupling, the result is that the conduction band Bloch functions are modified from Eq. S17: The new Bloch functions which reflect orthorhombic symmetry are,

$$|u_1^e\rangle = -\,\mathcal{C}_c\,|c\rangle|\uparrow\rangle - (\mathcal{C}_a\,|a\rangle + \,i\mathcal{C}_b\,|b\rangle)|\downarrow\rangle\,,$$
$$|u_2^e\rangle = -(\mathcal{C}_a|a\rangle - \,i\,\mathcal{C}_b|b\rangle)|\uparrow\rangle + \mathcal{C}_c|c\rangle|\downarrow\rangle\,. \qquad \text{(S25)}$$

In these expression the lower case symbols $|a\rangle, |b\rangle\,, |c\rangle$ denote orbital functions that transform like *x*, *y*, *z* under rotations as before, but where the axes *x* ,*y* are aligned to the ***a, b*** primitive unit vectors of the orthorhombic lattice. Note the orthorhombic primitive vectors ***a, b*** are rotated by 45 degrees about the c axis relative to the primitive vectors of the cubic lattice. The factors $\mathcal{C}_a, \mathcal{C}_b, \mathcal{C}_{\mathrm{c}}$ are c-numbers that reflect the effect of the crystal field splitting. These can be written approximately in terms of two phase angles, $\theta$ and $\phi$, determined by the crystal fields, $\delta$ and $\zeta$, as [1],

$$\mathcal{C}_a \approx \mathcal{C}_a(\theta,\varphi) = \frac{\cos\phi\cos\theta - \sin\phi}{\sqrt{2}}\,,$$
$$\mathcal{C}_b \approx \mathcal{C}_b(\theta,\varphi) = \frac{\cos\phi\cos\theta + \sin\phi}{\sqrt{2}}\,,$$
$$\mathcal{C}_c \approx \mathcal{C}_c(\theta,\varphi) = \cos\phi\sin\theta \qquad . \qquad \text{(S26)}$$

In these expressions, the phase angle $\theta$ is given in terms of the spin orbit coupling split-off parameter, $\Delta$, and the tetragonal strain $\delta$ by [15]:

$$\tan 2\theta = \frac{2\sqrt{2}\;\Delta}{\Delta - 3U_d\delta}, \qquad \theta \leq \frac{\pi}{2}, \qquad \text{(S27)}$$

while the phase angle $\varphi$ is determined by [1],

$$\tan 2\varphi = \frac{-4\,U_d\,\zeta\;\cos\theta}{\Delta + U_d\,\delta + \sqrt{\Delta^2 - \frac{2}{3}U_d^2\Delta\,\delta + \,U_d^2\delta^2}}\,. \qquad \text{(S28)}$$

Using the electron-hole pair basis,

$$|P_1\rangle = |u_1^e\rangle|u_1^h\rangle;\quad |P_2\rangle = |u_1^e\rangle|u_2^h\rangle;\quad |P_3\rangle = |u_2^e\rangle|u_1^h\rangle;\quad |P_4\rangle = |u_2^e\rangle|u_2^h\rangle\,, \qquad \text{(S29)}$$

we find the SR exchange Hamiltonian has the following representation:

$$\widetilde{H}_{Pair} = \frac{3}{2}\,\hbar\omega_{st} \times \begin{pmatrix} \mathcal{C}_a^2 + \mathcal{C}_b^2 & 0 & 0 & \mathcal{C}_b^2 - \mathcal{C}_a^2 \\ 0 & \mathcal{C}_c^2 & \mathcal{C}_c^2 & 0 \\ 0 & \mathcal{C}_c^2 & \mathcal{C}_c^2 & 0 \\ \mathcal{C}_b^2 - \mathcal{C}_a^2 & 0 & 0 & \mathcal{C}_a^2 + \mathcal{C}_b^2 \end{pmatrix}. \qquad \text{(S30)}$$

This Hamiltonian is diagonalized with the transformation [16],

$$\widetilde{H}_{XYZ} = \widetilde{M}_2^{\dagger}\widetilde{M}_1^{\dagger}\widetilde{H}_{Pair}\;\widetilde{M}_1\,\widetilde{M}_2\,, \qquad \text{(S31)}$$

where the unitary transformation matrices $\widetilde{M}_1$, $\widetilde{M}_2$ are given by,

$$\widetilde{M}_1 = \begin{pmatrix} 0 & 1 & 0 & 0 \\ \frac{-1}{\sqrt{2}} & 0 & \frac{1}{\sqrt{2}} & 0 \\ \frac{1}{\sqrt{2}} & 0 & \frac{1}{\sqrt{2}} & 0 \\ 0 & 0 & 0 & 1 \end{pmatrix} \quad \text{and} \quad \widetilde{M}_2 = \begin{pmatrix} 1 & 0 & 0 & 0 \\ 0 & \frac{-1}{\sqrt{2}} & \frac{i}{\sqrt{2}} & 0 \\ 0 & 0 & 0 & 1 \\ 0 & \frac{1}{\sqrt{2}} & \frac{i}{\sqrt{2}} & 0 \end{pmatrix}.$$

The first transformation ( $\widetilde{M}_1$) transforms the Hamiltonian to a basis of total angular momentum $F = J_e + J_h$ , taken in the order, $|F, F_z\rangle = |0,0\rangle, |1,1\rangle, |1,0\rangle, |1,-1\rangle$, while the second diagonalizes the Hamiltonian in a basis of exciton states $|\psi_{X_i}\rangle$ , taken in the order $|D\rangle, |A\rangle, |B\rangle, |C\rangle$, whose transition dipoles to the crystal ground state respectively vanish (*D*), or are aligned along the symmetry directions ***a, b, c*** of the orthorhombic crystal system for upper case *A*,*B*,*C* respectively. In this basis the exchange Hamiltonian is given by,

$$\widetilde{H}_{DABC} = \begin{pmatrix} E_D & 0 & 0 & 0 \\ 0 & E_A & 0 & 0 \\ 0 & 0 & E_B & 0 \\ 0 & 0 & 0 & E_C \end{pmatrix}. \tag{S32}$$

In this expression, the exciton eigenenergies are given by [1],

$$E_{X_i} = \frac{3}{2}\hbar\omega_{st}\, f_{X_i}\left(\frac{\Theta(L_e)}{\Theta_{bulk}}\right) \quad , \tag{S33}$$

where for each fine structure level the dimensionless parameter $f_{X_i}$ determines the relative energy order. These are given by,

$$f_D = 0 \ , \quad f_A = 2\,\mathcal{C}_a^2, \quad f_B = 2\,\mathcal{C}_b^2\ , \quad f_C = 2\,\mathcal{C}_c^2 \tag{S34}$$

The corresponding transition dipole matrix elements, $\boldsymbol{P}_{\boldsymbol{X}_i} = \langle\psi_{X_i}|\widehat{\boldsymbol{P}}|G\rangle$, evaluated between the exciton state $X_i$, described by state vector $|\psi_{X_i}\rangle$ and the crystal ground state $|G\rangle$, where $\widehat{\boldsymbol{P}}$ is the momentum operator, are given by $\boldsymbol{P}_{\boldsymbol{X}_i} = P\,\mathcal{O}\,\widetilde{\boldsymbol{p}}_{\boldsymbol{X}_i}$. Here, $P = -\mathrm{i}\,\langle s|\hat{p}_{\mathrm{a}}|a\rangle = -\mathrm{i}\,\langle s|\hat{p}_{\mathrm{b}}|b\rangle = -\mathrm{i}\,\langle s|\hat{p}_{\mathrm{c}}|c\rangle$ is the Kane momentum matrix element [17], $\mathcal{O}$ is an overlap integral for the exciton envelope wavefunction, given by, $\mathcal{O} = \int_V f(\boldsymbol{r}_e, \boldsymbol{r}_h) d^3 r$, where the integral is taken over the QD volume, *V*, and $\widetilde{\boldsymbol{p}}_{\boldsymbol{X}_i}$ are dimensionless vectors describing the orientation of the transition dipole which is determined by the crystal structure. The transition dipoles are aligned to the symmetry axes of the orthorhombic crystal, with magnitudes given by [1],

$$\widetilde{\boldsymbol{p}} = 0 \ , \quad \widetilde{\boldsymbol{p}}_A = \sqrt{2}\,\mathcal{C}_a\hat{a}\ , \quad \widetilde{\boldsymbol{p}}_B = \sqrt{2}\,\mathcal{C}_b\hat{b}\ , \quad \widetilde{\boldsymbol{p}}_C = \sqrt{2}\,\mathcal{C}_c\hat{c}\ . \tag{S35}$$

Here, $\hat{a}, \hat{b}$, and $\hat{c}$ are unit vectors aligned to the primitive orthorhombic lattice vectors, ***a, b, c***.

It is useful to express the energies and the oscillator strengths in terms of the orthorhombic and tetragonal strain components. Following Ref.[1], we linearize expressions Eq. S26 to Eq. S28 in the strains, which is a valid procedure provided that $U_d\delta \ll \Delta$ and $U_d\xi \ll \Delta$. The resulting expressions for the dimensionless functions $f_{X_i}$ are given for each fine structure level in terms of the tetragonal and orthorhombic strains, $\delta, \zeta$ by,

$$f_D = 0 \ ,$$

$$f_A = \left(\frac{2}{3} - \frac{4}{9}\frac{U_d\delta}{\Delta_{SO}} + \frac{4}{3}\frac{U_d\zeta}{\Delta_{SO}}\right),$$

$$f_B = \left(\frac{2}{3} - \frac{4}{9}\frac{U_d\delta}{\Delta_{SO}} - \frac{4}{3}\frac{U_d\zeta}{\Delta_{SO}}\right),$$
$$f_C = \left(\frac{2}{3} + \frac{8}{9}\frac{U_d\delta}{\Delta_{SO}}\right). \tag{S36}$$

As shown in Ref.[1]. The relative oscillator strengths for each fine structure level are also proportional to the functions $f_{X_i}$, as seen in Eq. S34 to S35. These functions are plotted versus strain in Supplementary Fig. 24.

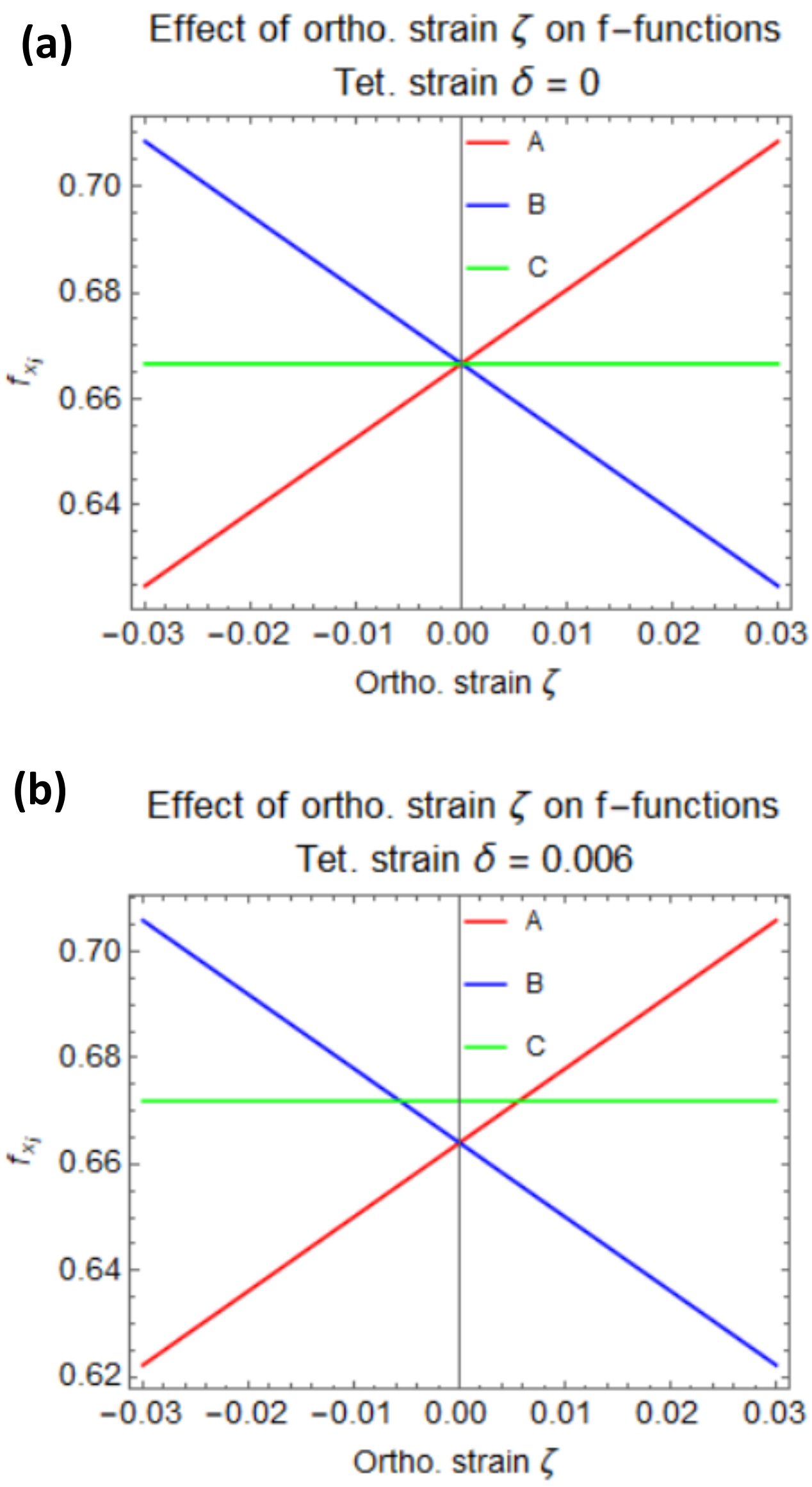

Supplementary Figure 24. Effect of orthorhombic strain on the energies of the bright excitons

Excitons A, B, C have transition dipole along the primitive orthorhombic lattice vectors, ***a, b, c***; their relative energies are determined via the $f_{X_i}$ functions. Plots are calculated with a deformation potential $U_d$ = 1.49 eV determined by fitting described in Supplementary Text 5. Panel (a) shows the $f_{X_i}$ functions versus orthorhombic strain with tetragonal strain set to zero; panel (b) shows the effect of non-zero tetragonal strain $\delta$= +0.006.

### Supplementary Text 4.3. Long range exchange in orthorhombic QDs

In addition to the SR exchange interaction discussed above we must also account for the "long-range" (LR) exchange interaction [13, 18] which contributes to the fine structure splitting of bound excitons [19] and in semiconductor QDs [20, 22, 23]. The LR exchange energy can be evaluated as the Coulomb energy associated with the exciton polarization [23]. In perovskite QDs with exciton states $X_i$, running over $|D\rangle, |A\rangle, |B\rangle, |C\rangle$, the LR exchange corrections are given in general by a 4×4 matrix whose elements are,

$$H^{LR}_{X_i,X_j} = \int_{V_1} dV_1 \int_{V_2} dV_2 \; \left[-\nabla_1 \cdot \boldsymbol{\mathcal{P}}_{X_i}(\boldsymbol{r_1})\right]^* \frac{1}{\epsilon_\infty |\boldsymbol{r_1} - \boldsymbol{r_2}|} \left[-\nabla_2 \cdot \boldsymbol{\mathcal{P}}_{X_j}(\boldsymbol{r_2})\right] \tag{S37}$$

Here, $\epsilon_\infty$ is the high frequency dielectric constant while polarization $\boldsymbol{\mathcal{P}}_{X_i}(\boldsymbol{r})$ is the transition dipole density associated with exciton state $X_i$, given by [1],

$$\boldsymbol{\mathcal{P}}_{X_i}(r_e) = i\frac{\hbar e}{m_0 E_g}\int d^3 r_h\, f(r_e, r_h)\, \tilde{\boldsymbol{p}}_{X_i}\, \delta(r_e - r_h) = i\frac{\hbar e}{m_0 E_g} f(r_e, r_e)\, \tilde{\boldsymbol{p}}_{X_i} \tag{S38}$$

We now consider the evaluation the LR exchange corrections for several cases. First we will review the result of evaluating Eq. S37 in the case that the exciton transition dipoles are oriented perpendicular to the nanocrystal facets. This situation has been assumed in all previous calculations to date of the LR exchange corrections to the exciton fine structure of perovskite QDs [1,11, 22, 24, 25, 26]. This would be the situation if the QD bounding facets are the $\{100\}_o$, $\{010\}_o$, and $\{001\}_o$ planes in the orthorhombic system. We will then evaluate Eq. S37 for the more general situation that the transition dipoles in the (***a,b***) plane are *not* perpendicular to the QD bounding facets. This is motivated by the recent understanding of the facet structure of $CsPbBr_3$ nanoplatelets, revealed by Bertolotti et al. [2] and more recently by Schmidt et al. [3], that the nanoplatelet bounding facets are the pseudocubic planes $\{100\}_c$ referred to the cubic lattice system. The distinction is that in the latter case, the transition dipoles in the absence of LR exchange corrections make at angle of 45 degrees with the lateral bounding facets in the (***a,b***) as shown in Supplementary Fig. 25. The bounding facet in the ***c*** direction is the same in the orthorhombic and pseudocubic systems [3].

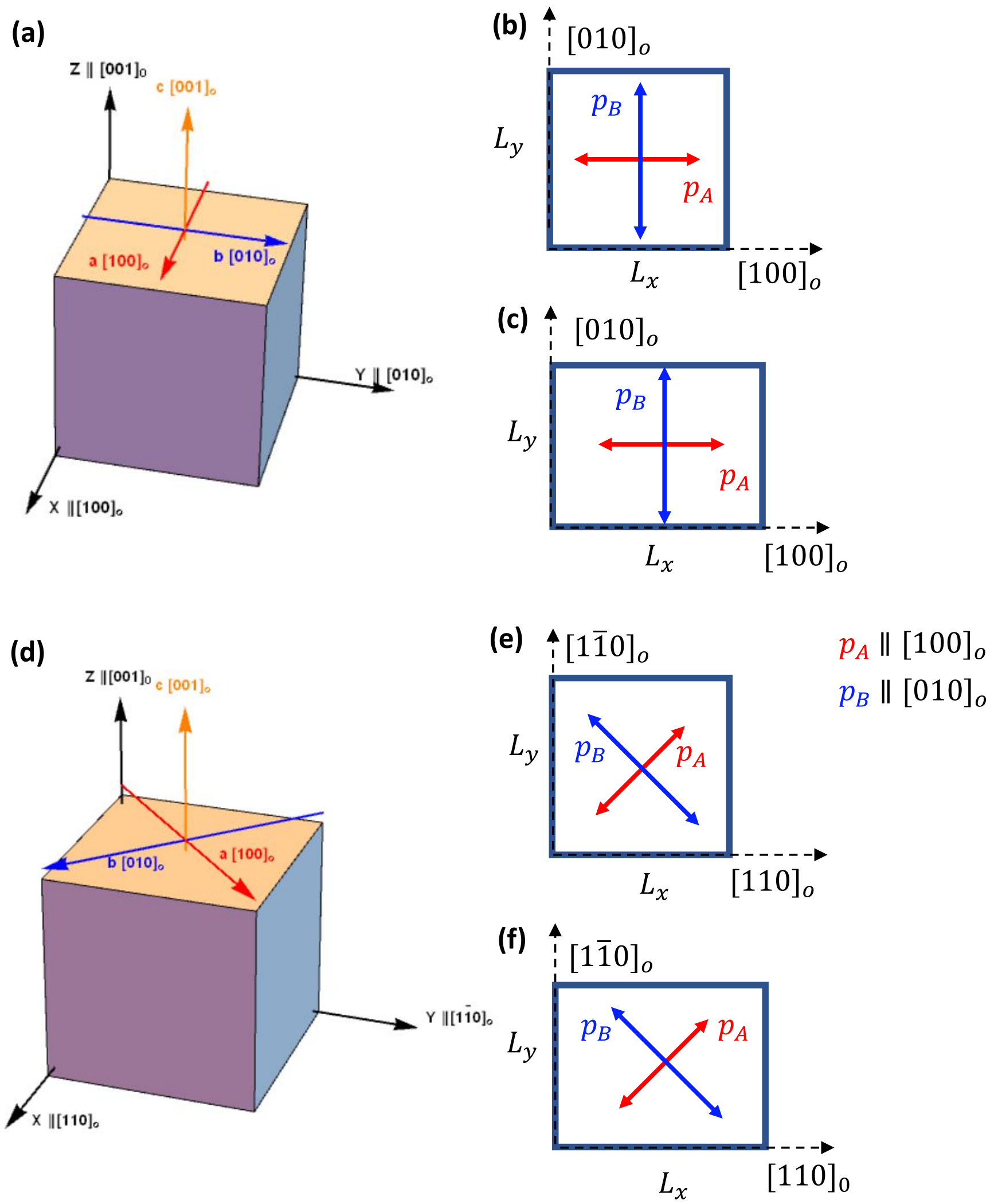

Supplementary Figure 25. Facet structure and orientation of exciton transition dipoles for $CsPbI_3$ QDs.

The orthorhombic facet model is depicted in panels (a-c) while the pseudocubic facet model is shown in panels (d-f). Panels (b) and (c) depict the orientation of the exciton transition dipoles $p_A$ and $p_B$ in the orthorhombic facet model where the facets are assumed to be perpendicular to the ***a, b*** axes of the orthorhombic nanocrystal. The edge lengths $L_x = L_y$ and $L_x \neq L_y$, in panels (b) and (c) respectively, Panels (e) and (f) depict the situation when the facets are perpendicular to the pseudocubic lattice vectors. In this case the transition dipoles are not perpendicular to the QD bounding facets in the (***a,b***) plane, making at angle of 45 degrees to the facets in the absence of LR exchange corrections. Note the magnitude of the transition dipoles associated with exciton states *A*, *B*, *C*, respectively aligned to the ***a, b, c*** lattice vectors, are unequal, see Eq. S35.

### Supplementary Text 4.3.1. LR exchange corrections for QD with orthorhombic bounding facets

We first recap the result corresponding to a cube shaped QD with facets defined by $\{100\}_o$, $\{010\}_o$, and $\{001\}_o$ crystal planes in the orthorhombic system as depicted in Supplementary Fig. 18(a) and 25(a). It was shown in [1] that in that case, the LR exchange interaction does not mix the $D, A, B, C$ exciton states so that the LR exchange Hamiltonian is diagonal. The resulting exciton level energies are expressed as,

$$E_{X_i} = \frac{3}{2}\left\{\hbar\omega_{ST} + \frac{\hbar\omega_{LT}}{3}\right\} f_{X_i}\left(\frac{\Theta(L_e)}{\Theta_{bulk}}\right) . \quad \text{(S39)}$$

Here the size dependence enters the expression through the exchange overlap factor, $\Theta(L_e)$, where $L_e = (V)^{1/3}$ is the effective edge length written in terms of the QD volume $V$, just as appeared in Eq. S33 for the SR exchange energy. The LR exchange energy correction is parameterized in terms of the bulk longitudinal-transverse (LT) exciton splitting $\hbar\omega_{LT}$, given by [1],

$$\hbar\omega_{LT} = \frac{4}{3\epsilon_\infty a_X^3}\frac{E_p}{m_0}\left(\frac{\hbar e}{E_g}\right)^2, \quad \text{(S40)}$$

where $E_p = 2P^2/m_o$ is the Kane energy, defined in terms of the Kane momentum matrix element $P$ previously discussed, $E_g$ is the bandgap, and $a_X$ is the bulk exciton radius. A further refinement derived in Ref. [11] is to add a correction associated with the interaction of the exciton polarization, Eq. S38, with surface image charges: This correction can be expressed in terms of the parameter $\kappa = \epsilon_\infty^{NC}/\epsilon_\infty^{med}$, the ratio of the high frequency dielectric constant in the QD to that of the surrounding medium:

$$E_{X_i} = \frac{3}{2}\left\{\hbar\omega_{ST} + \frac{\hbar\omega_{LT}}{3}\left[1 + \frac{12}{\pi^2}\left(\frac{\kappa - 1}{\kappa + 2}\right)\right]\right\} f_{X_i}\left(\frac{\Theta(L_e)}{\Theta_{bulk}}\right), \quad \text{(S41)}$$

A more significant correction occurs when the QD edge lengths are unequal as shown for the (***a,b***) cross sectional plane in Supplementary Fig. 25 (b). In this situation, it remains the case that the LR exchange Hamiltonian is diagonal: That is, even with unequal edge lengths in the X, Y and Z directions, the *A*, *B*, and *C* bright excitons do not couple to each other. The resulting expressions for the energy were derived in Ref. [1]:

$$E_{X_i} = \frac{3}{2}\left\{\hbar\omega_{ST} + \frac{\hbar\omega_{LT}}{3}\left[1 + \frac{12}{\pi^2}\left(\frac{\kappa - 1}{\kappa + 2}\right)\right]\mathcal{A}_{X_i}\right\} f_{X_i}\left(\frac{\Theta(L_e)}{\Theta_{bulk}}\right). \quad \text{(S42)}$$

Here, the $\mathcal{A}_{X_i}$ are dimensionless shape functions defined as,

$$\mathcal{A}_{X_i} = \mathcal{A}_{X_i}(L_x, L_y, L_z) \equiv \frac{3}{4\pi}\frac{\Omega\, I_{X_i}(L_x, L_y, L_z)}{\Theta(L_e)}, \quad \text{(S43)}$$

which are written in terms of Coulomb integrals $I_{X_i}(L_x, L_y, L_z)$ given by,

$$I_{X_i} = \iiint_{-\frac{L_x}{2},-\frac{L_y}{2},-\frac{L_z}{2}}^{\frac{L_x}{2},\frac{L_y}{2},\frac{L_z}{2}} \mathrm{d}^3\boldsymbol{r}_1 \iiint_{-\frac{L_x}{2},-\frac{L_y}{2},-\frac{L_z}{2}}^{\frac{L_x}{2},\frac{L_y}{2},\frac{L_z}{2}} \mathrm{d}^3\boldsymbol{r}_2 \left[\frac{\mathrm{d}f(\boldsymbol{r_1},\boldsymbol{r_1})}{\mathrm{d}\boldsymbol{r}_{1,i}}\right]^* \frac{1}{|\boldsymbol{r_1}-\boldsymbol{r_2}|} \left[\frac{\mathrm{d}f(\boldsymbol{r_2},\boldsymbol{r_2})}{\mathrm{d}\boldsymbol{r}_{2,\mathrm{i}}}\right]. \qquad \text{(S44)}$$

Here, the derivatives in the integrand are taken with respect to the $i$-components of the position vectors $\boldsymbol{r}_{1,2}$. The dependence of the $\mathcal{A}_{X_i}$ functions for QDs with orthorhombic bounding facets is illustrated in Supplementary Fig. 26 for QDs with edge length $L_x = L_y \neq L_z$, panel (a), and for a biaxial length distortion in panel (b). Examination of Supplementary Fig. 26 shows that elongation (shortening) of the NC along a given direction causes the LR exchange energy correction of the exciton whose transition dipole is parallel to the axis of elongation (shortening) to decrease (increase). It can be shown that the average of the anisotropy functions $\mathcal{A}_{X_i}$ over the three bright excitons is always equal to 1.

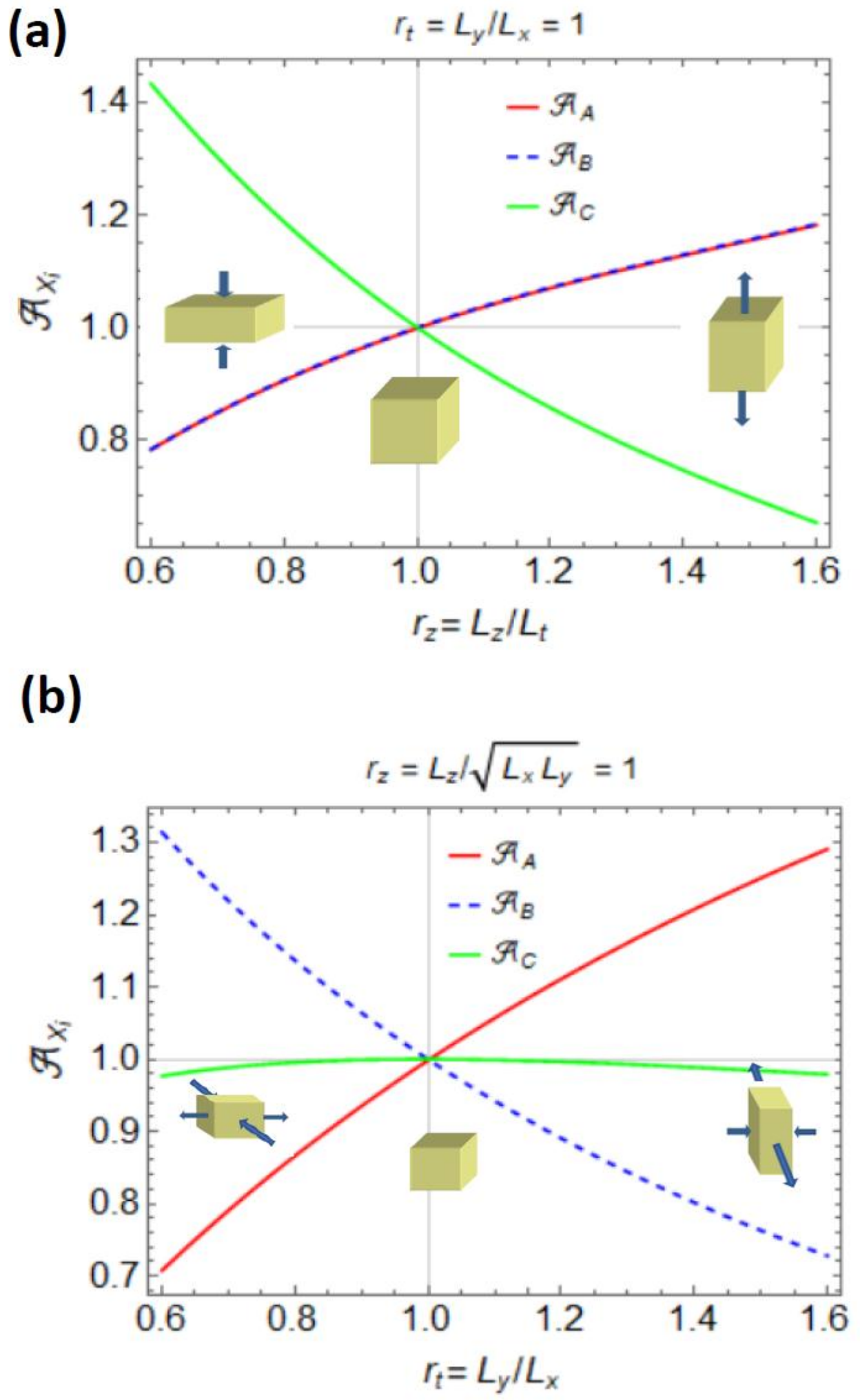

Supplementary Figure 26. Anisotropy functions for uniaxially/biaxially distorted QDs bounded by orthorhombic facets.

Panel (a) shows the anisotropy functions $\mathcal{A}_{X_i}$ with elongation or shortening of the QD edge length along the Z direction, taken as parallel to the orthorhombic $\boldsymbol{c}$ axis, with $L_y = L_x$. In this case the functions $\mathcal{A}_A = \mathcal{A}_B$ by symmetry; all functions $\mathcal{A}_{X_i} = 1$ for the cube shape corresponding to $r_z = L_z/L_x = 1$. Panel (b) shows the $\mathcal{A}_{X_i}$ functions for biaxial edge length distortion corresponding to Eq. S2 with $r_z = L_z/\sqrt{L_x L_y} = 1$. In this case the degeneracy of the $\mathcal{A}_{X_i}$ functions is broken except for the configuration corresponding to the prefect cube, $r_t = L_y/L_x = 1$.

## Supplementary Text 4.3.2. LR exchange corrections for QDs with pseudocubic bounding facets

For the situation where the QD bounding facets belong to the $\{100\}_c$ pseudocubic planes, as depicted in Supplementary Fig. 18(b) and 25(c), the situation is different that developed in the last section. While the *C* exciton whose transition dipole is aligned to the orthorhombic $\boldsymbol{c}$ axis remains uncoupled to the *A* and *B* excitons whose transition dipoles are parallel to the orthorhombic ***a, b*** primitive vectors, the transition dipoles of the *A* and *B* excitons are not perpendicular to the bounding facets $(100)_c$ and $(010)_c$ as shown in Supplementary Fig. 25. Consequently, the LR exchange corrections involve coupling between the *A* and *B* excitons. To handle this situation, we now define the following new coupled anisotropy functions:

$$\mathcal{A}_{X_iX_j} = \mathcal{A}_{X_iX_j}(L_x, L_y, L_z) \equiv \frac{3}{4\pi}\frac{\Omega\, I_{X_iX_j}(L_x, L_y, L_z)}{\Theta(L_\mathrm{e})}, \quad \text{(S45)}$$

where similar to before, the integrals $I_{X_iX_j}$ are given by,

$$I_{X_iX_j} = \iiint_{-\frac{L_x}{2}-\frac{L_y}{2}-\frac{L_z}{2}}^{\frac{L_x}{2}\frac{L_y}{2}\frac{L_z}{2}} \mathrm{d}^3\mathbf{r}_1 \iiint_{-\frac{L_x}{2}-\frac{L_y}{2}-\frac{L_z}{2}}^{\frac{L_x}{2}\frac{L_y}{2}\frac{L_z}{2}} \mathrm{d}^3\mathbf{r}_2\ [\boldsymbol{\nabla}_1 \cdot (f(\boldsymbol{r_1},\boldsymbol{r_1})\hat{\mathrm{p}}_i)]^* \frac{1}{|\boldsymbol{r_1}-\boldsymbol{r_2}|}[\boldsymbol{\nabla}_2 \cdot (f(\boldsymbol{r_2},\boldsymbol{r_2})\hat{\mathrm{p}}_j)]. \quad \text{(46)}$$

In this expression the terms $\hat{p}_i$ are unit vectors that give the direction of the transition dipole associated with exciton $X_i$ in the absence of LR exchange corrections (given in Eq. S35). For clarity we now express the transition dipoles of the *A*, *B* and C excitons, whose transition dipoles are parallel to the orthorhombic ***a, b, c*** axes, in terms of unit vectors in the pseudocubic crystallographic system, which are parallel to the NC facet edges. These are given by the unit vectors,

$$\hat{\mathrm{a}} = (\hat{\mathrm{x}} + \hat{\mathrm{y}})/\sqrt{2}\ ,\ \hat{\mathrm{b}} = (\hat{\mathrm{x}} - \hat{\mathrm{y}})/\sqrt{2}\ ,\ \hat{\mathrm{c}} = \hat{\mathrm{z}}\ . \quad \text{(S47)}$$

It is clear on inspection of Eq. S46 that cross terms between the *A* and *C* excitons, and between the *B* and *C* excitons, given by $\mathcal{A}_{A,C}$ and $\mathcal{A}_{\mathrm{B,C}}$, vanish by symmetry. On the other hand, $\mathcal{A}_{A,A}$ and $\mathcal{A}_{B,B}$ and $\mathcal{A}_{C,C}$ are non-zero. Significantly, the cross coupling between the A and B excitons, given by $\mathcal{A}_{A,B} = \mathcal{A}_{B,A}$, are also in general non-zero.

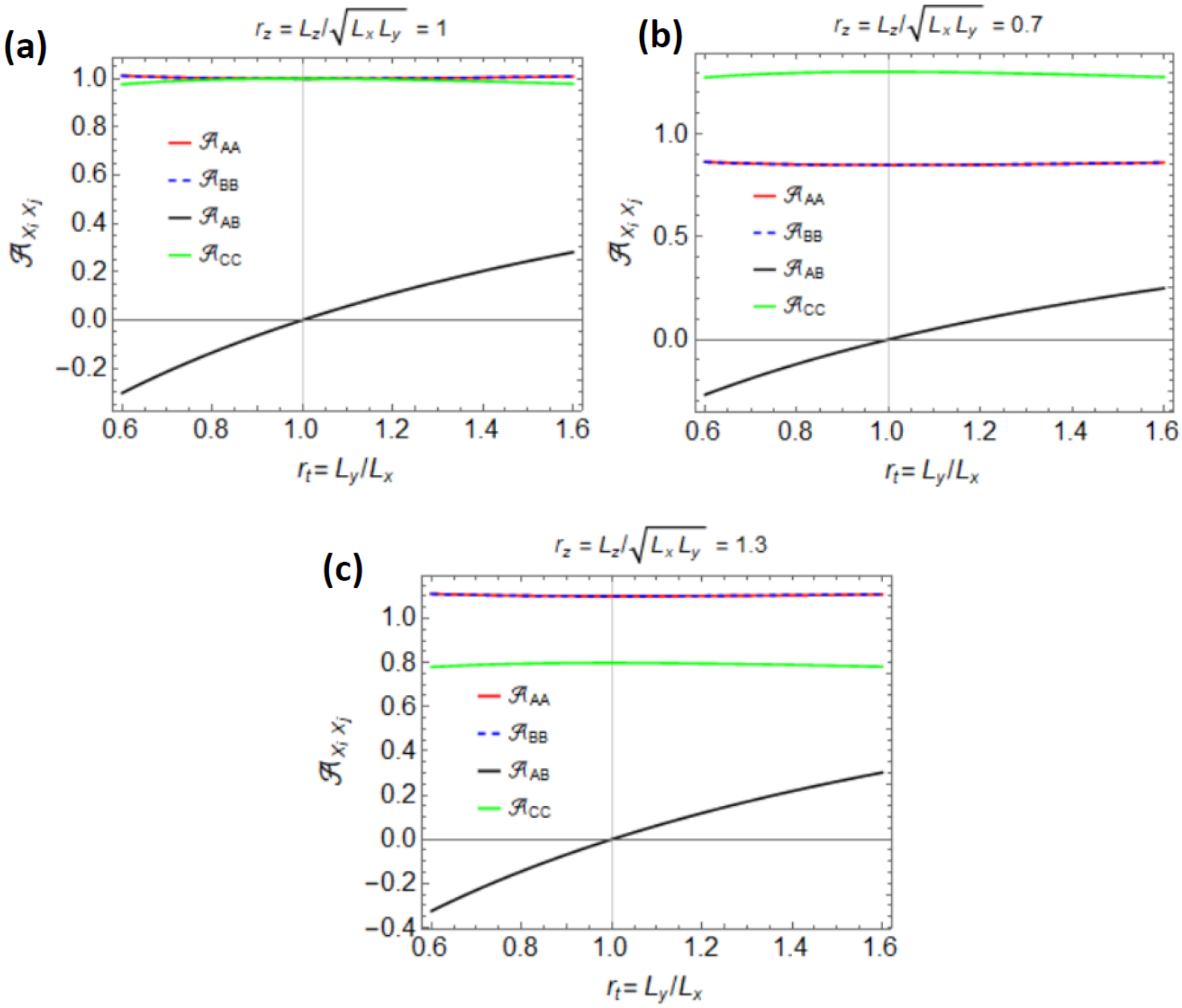

Supplementary Figure 27. Anisotropy functions versus transverse facet edge length ratio for QDs with pseudocubic facets.

Individual edge lengths are given by Eq. S2, with $L_e = \left(L_x L_y L_y\right)^{1/3}$ , and $r_t = L_y/L_x$. The $\mathcal{A}_{X_i X_j}$ functions are shown in panel (a) calculated for z edge length ratio $r_z = 1$ so that the edge length along z is fixed at $L_z = L_e$. In this case, $\mathcal{A}_{A,A} = \mathcal{A}_{B,B}$, and at $r_t = 1$, the functions $\mathcal{A}_{A,A} = \mathcal{A}_{B,B} = \mathcal{A}_{C,C} = 1$ are degenerate while the cross coupling term $\mathcal{A}_{A,B}$ vanishes at this point. In panels (b) and (c), we show the dependence of the $\mathcal{A}_{X_i X_j}$ functions versus transverse ratio $r_t$ for fixed $r_z = L_z/\sqrt{L_x L_y}$ set to 0.7 and 1.3, respectively; in these cases the degeneracy $\mathcal{A}_{A,A} = \mathcal{A}_{B,B}$ is preserved while the degeneracy with $\mathcal{A}_{C,C}$ is broken.

In Supplementary Fig. 27 we show exemplary plots of the $\mathcal{A}_{X_i X_j}$ functions for various biaxial length distortions of a cuboidal QD given by Eq. S2. In panel (a) we set the z edge length ratio $r_z = 1$ so that the edge length along z is fixed at $L_z = L_e$ , and vary the transverse edge length ratio $r_t = L_y/L_x$. For comparison, in panels (b) and (c) we show the dependence of these functions versus transverse ratio $r_t = L_y/L_x$ for fixed $r_z = L_z/\sqrt{L_x L_x}$ set to 0.7 and 1.3, respectively. We see in all cases that for $r_t = 1$ the cross-coupling term $\mathcal{A}_{A,B} = 0$. Further, for all parameters, $\mathcal{A}_{A,A} = \mathcal{A}_{B,B}$ which we expect by symmetry.

Putting it all together the exciton fine structure in the case of pseudocubic facets is given by the following expressions: The dark *D* exciton, and the bright $\hat{c}$-polarized *C* exciton have energies which are the same as given in the orthorhombic facet model. These are given by,

$$E_{\mathrm{D}} = 0\,, \tag{S48}$$

$$E_{\mathrm{C}} = \frac{3}{2}\left\{\hbar\omega_{ST} + \frac{\hbar\omega_{LT}}{3}[1+g_\kappa]\mathcal{A}_{\mathrm{C}}\right\} f_{\mathrm{C}}\left(\frac{\Theta(L_e)}{\Theta_{bulk}}\right), \tag{S49}$$

where, $g_\kappa = \frac{12}{\pi^2}\left(\frac{\kappa-1}{\kappa+2}\right)$. The remaining two excitons are formed from the coupled *A*, *B* excitons and are determined by diagonalizing the following Hamiltonian:

$$\widetilde{\mathrm{H}}_{A,B}{}^{\mathrm{ex}} = \frac{3}{2}\left\{\hbar\omega_{ST}\begin{pmatrix}\mathrm{f_A} & 0\\ 0 & \mathrm{f_B}\end{pmatrix} + \frac{\hbar\omega_{LT}}{3}[1+g_\kappa]\begin{pmatrix}\mathrm{f_A}\mathcal{A}_{\mathrm{A,A}} & \sqrt{\mathrm{f_A f_B}}\mathcal{A}_{\mathrm{A,B}}\\ \sqrt{\mathrm{f_A f_B}}\,\mathcal{A}_{\mathrm{BA}} & \mathrm{f_B}\mathcal{A}_{\mathrm{BB}}\end{pmatrix}\right\}\left(\frac{\Theta(L_e)}{\Theta_{bulk}}\right). \tag{S50}$$

The coupled eigenstates correspond to the pure *A*,*B* excitons only in the case that the NC edge lengths $L_x, L_y$ along the pseudo-cubic axes $[100]_c$ and $[010]_c$, respectively, are equal: $L_x = L_y$, since in that configuration, the cross coupling term $\mathcal{A}_{A,B}$ $\mathcal{A}_{B,A}$ vanishes. In all other cases, the exciton states are a superposition of the *A*, *B* excitons with transition dipoles whose orientation relative to the NC facets changes with the aspect ratio $L_x/L_y$. We label these states as $\alpha$ and $\beta$ to reflect their general mixed A/B character but note that generally these states are not pure *A* or *B*.

To illustrate this point, we show the calculated energies and the orientation of the transition dipoles of the $\alpha$ and $\beta$ states in Supplementary Fig. 28. As we will discuss in the next section, materials parameters used in this calculation, shown in in Supplementary Table 10, reflect measured values from the literature with one exception, the strain deformation potential $U_d$. This is determined by fitting to the FSS data shown in Supplementary Fig. 23(a) using the strain versus temperature data in Supplementary Fig. 23(b). In the plot, the orthorhombic strain is taken $\zeta$ = -0.03 and the tetragonal strain $\delta$ = +0.0056, reflecting the values at *T* = 80 K determined from the empirical fits in Supplementary Fig. 23. In the figure, the energies of the three bright exciton states $\alpha$, $\beta$, and *C* are shown as a function of the edge length ratio $r_t = L_y/L_x$, calculated for a QD with effective edge length , $L_e = L_z$ = 7.9 nm with $L_x, L_y, L_z$ related by Eq. S2. The orientation of the transition dipoles $\boldsymbol{p}_\alpha$and $\boldsymbol{p}_\beta$ are at 45 degrees to the QD facets if the edge lengths $L_x = L_y$; otherwise their orientations rotate to align with the long and short axes of the NC.

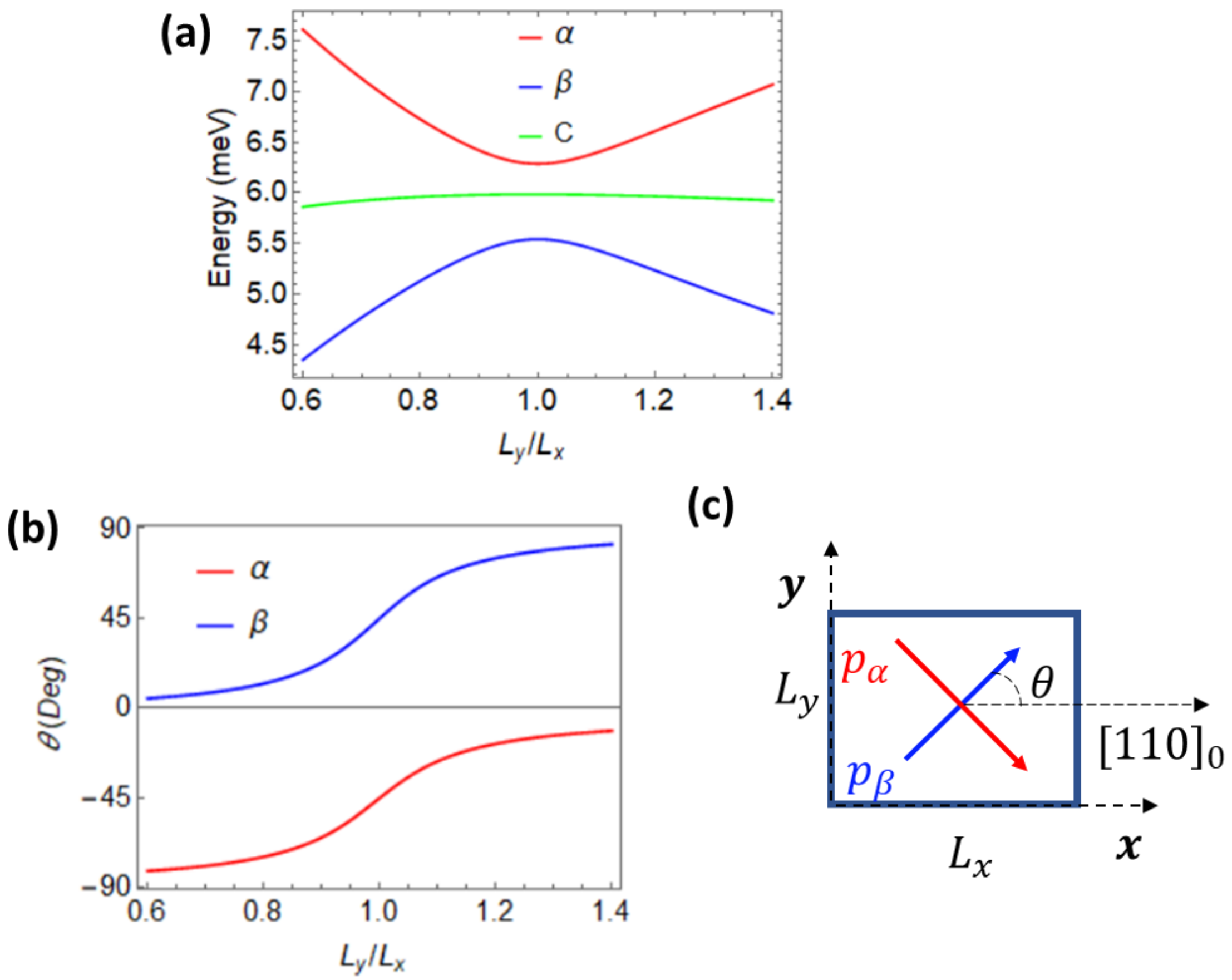

Supplementary Figure 28. Fine structure versus the transverse facet edge length ratio for QDs with pseudocubic facets.

For these calculations, $L_e = L_z$ = 7.9 nm, with edge lengths given by Eq. S2 . Exciton energies are shown versus $r_t = L_y/L_x$ in panel (a) while the angle of the exciton transition dipole for the states $\alpha$ and $\beta$ relative to the x-axis, defined here as parallel to the $[110]_o$ direction in the orthorhombic crystal system, is shown in panel (b). A schematic defining the angle is given in panel (c). The orthorhombic strain is taken as-0.03 and the tetragonal strain as +0.0056, reflecting the values at $T$ = 80 K determined from the empirical fits in Supplementary Fig. 23. All other parameters are as given in Supplementary Table 10.

## Supplementary Text 5. Exciton fine structure in orthorhombic versus pseudocubic facet models.

To illustrate the difference between the pseudocubic and the orthorhombic facet models (see Supplementary Fig. 18 and 25) we show in Supplementary Fig. 29 the bright exciton energies calculated in the orthorhombic facet model as compared with the pseudo-cubic facet model for different biaxial length distortions about a perfect cube shape. In the figure, the bright exciton in the pseudocubic facet model are shown plotted with solid lines and compared to results calculated with the orthorhombic facet model plotted as dot-dashed lines. The $C$ exciton energies are identical for the two models. For all of these calculations, $L_e = \left(L_x L_y L_z\right)^{1/3} =$ 7.9 nm, with edge lengths $L_x, L_y, L_z$ given by Eq. S2.2. Panel (a) and panel (b) show the energies versus the facet edge length ratio $r_t = L_y/L_x$ for fixed $r_z = L_z/\sqrt{L_x L_y}$, set to $r_z$ =1 in panel (a) and $r_z$ =1.15 in panel (b). Panels (c) and (d) show the exciton energies versus the facet edge length ratio $r_z = L_z/\sqrt{L_x L_y}$ for fixed $r_t = L_y/L_x$, set to $r_t$ =1 in panel (c) and $r_t$ =1.15 in panel (d). The orthorhombic strain is again taken $\zeta$ = -0.03 and the tetragonal strain $\delta$ = +0.0056, reflecting the values at $T$ = 80 K determined from the empirical fits in Supplementary Fig. 23. All other material parameters are as given in Supplementary Table 10.

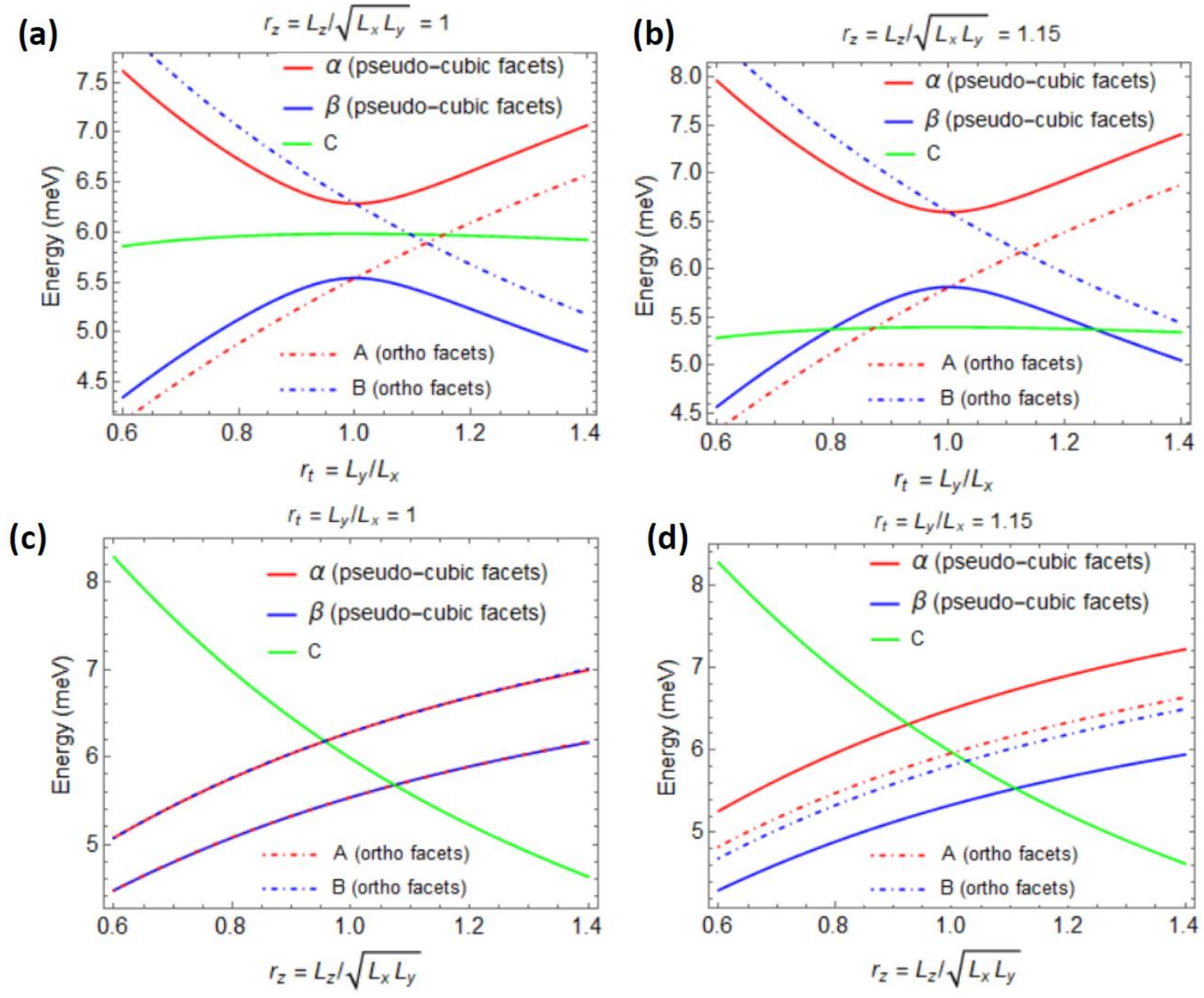

Supplementary Figure 29**.** Comparison of fine structure in QDs bounded by pseudocubic facets versus orthorhombic facets.

In the figure panels, the fine structure is shown for various biaxial length distortions. The bright exciton energies in the pseudocubic facet model are plotted with solid lines and compared to results calculated with the orthorhombic facet model plotted as dot-dashed lines. The $C$ exciton energies are identical for the two models and are shown in solid green. For all of these calculations, $L_e =$ 7.9 nm, with edge lengths given by Eq. S2. Panel (a) and panel (b) show the energies versus the facet edge length ratio $r_t = L_y/L_x$ for fixed $r_z = L_z/\sqrt{L_x L_y}$, set to $r_z$ =1 in panel (a) and $r_z$ =1.15 in panel (b). Panels (c) and (d) show the exciton energies versus the facet edge length ratio $r_z = L_z/\sqrt{L_x L_y}$ for fixed $r_t = L_y/L_x$, set to $r_t$ =1 in panel (c) and $r_t$ =1.15 in panel (d). The orthorhombic strain is taken$as$ -0.03 and the tetragonal strain as +0.0056, reflecting the values at $T$ = 80 K determined from the empirical fits in Supplementary Fig. 23. All other parameters are as given in Supplementary Table 10.

**The plots demonstrate the key difference between the *orthorhombic facet model* and the *pseudocubic facet model*:** Since the bright excitons $\alpha$ and $\beta$ whose transition dipoles lie in the plane spanned by the orthorhombic ***a, b*** primitive vectors are *coupled* via long-range exchange, these two exciton states are *always* separated by an energy gap, even in the presence of variations of the edge length ratios from 1:1:1 characteristic of a perfect cube shape. Such an energy gap does *not* appear in the orthorhombic facet model: With a distribution of the edge length ratios, the energy difference between any two bright exciton states considered across the QD distribution is continuously distributed from zero.

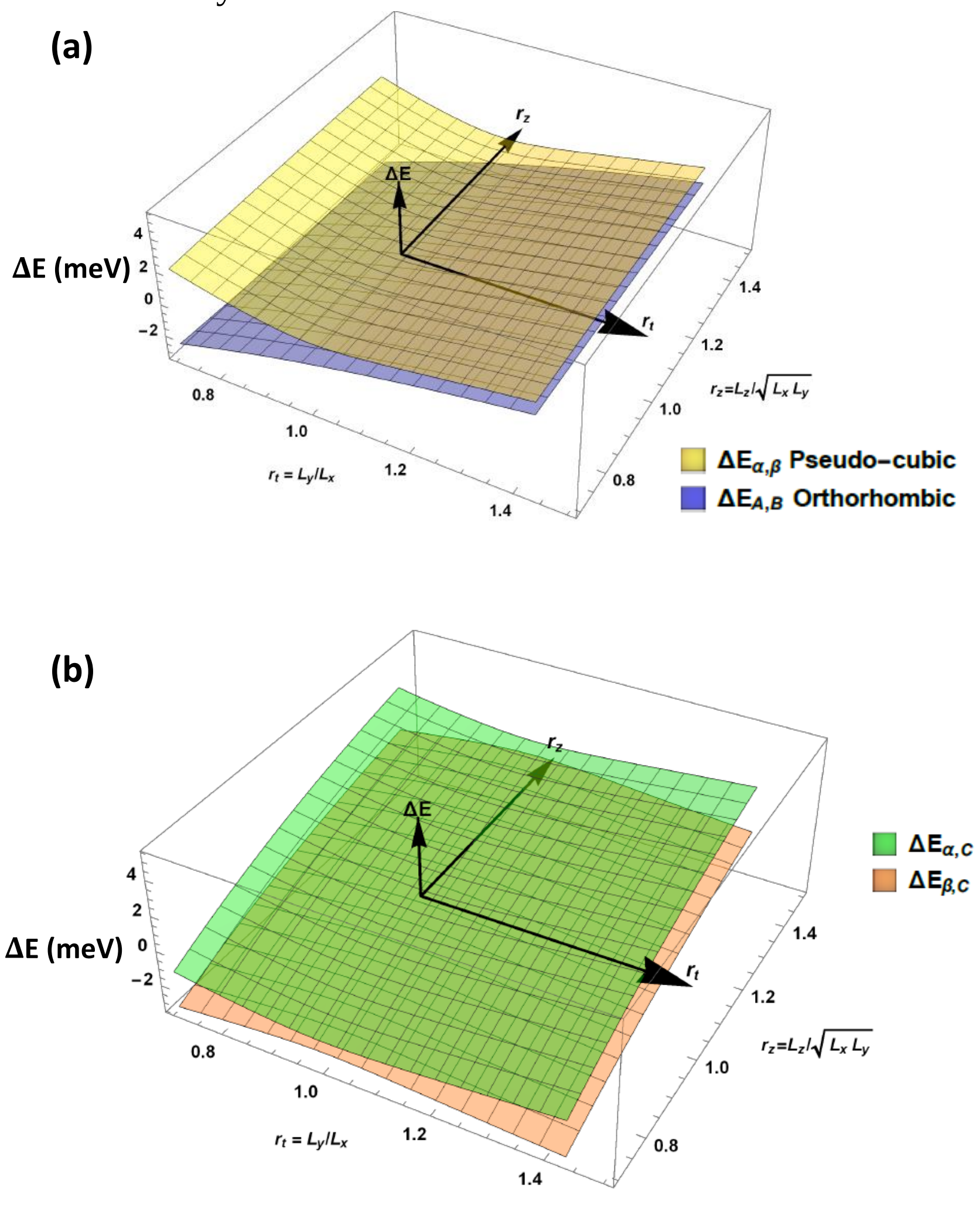

Supplementary Figure 30. Dependence of FSS on QD shape for pseudocubic versus orthorhombic facet models.

In panel (a), the energy spacing $\Delta E_{\alpha,\beta}$ between the $\alpha$ and $\beta$ levels in the *pseudocubic facet* model is plotted versus the biaxial shape distortions $r_t = L_y/L_x$ and $r_z = L_z/\sqrt{L_x L_y}$. Also shown for comparison in panel (a) is the energy separation between the *A* and *B* excitons in the *orthorhombic facet model*. The energy spacing has a minimum at $r_t$ =1 in the pseudocubic model but varies continuously through zero for the orthorhombic facet model. Panel (b) shows the energy separation between the $\alpha$ and $\beta$ levels and the $C$ exciton level in the pseudocubic facet model; these energy surfaces also cross zero; the corresponding surfaces in the orthorhombic facet model are not shown but are qualitatively similar. The orthorhombic strain is taken as -0.03 and the tetragonal strain as +0.0056, reflecting the values at $T$ = 80 K determined from the empirical fits in Supplementary Fig. 23. All other parameters are as given in Supplementary Table 10.

This point is demonstrated explicitly in Supplementary Fig. 30, which shows the exciton level splitting versus shape distortion for the pseudo-cubic facet model versus the orthorhombic facet model. Panel (a) in Supplementary Fig. 30 shows a three-dimensional plot of the energy separation between the $\alpha$ and $\beta$ levels in the pseudocubic facet model plotted versus the biaxial shape distortions $r_t = L_y/L_x$ and $r_z = L_z/\sqrt{L_x L_y}$. The energy spacing has a *minimum* at $r_t$ =1 in the pseudocubic model owing to the level repulsion between the *A*,*B* excitons due to LR exchange coupling. By contrast, in the orthorhombic facet model there is no such minimum: As expected from the plots in Supplementary Figures 3-29, the energy separation goes through zero. Panel (b) in Supplementary Fig. 30 shows the energy separation between the between the $\alpha$ and $\beta$ levels and the $C$ exciton level in the pseudocubic facet model; these energy surfaces also cross zero. The corresponding surfaces showing the energy separation between the $A$ and $C$ and the $B$ and $C$ excitons in the orthorhombic facet model are not shown but are qualitatively similar.
The discussion here and above is critical to the transient absorption measurements described in the main text which show clear signatures of quantum beating. Given the significant shape dispersion exhibited in Supplementary Figs. 18 and 20, any quantum beating signatures in the *orthorhombic facet model* should be completely obscured by shape inhomogeneity, since the energy separation between any two bright exciton levels varies continuously with deviation from perfect cube shape and goes through zero. Similarly, in the pseudocubic facet model, quantum beating between the bright $C$ exciton and the bright $\alpha$ and $\beta$ excitons also will be obscured completely by shape inhomogeneity.

By contrast, owing to the existence of the avoided crossing gap between the $\alpha$ and $\beta$excitons, shown in Supplementary Fig. 30, quantum beating between these states survives averaging over the shape distribution. Indeed, we will show in Supplementary Text 6 that the quantum beating observed reflects the FSS between states $\alpha$ and $\beta$ for those QDs whose shape places the splitting at the minimum in Supplementary Fig. 30. Moreover, the weak dependence of the energy separation between the $\alpha$ and $\beta$ levels on the distortion parameter $r_z = L_z/\sqrt{L_x L_y}$ leads to the quantum beating being dominated by the splitting at the most probable QD shape configuration which corresponds to the cube shape. We take advantage of this fact in the next section to fit the measured fine structure splitting within the quasicubic model using only measured parameters from the literature and the measured lattice constant data over temperature; the single fit parameter required is the strain deformation potential, $U_d$.

## Supplementary Text 5.1. Model parameters and fitting procedure

As noted above and to be fully substantiated in our discussion of the transient absorption in Supplementary Text 6, the quantum beating signature observed in $CsPbI_3$ QDs is dominated by the most probable shape configuration, corresponding to the cube shape, with length ratios (see Eq. S2) $r_t = L_y/L_x = 1$, and $r_z = L_z/\sqrt{L_x L_y} = 1$. In the pseudocubic facet model, this configuration corresponds to a minimum in the FSS between the $\alpha$ and $\beta$ bright excitons. Measurement of quantum beating in co-circular polarized and counter-circular polarized degenerate transient absorption reveals temperature-dependent fine structure splitting (FSS) (see Fig. 3 in the main text and Supplementary Fig. 23 below), which is roughly correlated with the temperature-dependence of the measured lattice constants which break the cubic symmetry. The symmetry breaking is quantified in terms of the temperature-dependent tetragonal and orthorhombic strains, $\delta$ and $\zeta$, calculated from the measured XRD data for the $L$= 7.9 nm NCs and displayed in Supplementary Fig. 23 panel (b).

To test the posited connection between the temperature-dependent lattice strains and the temperature-dependent FSS measured via transient absorption, we performed a fit of the FSS energies measured in TA versus temperature using the quasicubic model, Eq. S50, applied to the average NC shape which is a cube with equal edge lengths, i.e., $r_t = r_z = 1$. The temperature-dependent strains, displayed in Supplementary Fig. 23, enter into the model via the strain-induced symmetry breaking reflected in Eq. S36. The only unknown parameter is the strain deformation potential, $U_d$, which is determined by fit. All other parameter values, summarized in Supplementary Table 10, are taken from measurements or calculations in the literature for bulk or thin-film $CsPbI_3$ or the closely related material methyl-ammonium lead iodide, $MAPbI_3$, where parameters specific to $CsPbI_3$ are not known. Notably, the $CsPbI_3$ bulk bandgap, exciton radius, effective dielectric constant, and reduced effective mass are known for the magneto-transmission measurements reported on thin-film $CsPbI_3$ in Yang *et al.*, Ref. [8]. From the reduced effective mass, we are able to extract the Kane energy parameter $E_p$ using [11],

$$E_p = \frac{3}{2}\frac{m_0}{\mu}\; E_g \qquad \text{(S51)}$$

Given the Kane energy we calculate the LT splitting parameter, $\hbar\omega_{LT}$ using S39. For this calculation we require the exciton radius, $a_x$ = 4.64 nm, determined from the parameters measured by Yang *et al.*, Ref. [8], and the high frequency dielectric constant of $CsPbI_3$, $\epsilon_\infty^{NC}$=5.0, calculated for $CsPbI_3$ by Sapori *et al.* and reported in Ref. [27]. Two key remaining parameters needed are the spin orbit coupling split-off parameter, $\Delta$, and the short-range exchange constant, $\hbar\omega_{ST}$. We use the value of $\Delta$ = 1.42 eV reported for $MAPbI_3$ in Ref. [27], while the bulk singlet-triplet constant $\hbar\omega_{ST}$ = 0.127 meV is determined from the value reported for $MAPbI_3$ in Ref. [27], by assuming the bulk exchange constant $C_{ex}$ is the same for the two materials and adjusting for the different exciton radius and unit cell volume in $CsPbI_3$ using $\hbar\omega_{st} = 2/3\, C_{ex}\, \Omega/(\pi a_x^3)$. Finally, for the outside dielectric constant we use the high frequency value for oleic acid (cis-9-octadecenoic acid) from Ref. [29].

Supplementary Table 10. Summary of material parameters for $CsPbI_3$ QDs.

The strain deformation potential, $U_d$, is determined by fitting as described in the text. All other parameters are taken from measurements or calculations in the literature for bulk or thin-film $CsPbI_3$ or methyl-ammonium lead iodide, $MAPbI_3$, where parameters specific to $CsPbI_3$ are not known.

| Parameter | Symbol | Value | Comment/Source |
|---|---|---|---|
| $CsPbI_3$ bulk bandgap | $E_g$ | 1.723 eV | Ref.[8] |
| $CsPbI_3$ bulk exciton reduced mass | $\mu$ | 0.114 | Ref.[8] |
| $CsPbI_3$ Kane energy | $E_p$ | 22.7 eV | Eq. S51 |
| Exciton effective relative dielectric constant | $\epsilon_{eff}$ | 10 | Ref.[8] |
| $CsPbI_3$ bulk exciton radius | $a_X$ | 4.64 nm | Ref.[8] |
| $CsPbI_3$ bulk exciton binding energy | $B_X$ | 15.5 meV | Ref.[8] |
| $CsPbI_3$ high frequency relative dielectric constant | $\epsilon_\infty^{NC}$ | 5.0 | Ref. [27] |
| SR exchange constant | $C_{ex}$ | 256 meV | Value for orthorhombic $MAPbI_3$, Ref. [27] |
| Singlet-triplet splitting, bulk $CsPbI_3$ | $\hbar\omega_{ST}$ | 0.127 meV | Eq. S15 |
| LT splitting, bulk $CsPbI_3$ | $\hbar\omega_{LT}$ | 2.237 meV | Eq. S40 |
| Outside dielectric constant | $\epsilon_\infty^{med}$ | 2.126 | $\epsilon_\infty^{med} = n^2$; refractive index for oleic acid (cis-9-octadecenoic acid) from Ref. [29] |
| Spin orbit coupling split-off parameter | $\Delta$ | 1.42 eV | Value for $MAPbI_3$, Ref. [28] |
| Strain deformation potential | $U_d$ | 1.485 eV | This work |

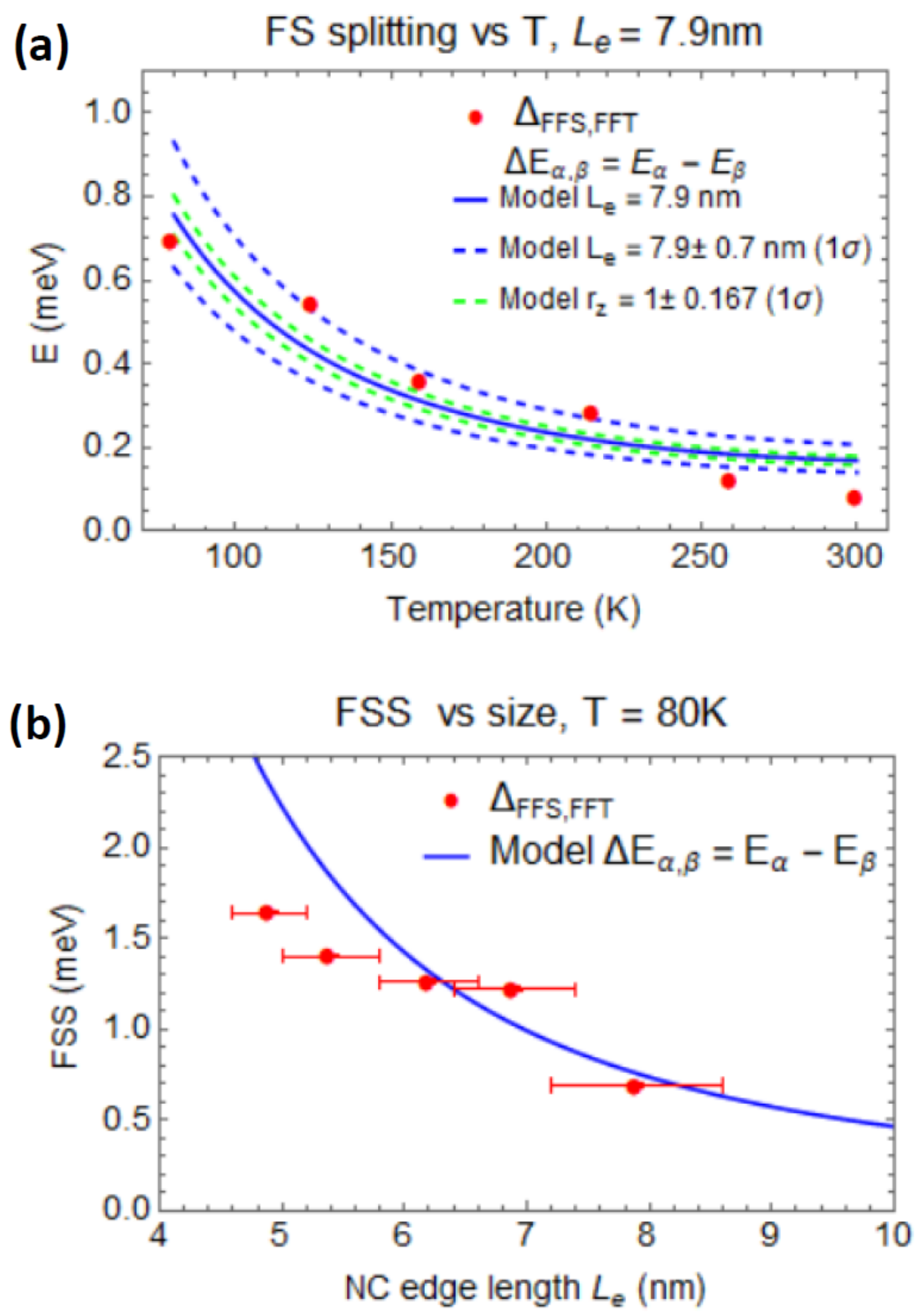

Supplementary Figure 31. Single-parameter fit to measured temperature-dependent exciton FSS.

Panel (a) shows with red markers the measured temperature-dependent fine structure splitting (FSS) for QDs with effective edge length $L_e = 7.9$ nm, determined by FFT of the quantum beating traces. The model fit, plotted with the solid blue line, shows the energy difference between the $\alpha$ and $\beta$ bright excitons calculated for a cube-shaped QD, representing the average shape determined by TEM imaging, see Supplementary Fig. 20. Dashed blue lines show the model calculated at 1 standard deviation ($\sigma$) size variation from the average, $L_e = 7.9$, while dashed green lines show the model calculated at 1 standard deviation variation in the edge length ratio $r_z = L_z/\sqrt{L_x L_y}$ about the average, $r_z = 1$. The single fit parameter is the strain deformation potential, $U_d$, determined as 1.485 eV, which relates the measured strain versus temperature (see Supplementary Fig. 23) to the symmetry breaking of the bright exciton reflected in Eq. S36. Panel (b) applies the model to calculate the size-dependent FSS at $T$ = 80 K. Red markers show the measured FSS determined by FFT of the TA traces, with horizontal error bars reflecting the 1 sigma size variation.

The resulting single-parameter fit to the FSS splitting over temperature for QDs with effective edge length $L_e = 7.9$ nm is shown in Supplementary Fig. 31, panel (a). The measured temperature-dependent FSS over temperature, shown as red points, is determined by fast Fourier transform (FFT) of the temperature-dependent quantum beating traces for QDs with effective edge length $L_e = 7.9$ nm. The model fit, plotted with the solid blue line, shows the energy difference between the $\alpha$ and $\beta$ bright excitons calculated for a cube-shaped QD with size $L_e = 7.9$ nm, representing the average shape and size determined by TEM imaging, see Supplementary Fig. 20. The single fit parameter, the strain deformation potential, $U_d$, is determined as 1.485 eV, which relates the measured strain versus temperature (see Supplementary Fig. 23) to the symmetry breaking of the bright exciton over temperature reflected in Eq. S36. The dashed blue lines in panel (a) show the model calculation at 1 standard deviation size variation from the average, while the dashed green lines show the model calculation at 1 standard deviation in the edge length ratio $r_z = L_z/\sqrt{L_x L_y}$ about the average, $r_z = 1$. (The effect of variation in the edge length ratio $r_t = L_y/L_x$ is separately addressed in Supplementary Text 6). These plots show that the fit is insensitive to deviations from the average cube shape, validating the fit approach. The agreement between the model and the data over temperature is quite reasonable. Re-stating the fitted deformation potential in terms of an effective crystal field [1] resulting from the strains at $T$ = 80 K, we find the values $U_d\zeta = -45.2$ meV for the orthorhombic crystal field, and $U_d\delta = +8.3$ meV for the tetragonal crystal field. These values are similar in magnitude to the values computed using density functional theory for orthorhombic $CsPbBr_3$, namely, $U_d\zeta = -45.5$ meV and $U_d\delta = -5.5$ meV [1] lending confidence in the result and in the reasonableness of the deformation potential found.

Supplementary Fig. 31 panel (b) shows application of the model (blue line) to calculate the *size-dependent* FSS at $T$ = 80 K. Red markers show the measured FSS determined by FFT of the TA traces for NCs of different sizes at $T$ = 80 K, with horizontal error bars reflecting the 1 standard deviation size variation. The size dependence observed in the experiment is clearly captured by the model, which can be traced to the size dependence of the exchange overlap integral, Eq. S12 and shown in Supplementary Fig. 22.

## Supplementary Text 6. Modelling the transient absorption (TA) measurements

In this section we analyze the degenerate transient absorption (TA) measurement on orthorhombic $CsPbI_3$ QDs performed using circularly polarized pump and probe pulses, utilizing the fine structure model developed in Methods. In these experiments, a short optical pump pulse that is spectrally broad in comparison to the exciton fine structure splitting is incident on the sample with wave vector $\boldsymbol{K}$ at time $t = 0$. At a later time $t$ a probe pulse arrives that is also spectrally broad with respect to the fine structure splitting. We first analyze the selection rules for the absorption of the circularly polarized pump and probe pulses and then proceed to apply the fine structure model to describe the quantum beating phenomenon.

## Supplementary Text 6.1. Overview of selection rules for TA with circular polarized light

With respect to the wave vector, the pump and probe pulses are arranged to have either positive or negative helicity, meaning that their angular momentum projection along the direction $\hat{\boldsymbol{K}}$ is $\pm 1$ in units of $\hbar$. For example, for light propagating along the + $\hat{z}$ direction, the circular polarization vectors with positive and negative helicity are respectively given by:

$$\boldsymbol{\sigma}^+ = 1/\sqrt{2}\ (\hat{x} + i\,\hat{y}),$$
$$\boldsymbol{\sigma}^- = 1/\sqrt{2}\ (\hat{x} - i\,\hat{y}). \qquad \text{(S52)}$$

We first consider the absorbance of the sample at a time $t < 0$ before the arrival of the pump pulse. Then, given the QD density $N$, the sample path length, $d$ and the cross section, $\chi$, for absorption by a QD, the absorbance for either $\boldsymbol{\sigma}^{\pm}$ polarization is given by,

$$A(t<0) = \frac{1}{Ln(10)}\,\chi\,N\,d\,. \qquad \text{(S53)}$$

The absorbance is the same for either polarization as the samples are non-chiral and there is no applied magnetic field. Next, we consider the absorbance after the arrival of a pump pulse of $\boldsymbol{\sigma}^+$polarization. We assume that a fraction $p$ of the NCs have absorbed a pump photon while fraction $(1-p)$ have not. Then after arrival of the pump pulse at t=0, subsequent absorption of the sample can be considered according to two subgroups of QDs: 1) Absorption due to the fraction (1-$p$) of the QDs that have *not* absorbed a pump photon, which we label the cold absorbers; and 2), absorption by the fraction $p$ of the QDs that *did* absorb a pump photon. The absorption by the cold absorbers is given by,

$$A_{cold}(t>0) = \frac{1}{Ln(10)}\,\chi\,N\,d\,(1-p)\,. \qquad \text{(S54)}$$

Now we consider the absorption by the fraction, $p$, of the QDs that *did* absorb a pump photon. We know there are four exciton fine structure levels of the system, namely the optically inactive dark D state and the three optically active bright states. To analyze the dynamics it is easiest to first describe the relevant processes in a basis of exciton total angular momentum, $F$, and its projection $F_{\hat{K}}$ along the axis $\hat{\boldsymbol{K}}$ which is parallel to $\boldsymbol{K}$, as shown in Supplementary Fig. 32. Absorption of a photon with polarization $\boldsymbol{\sigma}^+$, which has angular momentum projection $J_{\hat{K}} = +1$, promotes a NC from the QD ground state $G$ to the exciton state $F_{\hat{K}} = +1$ by conservation of angular momentum as shown in the figure. If a subsequent probe pulse arrives that is also polarized $\boldsymbol{\sigma}^+$ ("co-helical pump/probe") then two processes are possible in principle: i) Stimulated emission from the +1 exciton state can occur, proportional to its probability of occupation, $f_{+1}(t)$ , and ii) absorption from the -1 exciton state into the bi-exciton state can occur, proportional to the probability of occupation, $f_{-1}(t)$, of the -1 exciton state, weighted by a factor $\eta_{XX} < 1$ reflecting the energetic shift (and possibly different linewidth) of the transition from the exciton to the lowest energy biexciton, relative to the exciton absorption transition. We note that in the co-helical case considered, the +1 exciton state cannot absorb $\boldsymbol{\sigma}^+$probe light since the accessible bi-exciton state is the singlet; absorption into excited biexciton states is possible but is substantially blue shifted and is expected to have much lower transition probability [30]. On the other hand, if the probe pulse is counter-helical to the pump, i.e., $\boldsymbol{\sigma}^+$ pump with $\boldsymbol{\sigma}^-$ probe polarization**,** there are again two processes that can occur: iii) Stimulated emission from the -1 exciton state can occur proportional to the population $f_{-1}(t)$ in the -1 exciton state, and, iv) absorption from the +1 exciton state into the bi-exciton state, proportional to the population $f_{+1}(t)$ in the +1 exciton state, again weighted by a factor $\eta_{XX} < 1$ reflecting the energetic shift of the biexciton transition. Similar as before, we ignore that the -1 exciton state absorbs $\boldsymbol{\sigma}^-$ probe to populate the lowest energy singlet biexciton. Putting it together then, the absorbance $A_{+,\pm}(t)$ for the system after $\boldsymbol{\sigma}^+$ pump followed by a $\boldsymbol{\sigma}^\pm$ probe at time t, is given by,

$$A_{+,\pm}(t) = \frac{1}{Ln(10)} \{\chi\, N\, d\, (1-p) + \chi\, N\, d\, p\, \eta_{XX}\, f_{\mp 1}(t) - \chi\, N\, d\, p\, f_{\pm 1}(t)\} \ , \tag{S55}$$

where the first term reflects the absorption by the "cold" absorbers, the second term reflects absorption from the exciton to the bi-exciton state, and the third term reflects the stimulated emission process. The transient absorption $\Delta A_{+,\pm}(t) = A_{+,\pm}(t) - A(t<0)$ is thus given by,

$$\Delta A_{+,\pm}(t) = \frac{1}{Ln(10)}\ \chi\, N\, d\ \ \mathrm{p}\,\{\eta_{XX}\, f_{\mp 1}(t) - f_{\pm 1}(t) \ - 1\}, \tag{S56}$$

Normalizing we find,

$$\frac{\Delta A_{+,\pm}(t)}{A} = p\,(\eta_{XX}\, f_{\mp 1}(t) - f_{\pm 1}(t) \ - 1\,)\,. \tag{S57}$$

Eq. S57 is the basis for our analysis below of the quantum beating phenomenon. To proceed, we must now evaluate the time-dependent probabilities $f_{\pm 1}(t)$ of occupation of the exciton $\pm 1$ angular momentum states after excitation by the $\boldsymbol{\sigma}^+$ pump. The key to understanding the quantum beating observed in the TA measurements is that the $+1$ angular momentum state created by absorption of the $\boldsymbol{\sigma}^+$ pump is not a stationary state of an orthorhombic QD.

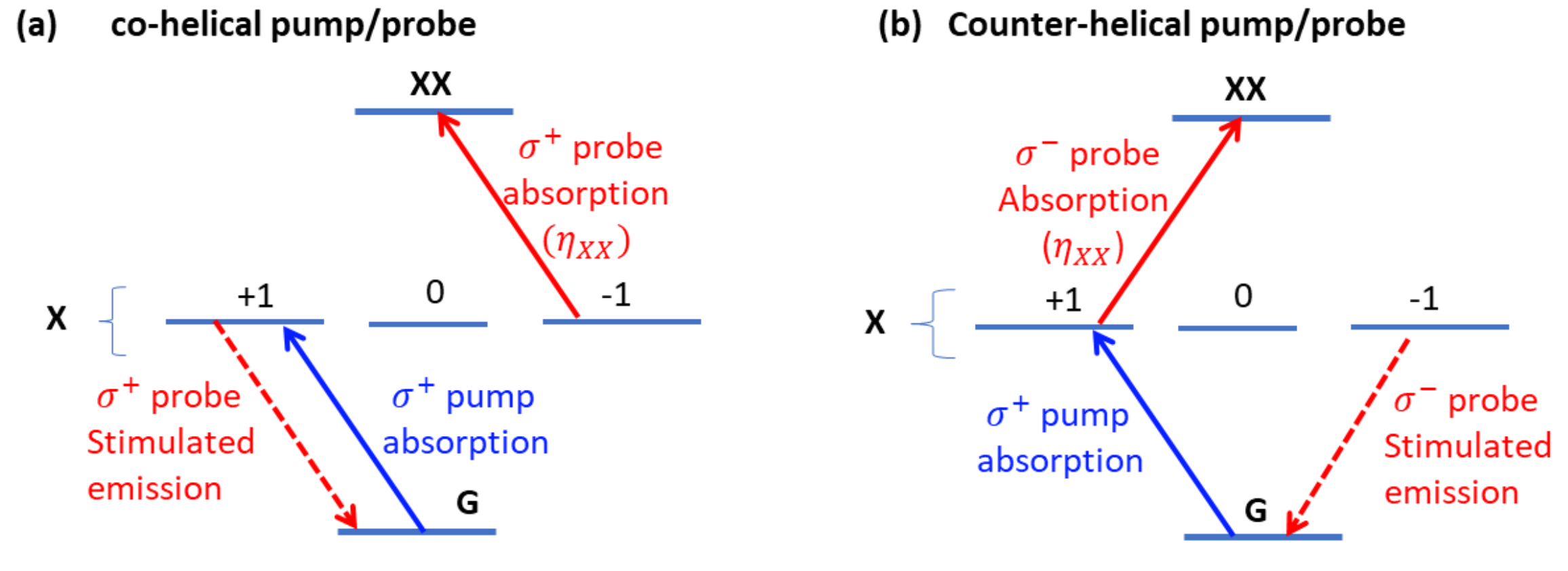

Supplementary Figure 32. Exciton level structure and optical selection rules for circular polarized light.

In both panel (a) and (b), a pump pulse with $\boldsymbol{\sigma}^+$polarization is absorbed, promoting the QD from its ground state, G, to the exciton (X) state with angular momentum projection +1 in the direction of the light wave vector. Panel (a) shows possible processes for $\boldsymbol{\sigma}^+$probe light while panel (b) shows the available processes for $\boldsymbol{\sigma}^-$ probe polarization. Processes include stimulated emission from the +1 X state back to the QD ground state G, as well as absorption from the X states to the singlet biexciton (XX) state. The factor $\eta_{XX} < 1$ weighting the biexciton excited state absorption reflects the energetic shift and different linewidth of the transition from the exciton to the lowest energy biexciton, relative to the exciton absorption transition, see text. In the following TA kinetics simulations, for simplify we actually use $\eta_{XX} = 0$, which can be justified by that, on all the TA spectra shown above, the red-side absorption feature (X to XX absorption) is well separated from the exciton bleach center. This means that the biexciton binding energy already exceeds the transition linewidth, so the contribution of the red-side absorption at the exciton bleach center can be ignored.

## Supplementary Text 6.2. Analysis of quantum beating in TA

As described above, when a pump photon of $\boldsymbol{\sigma}^+$ polarization and wave vector $\boldsymbol{K}$ is absorbed, the QD is promoted from its ground state $G$ to the exciton state with angular momentum projection $F_{\hat{\boldsymbol{K}}} = +1$, satisfying conservation of angular momentum. This exciton state, which we denote as $\Psi_{+1}$, is given by [31],

$$|\Psi_{+1}\rangle = \boldsymbol{\sigma}^+ \cdot \hat{\boldsymbol{p}} \; |G\rangle \,, \tag{S58}$$

where $\hat{\boldsymbol{p}}$ is the momentum operator and we omit normalization factors for clarity. Correspondingly, the exciton state $F_{\hat{\boldsymbol{K}}} = -1$, denoted as $\Psi_{-1}$, is given by

$$|\Psi_{-1}\rangle = \boldsymbol{\sigma}^- \cdot \hat{\boldsymbol{p}} \; |G\rangle \;. \tag{S59}$$

However, these exciton states are not stationary as noted above. To calculate the dynamic response of the system we must express the states $\Psi_{\pm 1}$ each as a superposition of the stationary fine structure levels described in Supplementary Text 4, Eqs. S49-S50. We index these levels $\psi_i$ with energy $E_i$ and a transition dipole for absorption from the crystal ground state given by the vector $\boldsymbol{p}_i$. In this basis, the exciton states with angular momentum $F_{\hat{\boldsymbol{K}}} = \pm 1$ are given by,

$$|\Psi_{\pm 1}\rangle = \sum_{\boldsymbol{i}} |\,\psi_i\rangle\langle\,\psi_i|\;(\boldsymbol{\sigma}^{\pm}\cdot\hat{\boldsymbol{p}})|G\rangle \quad \equiv \sum_{\boldsymbol{i}} C_i^{\pm}\;|\psi_i\rangle\,;$$

$$C_i^{\pm} \equiv (\boldsymbol{\sigma}^{\pm}\cdot\boldsymbol{p_i}) \tag{S60}$$

In this expression, $\boldsymbol{p_i} = \langle\psi_i|\hat{\boldsymbol{p}}|G\rangle$, where $G$ is the crystal ground state. The exciton state created by absorption of a pump photon with polarization $\boldsymbol{\sigma}^+$ then evolves in time as,

$$|\Psi_{+1}(t)\rangle = \sum_{\boldsymbol{i}} C_i^{+}\, e^{-i\,t\frac{E_i}{\hbar}}\,|\psi_i\rangle\,. \tag{S61}$$

We can now evaluate the time-dependent probabilities $f_{\pm 1}(t)$ of occupation of the exciton $\pm 1$ angular momentum states after excitation by the $\boldsymbol{\sigma}^+$ pump. These are given by,

$$f_{\pm 1}(t) \;=\; \left|\langle\Psi_{\pm 1}(0)|\Psi_{+1}(t)\rangle\right|^2 \quad . \tag{S62}$$

This expression can be reduced using the orthogonality of the basis set $\psi_i$:

$$f_{\pm 1}(t) \;=\; \left|\sum_{\boldsymbol{i}}[C_i^{\pm}]^* C_i^{+} e^{-i\,t\frac{E_i}{\hbar}}\right|^2 \quad . \tag{S63}$$

This expression can be reduced into a time-independent term and a time-dependent term:

$$f_{\pm 1}(t) = \sum_{i} \left| \left[ C_i^{\pm} \right]^* C_i^{+} \right|^2 + \sum_{i \neq j} \left[ C_i^{\pm} C_j^{+} \right]^* C_i^{+} C_j^{\pm} \; e^{-i\,t\frac{E_i - E_j}{\hbar}} \quad . \qquad \text{(S64)}$$

The time-dependent terms in the expression above are seen to have frequencies corresponding to the differences between the energies of the stationary fine structure levels, giving rise to quantum beating. Finally, we insert a phenomenological decoherence term, $\tau_{dec}$ and a lifetime term $\tau_1$ (which can derived formally using density matrix theory).

$$f_{\pm 1}(t) = e^{-t/\tau_1} \sum_{i} \left| \left[ C_i^{\pm} \right]^* C_i^{+} \right|^2 + e^{-t/\tau_{dec}} \sum_{i \neq j} \left[ C_i^{\pm} C_j^{+} \right]^* C_i^{+} C_j^{\pm} \; e^{-i\,t\frac{E_i - E_j}{\hbar}} \qquad \text{(S65)}$$

As a further refinement we can replace the exponential decoherence term in Eq. 65 with Gaussian damping, with time constant $\tau_g$, in the limit that the decoherence is due primarily to inhomogeneous broadening rather than other processes such as phonon scattering:

$$f_{\pm 1}(t) = e^{-t/\tau_1} \sum_{i} \left| \left[ C_i^{\pm} \right]^* C_i^{+} \right|^2 + e^{-(t/\tau_g)^2} \sum_{i \neq j} \left[ C_i^{\pm} C_j^{+} \right]^* C_i^{+} C_j^{\pm} \; e^{-i\,t\frac{E_i - E_j}{\hbar}} \qquad \text{(S66)}$$

In the modelling below, we employ Eq. S66 using the Gaussian form of the decoherence, reflecting the size related inhomogeneous broadening shown in Supplementary Fig. 35 below, which is otherwise not included in the model (although shape-related inhomogeneous broadening is accounted for explicitly in Fig. 5 and in Supplementary Figs 37-39 to follow).

To calculate the transient absorption, we insert the expression Eq S66 into Eq. S57, with the further addition that the last term on the right of Eq. S57, which reflects the total population of QDs that are in any excited exciton sub-level, must also reflect $\tau_1$ lifetime decay.

$$\frac{\Delta A_{+,\pm}(t)}{A} = p\,(\eta_{XX}\, f_{\mp 1}(t) - f_{\pm 1}(t) \;\; - e^{-t/\tau_1})\,. \qquad \text{(S67)}$$

Note that the sum in Eq. S65 or Eq. S66 is taken over the three bright levels whose energies are $E_i$ and whose transition dipole for absorption from the crystal ground state are given by the vectors $\boldsymbol{p}_i$. These quantities, which enter into our expression via Eq. S60, are given by Eq. S49, and by diagonalization of Eq. S50, using Eq. S35 for the transition dipoles of the uncoupled exciton levels.

For an inhomogeneous distribution of QDs with different edge lengths $L_x, L_y, L_z$, the energies $E_i$ and the transition dipoles $\boldsymbol{p}_i$ are functions of the size and shape of the NC. In this case we compute the TA signature by integrating the expression Eq. S67 against the distribution function, which we denote as $g(L_x, L_y, L_z)$. Note that because the distribution is explicitly accounted for by this $g$ function, the phenomenological damping time $\tau_g$ here is different from the $T_{dec}$ or $T_\delta$ defined in the empirical expression in Eq 3; the latter two includes the distributional dephasing due to shape and size variation and thus should be shorter than $\tau_g$. Moreover since the NCs can be oriented randomly with respect to the light wave vector ***K***, the model response must be averaged over the NC orientation. In our modeling we take the light vector to be directed in the positive z direction. While it is possible that these cuboid-shaped QDs lie with one of their six facets being flat along the substrate, that assumption may not be correct. Below, we therefore compute orientational averages corresponding to two scenarios: 1) assuming that all QDs lay flat on the substrate, with 1/6 of the QDs oriented respectively with one of *six* facets aligned with its normal parallel to the optical axis with equal probability; or, 2) assuming that the QD orientation is fully randomized in roll, elevation and azimuthal angles. We find that the frequency spectrum of the quantum beating is negligibly different between the two scenarios.

## Supplementary Text 6.3. Calculations of quantum beating in TA neglecting inhomogeneity

To demonstrate the TA model, in Supplementary Fig. 33 we show the quantum beating traces expected for $CsPbI_3$ QDs of a single size and shape, calculated for QDs with pseudocubic facets. Calculations are performed with the QDs oriented with the ***c***-axis parallel to the z-direction, which is the optical axis, in panel a, b, c, and compared to results obtained with the ***c***-axis oriented perpendicular to the z direction, in panels d, e, f. The effective size $L_e = (L_x L_y L_z)^{1/3}$ is set to 7.9 nm and a uniaxial shape is assumed with $L_x = L_y$ but $L_z = 0.9 L_x$. The figure shows that with the c-axis oriented parallel to the optical axis, a single beat frequency is expected, representing beating between the two exciton levels whose dipoles are oriented parallel to the ***a, b*** axes, while for the case with the c-axis perpendicular to the optical axis, all three bright fine structure levels are excited by the pump pulse, leading to *two* beat frequencies in the difference $(\sigma_+\sigma_- - \sigma_+\sigma_+)$ spectrum, contrary to what is measured (see main text Figs. 2,3).

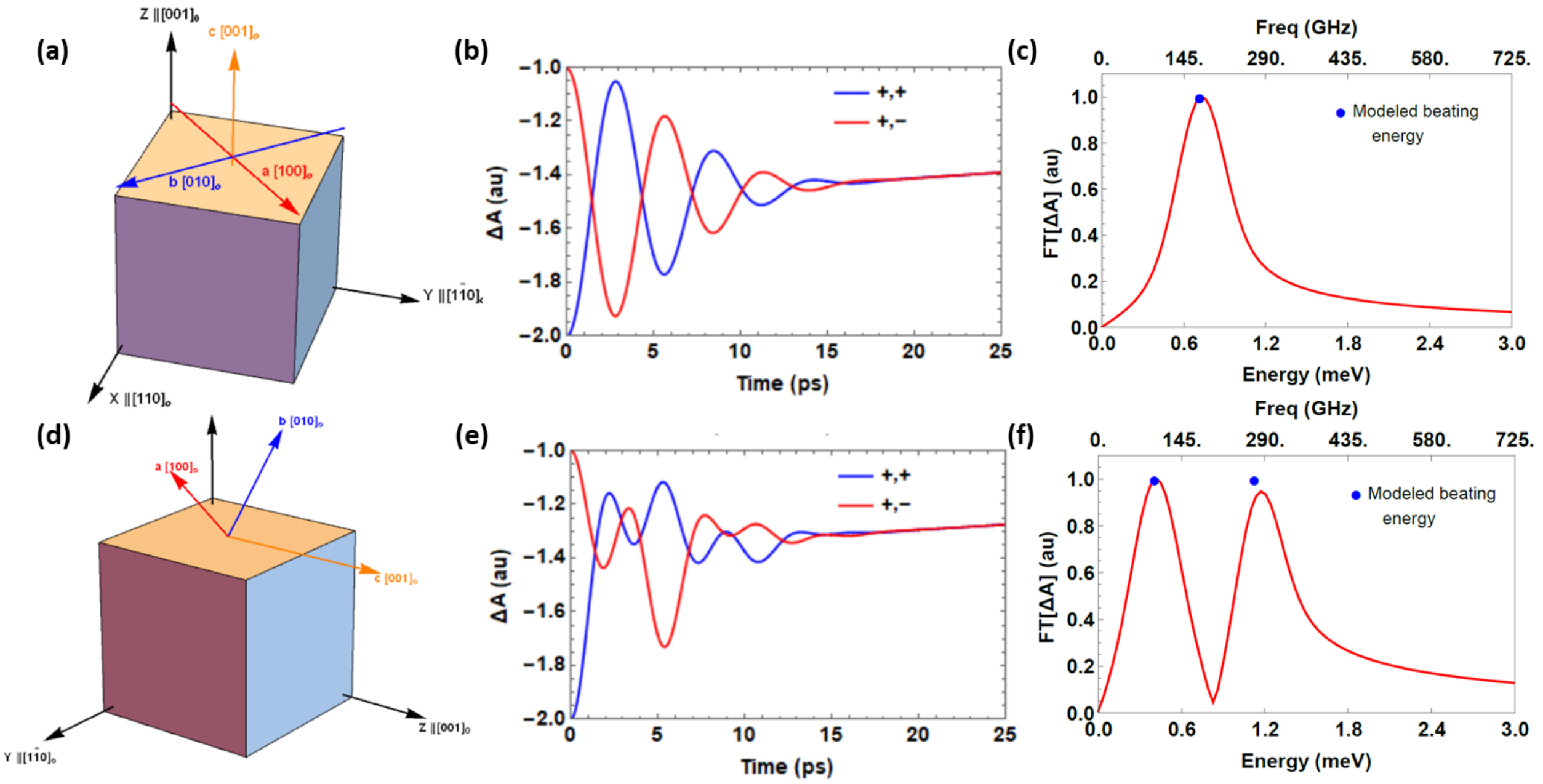

Supplementary Figure 33. Calculated TA and FT spectra for QDs with pseudocubic facets, no shape dispersion.

The QDs have effective size $L_e = 7.9$ nm and are assumed to have a single shape, with $L_x = L_y$ but $L_z = 0.9\,L_x$; the calculation is performed at $T$ = 80 K. Calculations performed with the NCs oriented with the c-axis parallel the optical axis, as depicted in panel (a), are shown in panels (b) and (c), while results obtained with the c-axis oriented perpendicular to the optical axis, depicted in panel (d), are shown in panels (e, f). The Fourier Transforms (FTs) are made on the TA difference trace, $\sigma_+\sigma_- - \sigma_+\sigma_+$ . Material parameters used for the calculation are given in Supplementary Table 10; the phenomenological $\tau_g$ parameter is set to 7.9 ps and the lifetime parameter $\tau_1 = 326.5\,ps$ to match experimentally measured TA traces for this size. For these calculations $\eta_{XX} = 0.$ The orthorhombic strain is taken as -0.03 and the tetragonal strain as +0.0056, reflecting the values at $T$ = 80 K determined from the empirical fits in Supplementary Fig. 23

By contrast, in the orthorhombic facet model, shown in Supplementary Fig. 34, with the QD facets perpendicular to the orthorhombic ***a, b, c*** axes, there is only one beat frequency for any QD orientation, since in this case the three orthogonal bright exciton transition dipoles are always aligned to the ***a, b, c*** primitive vectors and therefore orthogonal to the NC facets as shown in the figure. Superficially, the existence of just one beat frequency matches the experimentally measured TA so we might conclude that the NC bounding facets must be formed from the lowest index orthorhombic crystal planes.

However, this interpretation neglects the fact that the shape distribution of the NCs is *not* uniform. Modelling the TA traces must account for the experimentally measured size and shape distribution, described in the next section.

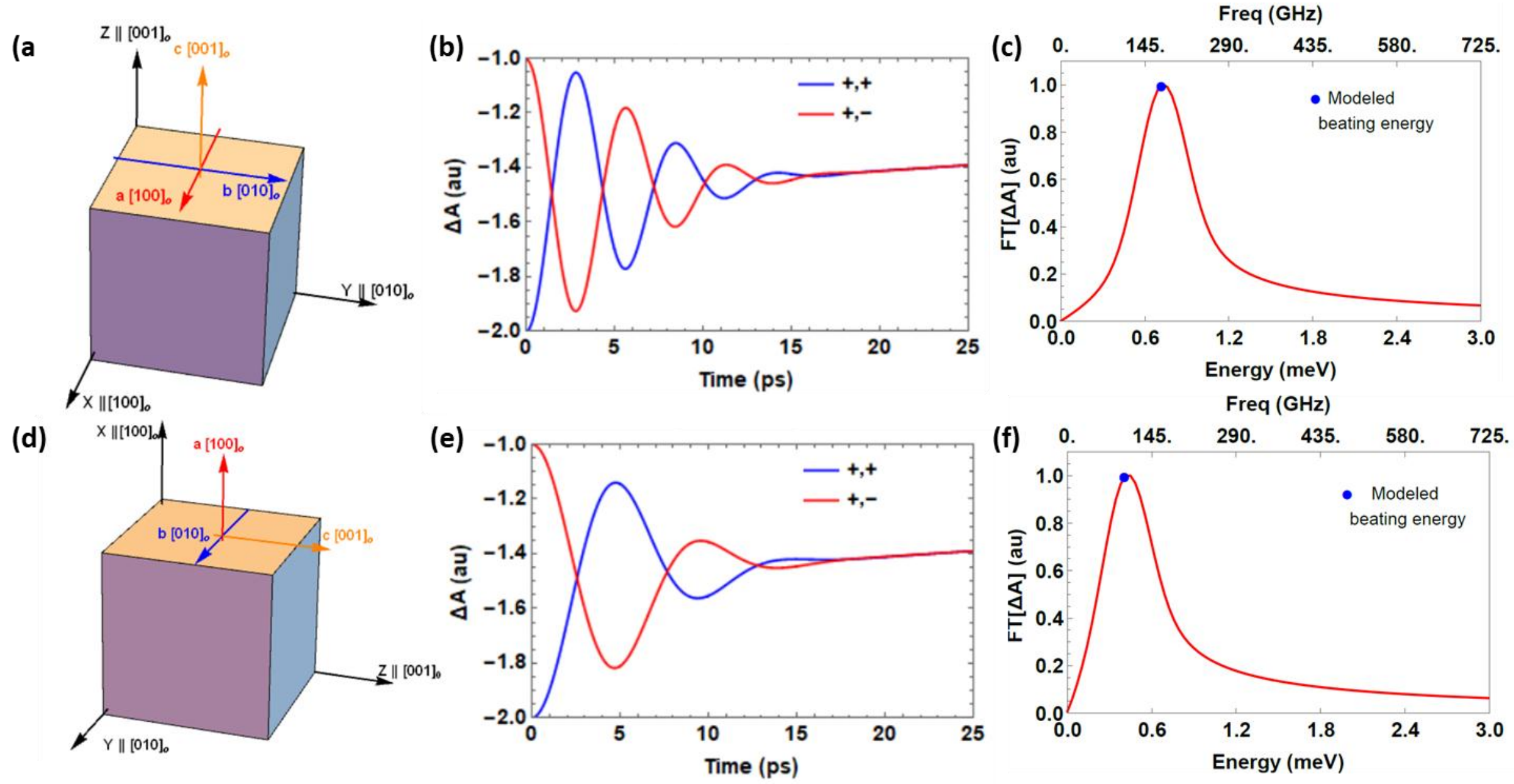

Supplementary Figure 34. Calculated TA and FT spectra for QDs with orthorhombic facets, no shape dispersion.

The QDs have effective size $L_e = 7.9$ nm and are assumed to have a single shape, with $L_x = L_y$ but $L_z = 0.9\,L_x$; the calculation is performed at $T$ = 80 K. Calculations performed with the QDs oriented with the c-axis parallel the optical axis, as depicted in panel (a), are shown in panels (b) and (c), while results obtained with the c-axis oriented perpendicular to the optical axis, depicted in panel (d), are shown in panels (e, f). The Fourier Transforms are made on the TA difference trace, $\sigma_+\sigma_- - \sigma_+\sigma_+$. Material parameters used for the calculation are given in Supplementary Table 10; the phenomenological $\tau_g$ parameter is set to 7.9 ps and the lifetime parameter $\tau_1 = 326.5\,ps$ to match experimentally measured TA traces for this size. For these calculations $\eta_{XX} = 0$. The orthorhombic strain is taken as -0.03 and the tetragonal strain as +0.0056, reflecting the values at $T$ = 80 K determined from the empirical fits in Supplementary Fig. 23

## Supplementary Text 6.4. Effect of QD size and shape inhomogeneity on TA

The $CsPbI_3$ QDs studied here were measured by TEM to determine the distribution of average size and to quantify the distribution of shapes as discussed in Supplementary Text 2 and quantified in Supplementary Fig. 20. Given the dependence of the fine structure energies on the QD volume and on the facet edge length ratios shown in Supplementary Texts 3-5, it is essential to consider the effect of the shape and size inhomogeneity on the level spacings of the QDs, since the sample inhomogeneous broadening should be expected to wash out the quantum beats observed in TA.

We first examine the effect of inhomogeneity in the average edge length. Supplementary Fig. 35 shows the effect of the size inhomogeneity on the distribution of splitting energies expected in a cube-shaped NC of average size 7.9 nm. Owing to the fairly weak dependence of the exchange overlap factor (see Supplementary Fig. 22) on the QD effective edge length, at the level of the size inhomogeneity shown, the energy width of the bright state energy separations is relatively small, less than 0.35 meV. In our calculations we include the effect of the distribution in the effective edge length $L_e = (L_x\, L_y\, L_z)^{1/3}$ within a phenomenological decoherence parameter $\tau_g$, see Eq. S66 .

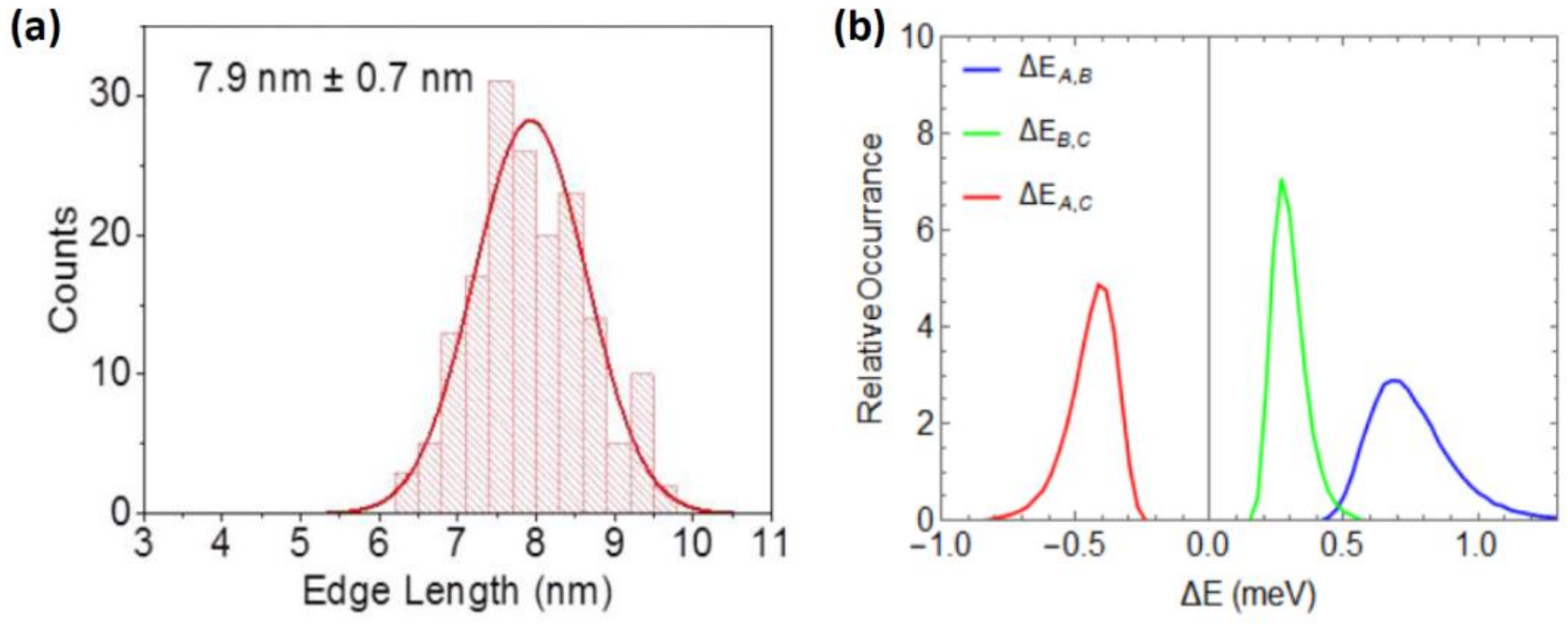

Supplementary Figure 35. Effect of QD size inhomogeneity on bright state FSS.

Panel (a) shows the effective size distribution for $CsPbI_3$ QDs with average edge length 7.9 nm. Panel (b) shows the resulting inhomogeneous distribution of the energy separations between bright exciton levels due to size variation in cube-shaped $CsPbI_3$ QDs with edge length distributed normally about an average of 7.9 nm with a standard deviation of 0.7 nm as in panel (a). Parameters used for the calculation are given in Supplementary Table 10; the orthorhombic strain is taken as -0.03 and the tetragonal strain as +0.0056, reflecting the values at $T$ = 80 K determined from the empirical fits in Supplementary Fig. 23. The width of the distribution of $\Delta E_{A,B}$, representing the energy difference between the bright excitons with transition dipoles oriented along the $\boldsymbol{a}$ and $\boldsymbol{b}$ lattice directions, is 0.35 meV full-width at half maximum (FWHM). This value equals the sum of the FWHM of the other two distributions shown, which correspond to the energy separation between the $A$ and $B$ excitons and the $C$ exciton (whose transition dipole is polarized along the $\hat{c}$ direction), which lies energetically between them

The more significant source of inhomogeneous broadening is due to the measured *shape distribution* reflected in Supplementary Fig. 20 panel (b). To assess the impact of shape inhomogeneity, we employ the QD shape model described in Supplementary Text 2, Eq. S2. In Supplementary Fig. 34 we show the distributions in the bright state energy separations that result due to variation in the transverse length ratio $r_t = L_y/L_x$ at fixed $L_z$, panel (a), and due to variation in the ratio $r_z = L_z/\sqrt{L_x L_y}$ for fixed $r_t = 1$, panel (c). We see from panel (a) that since the bright excitons $\alpha$ and $\beta$ in the *pseudocubic facet model* are *coupled* via long-range exchange, these two exciton states are *always* separated by an energy gap due to the avoided crossing with respect to the shape distortions from perfect cube shape as previously noted in Supplementary Text 5. This results in a sharply peaked distribution of the energy separation $\Delta E_{\alpha,\beta}$, the energy difference between the bright excitons between states $\alpha$ and $\beta$ in the pseudocubic facet model as shown in panel (b); the peak in the distribution occurs at the minimum energy corresponding to the cube shape. Such an energy gap does *not* appear in the *orthorhombic facet model*, for which the energies are plotted using dashed lines in panel (a). With the distribution of the edge length ratios, the energy difference between any two bright exciton states considered across the QD distribution is broadly distributed about zero in this case as indicated by the corresponding distribution of $\Delta E_{A,B}$ in panel (b). Finally, the distribution of the energy separations between the $\alpha$ and $\beta$ and the $C$ exciton, respectively denoted $\Delta E_{\alpha,C}$ and $\Delta E_{\beta,C}$ in panel (d), are broadly distributed due to the lack of an avoided crossing. As a result, quantum beating involving the $C$ exciton will be washed out by shape dispersity.

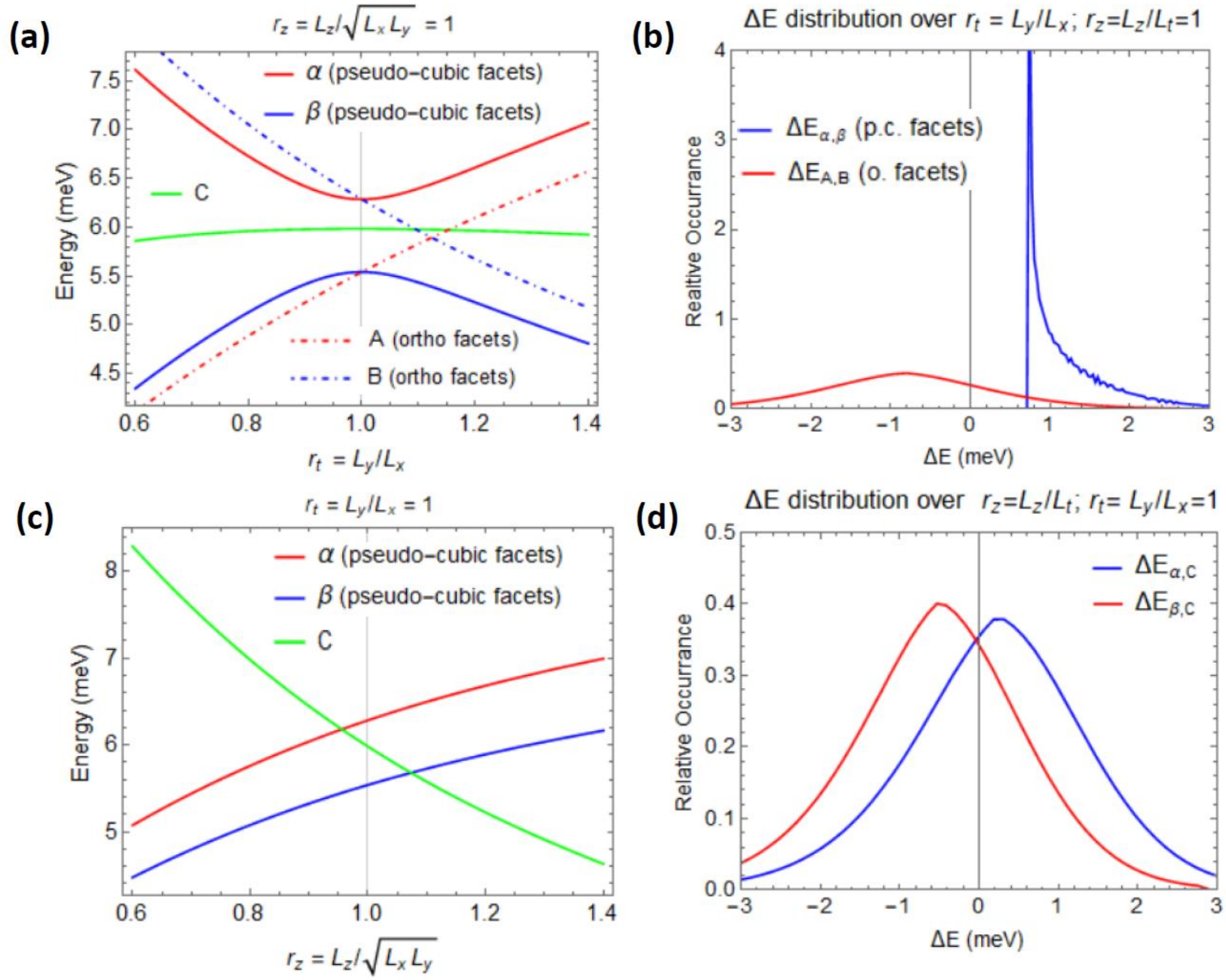

Supplementary Figure 36. Effect of QD shape inhomogeneity on bright state FSS.

The QD model shape comprises a biaxial distorted rectangular prism with edge lengths given according to Eq S2. Panel (a) shows the computed bright exciton energy levels versus the edge ratio $r_t = L_y/L_x$ at fixed $r_z = 1$, so that $L_z = L_e$, where the effective edge length $L_e = \left(L_x\, L_y\, L_z\right)^{1/3}$= 7.9 nm. Energies calculated in the pseudocubic facet model are shown in solid lines, while energies computed in the orthorhombic facet model are plotted in dashed lines. The energy of the $C$ exciton is the same in both models. The resulting distribution of $\Delta E_{\alpha,\beta}$, representing the energy difference between the bright excitons states $\alpha$ and $\beta$ in the pseudocubic facet model, is shown in panel (b) with a solid blue line ("p.c. facets"). The corresponding distribution in the orthorhombic facet model ("o. facets") is shown for comparison in red. In panel (c) the bright state energies are shown versus the edge ratio $r_z = L_z/\sqrt{L_x L_y}$ at fixed $r_t = L_y/L_x = 1$ in the pseudocubic facet model. The corresponding distribution of energy separations between $\alpha$ and $\beta$ and the $C$ exciton, respectively denoted $\Delta E_{\alpha,C}$ and $\Delta E_{\beta,C}$, are shown in panel (d). Parameters used for these calculation are given in Supplementary Table 10; the orthorhombic strain is taken as -0.03 and the tetragonal strain as +0.0056, reflecting the values at $T$ = 80 K determined from the empirical fits in Supplementary Fig. 23.

These points are illustrated in Supplementary Fig. 37 and Supplementary Fig. 38, which show the calculated shape-averaged transient absorption traces and their Fourier transforms in the *pseudocubic facet model* and the *orthorhombic facet model*, respectively, using the corresponding shape distributions modelled in Supplementary Fig. 36. In the top row of Supplementary Fig. 37 we show the shape-averaged TA traces and the Fourier transform (FT) of the difference trace, $\sigma_+\sigma_- - \sigma_+\sigma_+$ with the NC oriented with the c-axis parallel to the optical axis, while the bottom row of Supplementary Fig. 37 shows the corresponding traces with the c-axis oriented perpendicular to the optical axis. Quantum beating is seen in the configuration shown in Supplementary Fig. 37, top row, owing to the peak in the distribution of the energy spacing between the bright excitons states $\alpha$ and $\beta$ due to their avoided crossing. In the bottom row, there is no quantum beating in the difference trace because, with this QD orientation, the transition dipoles of the $\alpha, \beta$ excitons both lie in the Y-Z plane and thus interact equally with the probe light whether polarized $\sigma_+$ or $\sigma_-$; this quantum beat does however occur in the individual TA traces, panel (e).

In the corresponding calculation for the *orthorhombic facet model*, shown in Supplementary Fig. 38, the quantum beating is entirely washed out by shape dispersity as expected from the energy spacing distribution for the "o. facet model" shown in Supplementary Fig. 36 panel (b).

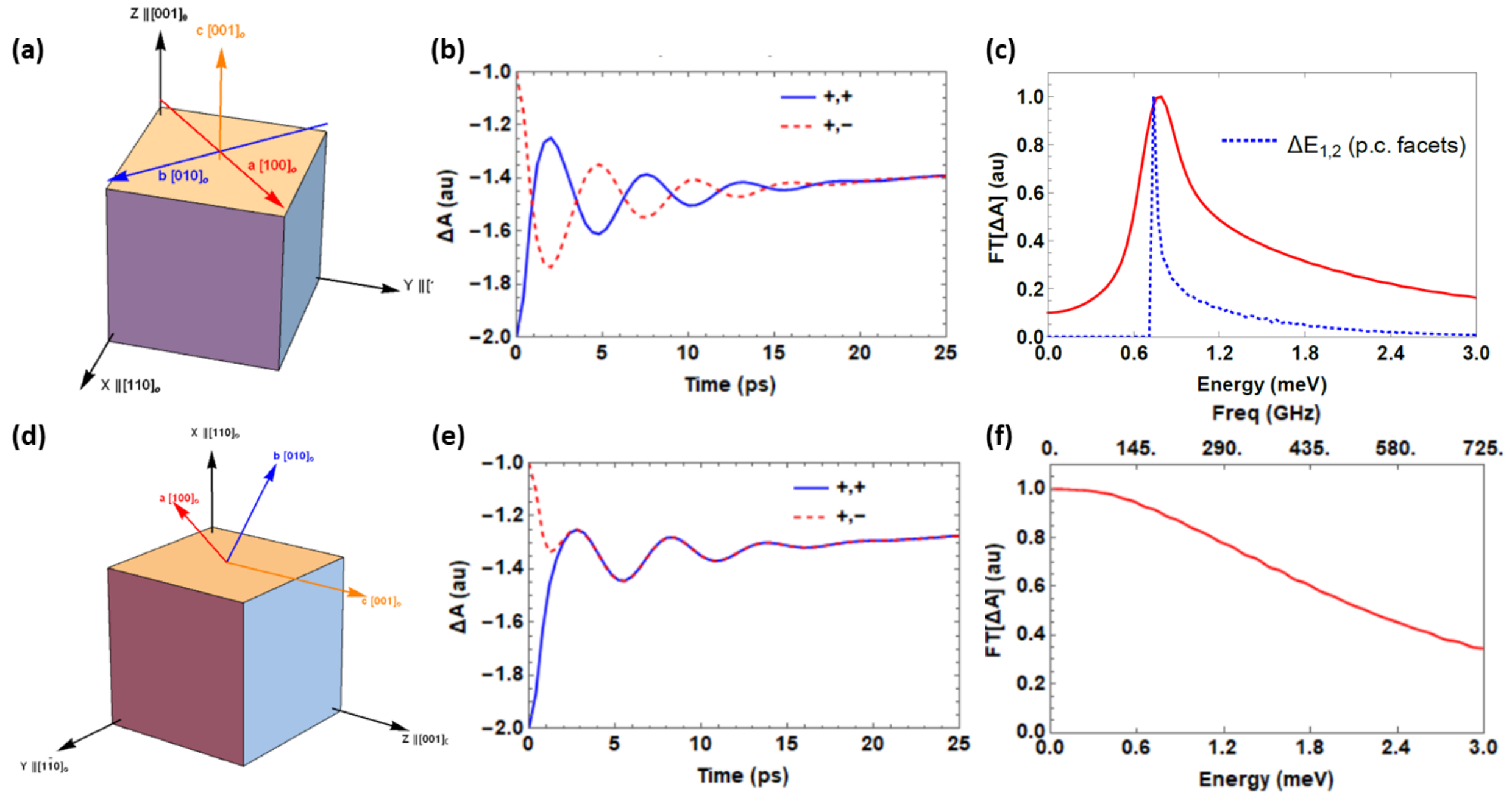

Supplementary Figure 37. Calculated TA and FT spectrum for QDs with pseudo-cubic facets, with shape dispersion.

Calculations are performed at $T$ = 80 K, with QDs of effective size $L_e = 7.9$ nm and with biaxial shape as reflected in Eq. S2 corresponding to the measured shape distribution in Supplementary Fig. 20 panel (b). Panel (a) shows the QD oriented with the c-axis parallel to the optical axis (parallel to Z in the panel). For this configuration the TA traces averaged over the distribution in $r_t = L_y/L_x$ is shown in panel (b) with the Fourier transform (FT) of the difference $\sigma_+\sigma_- - \sigma_+\sigma_+$ shown in panel (c). The peak in the FT spectrum corresponds to the peak in the distribution of $\Delta E_{\alpha,\beta}$, the energy difference between the bright excitons states $\alpha$ and $\beta$ as shown in Supplementary Fig. 36, reproduced in panel (c) for comparison. Panel (d) depicts the QDs with their c-axis oriented perpendicularly to the optical axis (shown parallel to X in the panel). For this configuration, the TA trace averaged over the distribution in $r_z = L_z/\sqrt{L_x L_y}$ is shown in panels (e) with FT spectrum of the difference $\sigma_+\sigma_- - \sigma_+\sigma_+$ shown in panel (f). In this geometry, the quantum beating in the difference trace, $\sigma_+\sigma_- - \sigma_+\sigma_+$, is completely washed out by shape inhomogeneity, see the distribution in Supplementary Fig. 36 panel(d), while there is beating for the individual $\sigma_+\sigma_-$ or $\sigma_+\sigma_+$ traces. Material parameters used for the calculation are given in Supplementary Table 10; the phenomenological decoherence parameter $\tau_g$ is set to 11.8 ps to match the width of the inhomogeneous average size distribution, 0.35 meV, shown in Supplementary Fig 35. The lifetime parameter $\tau_1 = 326.5\, ps$ was set to match experimentally measured TA traces for this size. For these calculations $\eta_{XX} = 0$. The orthorhombic strain is taken as -0.03 and the tetragonal strain as +0.0056, reflecting the values at $T$ = 80 K determined from the empirical fits in Supplementary Fig. 23.

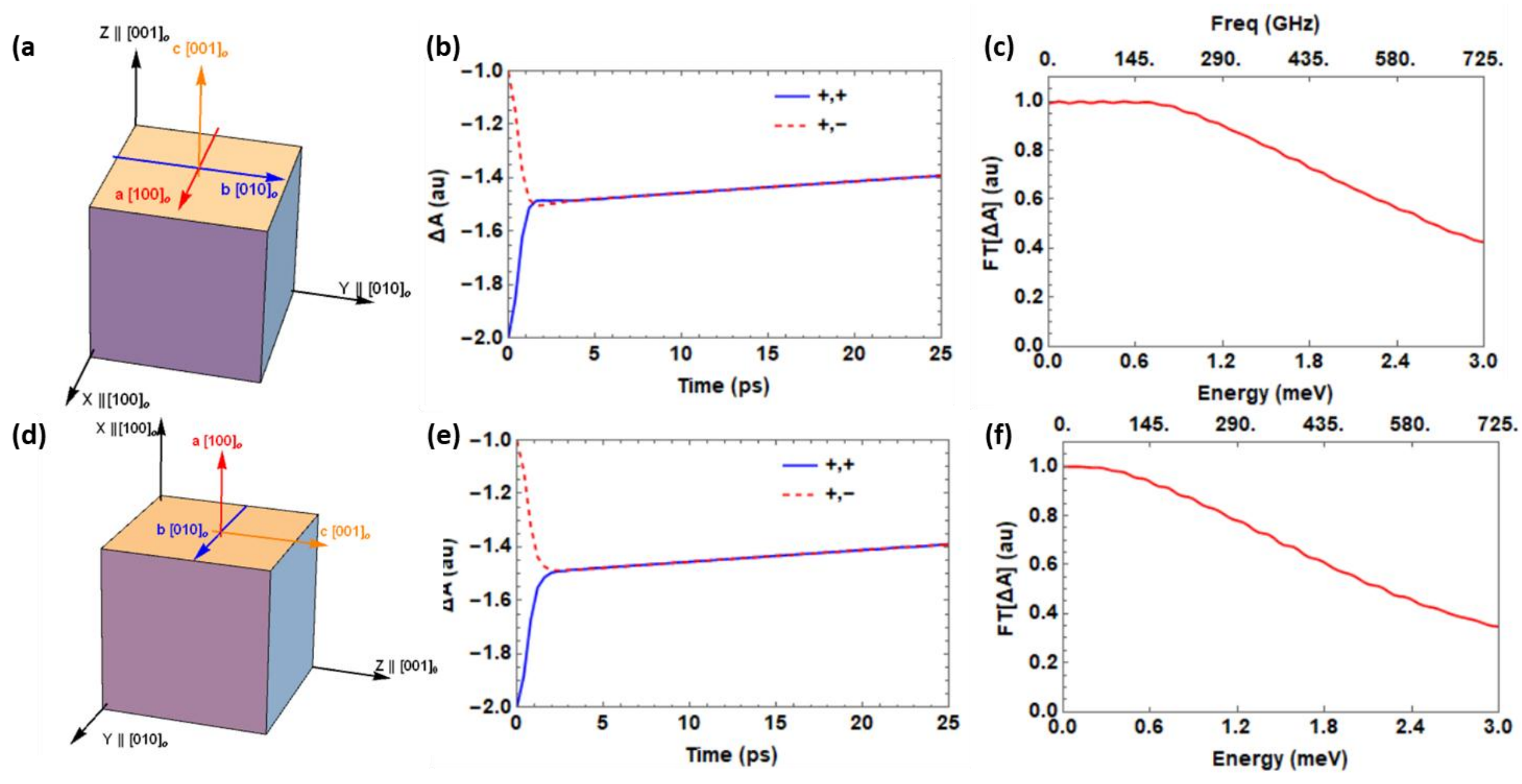

Supplementary Figure 38. Calculated TA and FT spectrum for QDs with orthorhombic facets, with shape dispersion.

Calculations are performed at $T$ = 80 K, with QDs of effective size $L_e = 7.9$ nm and with biaxial shape as reflected in Eq. S2 corresponding to the measured shape distribution in Supplementary Fig. 20 panel (b). Panel (a) shows the QD oriented with the c-axis parallel to the optical axis (parallel to Z in the panel). For this configuration the TA trace averaged over the distribution in $r_t = L_y/L_x$ is shown in panel (b) with the Fourier transform (FT) of the difference trace $\sigma_+\sigma_- - \sigma_+\sigma_+$ in panel (c). Panel (d) depicts the QDs with their c-axis oriented perpendicularly to the optical axis (shown parallel to X in the panel). For this configuration, the TA trace averaged over the distribution in $r_z = L_z/\sqrt{L_x L_y}$ is shown in panels (e) with FT spectrum of the difference $\sigma_+\sigma_- - \sigma_+\sigma_+$ shown in panel (f). In this geometry, the quantum beating is entirely washed out by shape inhomogeneity, see the distribution in Supplementary Fig. 36

panel(d). Material parameters used for the calculation are given in Supplementary Table 10; the phenomenological decoherence parameter $\tau_g$ is set to 11.8 ps to match the width of the inhomogeneous average size distribution, 0.35 meV, shown in Supplementary Fig 35. The lifetime parameter $\tau_1 = 326.5\,ps$ was set to match experimentally measured TA traces for this size. For these calculations $\eta_{XX} = 0$. The orthorhombic strain is taken as -0.03 and the tetragonal strain as +0.0056, reflecting the values at *T* = 80 K determined from the empirical fits in Supplementary Fig. 23.

In the TA measurements the QDs are expected to be randomly oriented with respect to the optical axis. However it is not clear experimentally whether the QDs lay flat on the substrate, i.e., with one third of the QDs oriented respectively with one of the *three* pseudo-cubic facets-called the facets X, Y or Z, aligned with its normal parallel/antiparallel to the optical axis (we label this the Z axis), with equal probability; or whether the QD orientation is fully randomized in roll, elevation and azimuthal angles. In Supplementary Fig. 39 we compare these two scenarios, and find that the frequency spectrum of the quantum beating is negligibly different between the two.

A general orientation can be described conceptually by starting with a QD whose c-axis is aligned parallel to the optical axis, Z, with its X and Y facets aligned to laboratory x, y directions. The orientation can then be specified by executing a roll rotation by angle $\phi$ about the c axis, followed by a rotation about the laboratory y-axis by elevation angle $\vartheta$, followed by an azimuthal rotation of the QD about the optical axis, Z, by azimuth angle $\gamma$. In evaluating the question at hand, the problem is simplified by the fact that the interaction of the QD with circularly polarized light incident along the Z direction is independent of the azimuthal angle, $\gamma$ , of the QD about the optical axis (since the electric field vector sweeps the azimuth angle at optical frequencies). In Supplementary Fig. 39 panel (a) we show schematically the first case, where the red arrows represent the orientation of the c-axis where for clarity we have set the azimuth angle equal to the roll angle. This situation, which can be described as a Lebedev quadrature expansion with precision p = 3 [32,33,34], corresponds to the six QD orientations with the c-axis aligned to +/- x, +/- y and +/- z directions, respectively, that is, aligned along the mutually orthogonal x, y and z directions. In this case when the c-axis aligned to +/- x, the QD X-facet is directed with its normal parallel or antiparallel to the optical axis; when the c-axis aligned to +/- y, the QD Y-facet is directed with its normal parallel or antiparallel to the optical axis. Panels (b) and (c) respectively show the transient absorption signature for $\sigma_+$ pump followed by either $\sigma_+$ or $\sigma_-$ probe at the exciton line, and the Fourier transform of the difference $\sigma_+\sigma_- - \sigma_+\sigma_+$. The peak in the FT spectrum in panel (c) corresponds to the peak in the distribution of $\Delta E_{\alpha,\beta}$, the energy difference between the bright excitons states $\alpha$ and $\beta$ as shown in Supplementary Fig. 36, which originates from the avoided crossing between these states. All other quantum beating signatures are obscured by the shape dispersity of the samples. These calculations were performed using the full "two-dimensional" shape distribution shown in Supplementary Fig. 20 panel (d).

In Supplementary Fig. 39 panel (d) we show schematically the second case of fully randomized orientation, where the averaging is performed using Lebedev quadrature with precision p = 11 [32,33,34], corresponding to averaging the TA response over 50 distinct equally separated points on the unit sphere representing the roll and elevation angles $\phi, \vartheta$ of the QDs. As in panel (a), the blue arrows in panel (d) represent the orientation of the c-axis where for clarity we have set the azimuth angle equal to the roll angle. At precision p11, Lebedev quadrature can be shown to be exact for averaging transient absorption, whose orientational dependence can be described by a polynomial of order p=8 over the surface of the unit sphere representing the roll and elevation angles $\phi, \vartheta$; convergence was verified numerically. Panels (e) and (f) respectively show the orientationally averaged transient absorption signature for $\sigma_+$ pump followed by either $\sigma_+$ or $\sigma_-$ probe at the exciton line, and the Fourier transform of the difference $\sigma_+\sigma_- - \sigma_+\sigma_+$. The peak in the FT spectrum in panel (d) again corresponds to the peak in the distribution of $\Delta E_{\alpha,\beta}$, the energy difference between the bright excitons states $\alpha$ and $\beta$ as shown in Supplementary Fig. 36, and we see that there is negligible difference between the results shown in Supplementary Fig. 39 panels (c) and (f).

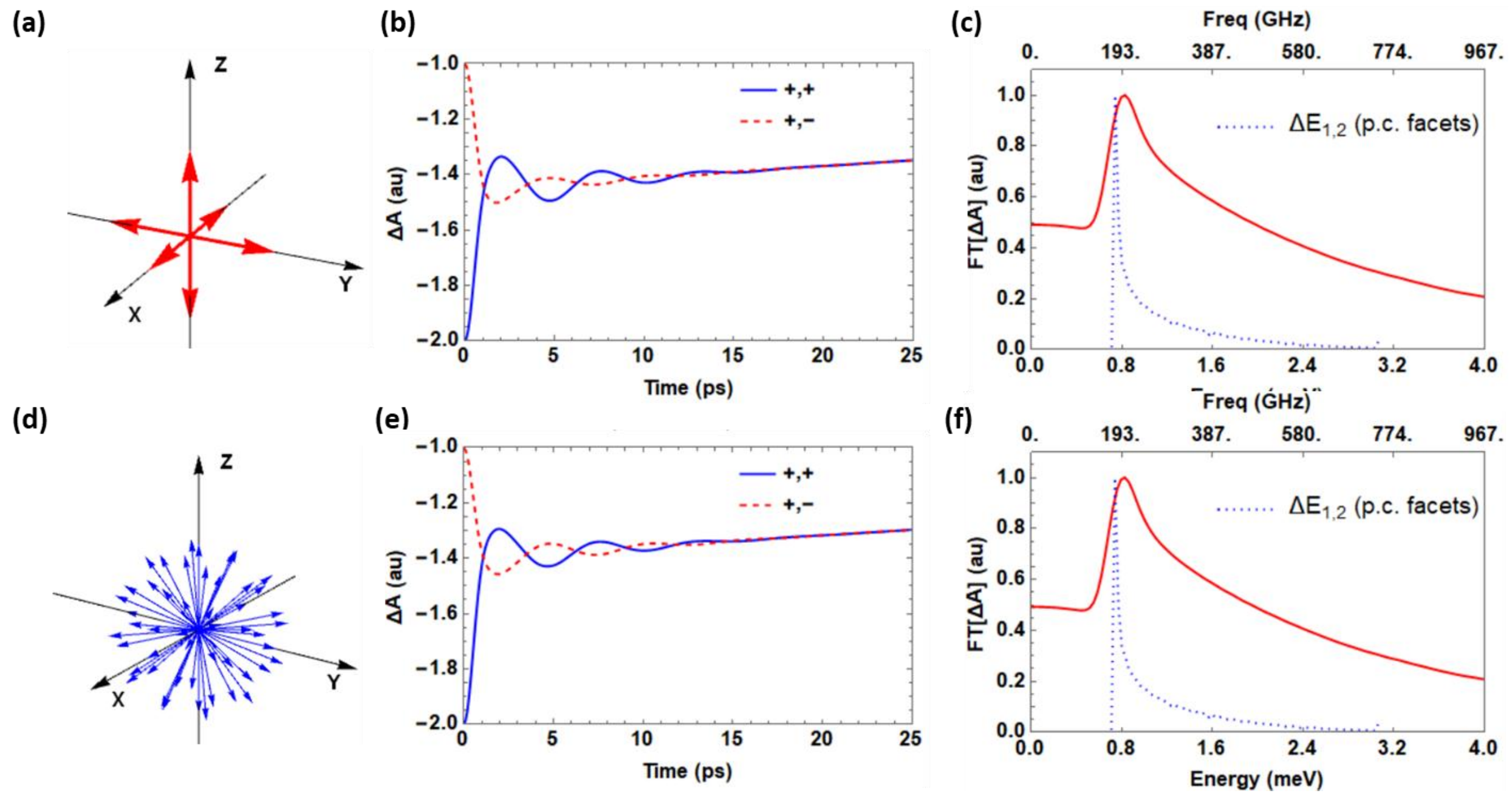

Supplementary Figure 39. Shape-averaged TA and FT spectrum for QDs with pseudocubic facets, all orientations.

Calculations are performed at $T$ = 80 K, with QDs of effective size $L_e = 7.9$ nm and with biaxial shape as reflected in Eq. S2 corresponding to averaging over the measured shape distribution in Supplementary Fig. 20 panel (b). Panel (a) depicts the possibility that the nanocrystals are randomly oriented during TA measurement, with 1/6 oriented with an X, Y or Z facet oriented normal to the optical axis, taken as the Z direction. The corresponding orientation- and shape-averaged TA traces are shown in panel (b) with the Fourier transform (FT) of the difference, $\sigma_+\sigma_- - \sigma_+\sigma_+$, shown in panel (c). Panel (d) represents the case that the orientation of the QDs is fully randomized in roll, elevation and azimuth angles. The corresponding orientation- and shape-averaged TA traces are shown in panel (e) with the Fourier transform of the difference, $\sigma_+\sigma_- - \sigma_+\sigma_+$, shown in panel (f). The peak in the FT spectra in panels (c) and (f) corresponds to the peak in the distribution of $\Delta E_{\alpha,\beta}$, the energy difference between the bright excitons states $\alpha$ and $\beta$ as shown in Supplementary Fig. 36, reproduced in panels (c,f) for comparison. All other beating frequencies are averaged out due to shape dispersity. The difference between the beat frequency distributions shown in panels (c,f) is negligible. Material parameters used for the calculation are given in Supplementary Table 10; the phenomenological $\tau_g$ decoherence parameter is to match the width of the inhomogeneous average size distribution, 0.35 meV, shown in Supplementary Fig 35. The lifetime parameter $\tau_1 = 326.5\,ps$ was set to match experimentally measured TA traces for this size. For these calculations $\eta_{XX} = 0$. The orthorhombic strain is taken as -0.03 and the tetragonal strain as +0.0056, reflecting the values at $T$ = 80 K determined from the empirical fits in Supplementary Fig. 23.

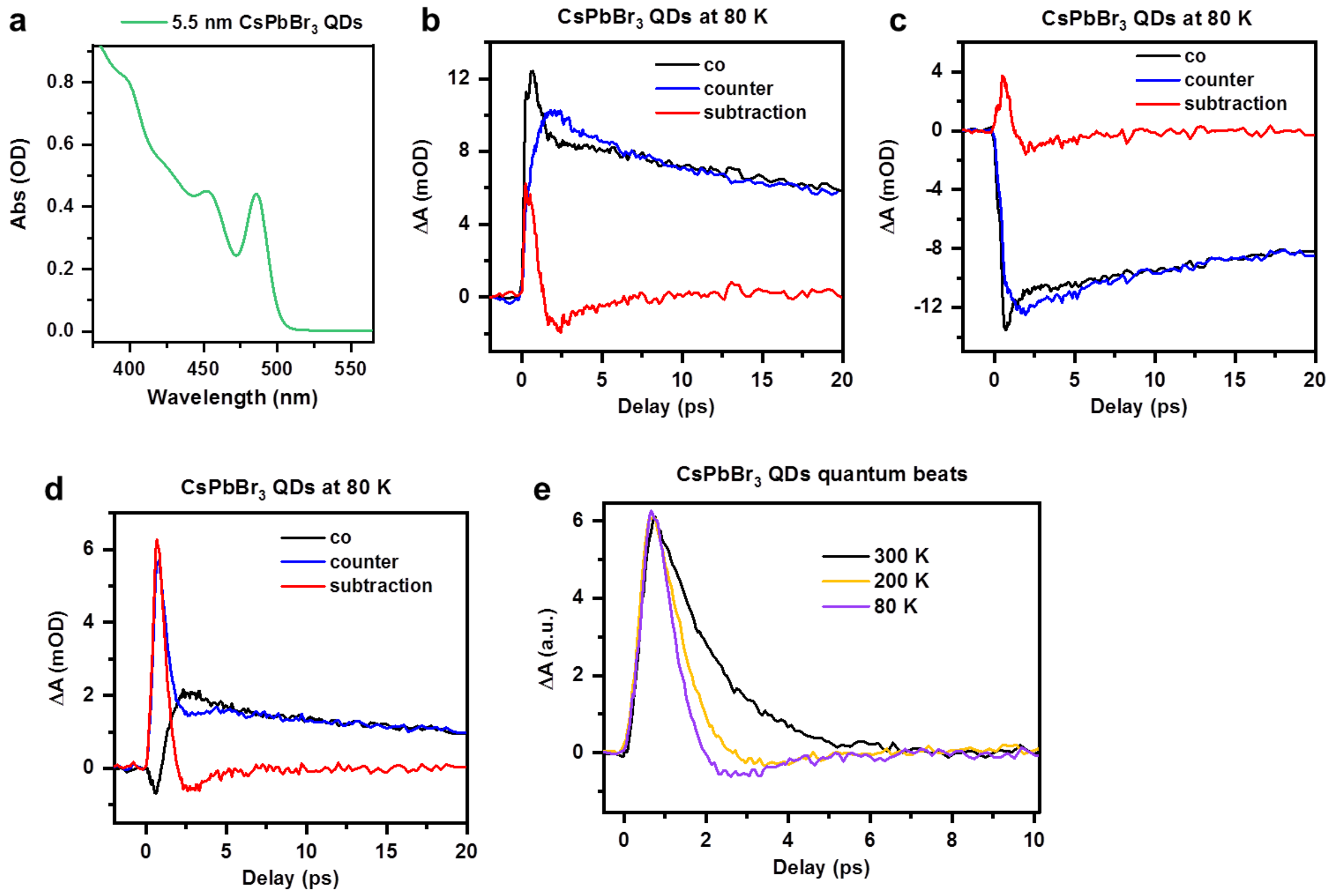

Supplementary Figure 40. Quantum beats in $CsPbBr_3$ QDs.

(a) Absorption spectrum of ~5.5 nm $CsPbBr_3$ QDs. (b-d) 80 K TA kinetics measured by co (black) and counter (blue) pump/probe and their subtraction (red), probed at (b) blue-side induced absorption, (c) exciton bleach and (d) red-side induced absorption. The subtracted kinetics consistently show signature of strongly-damped quantum beats. (e) Comparison of the quantum beating kinetics at varying temperatures.

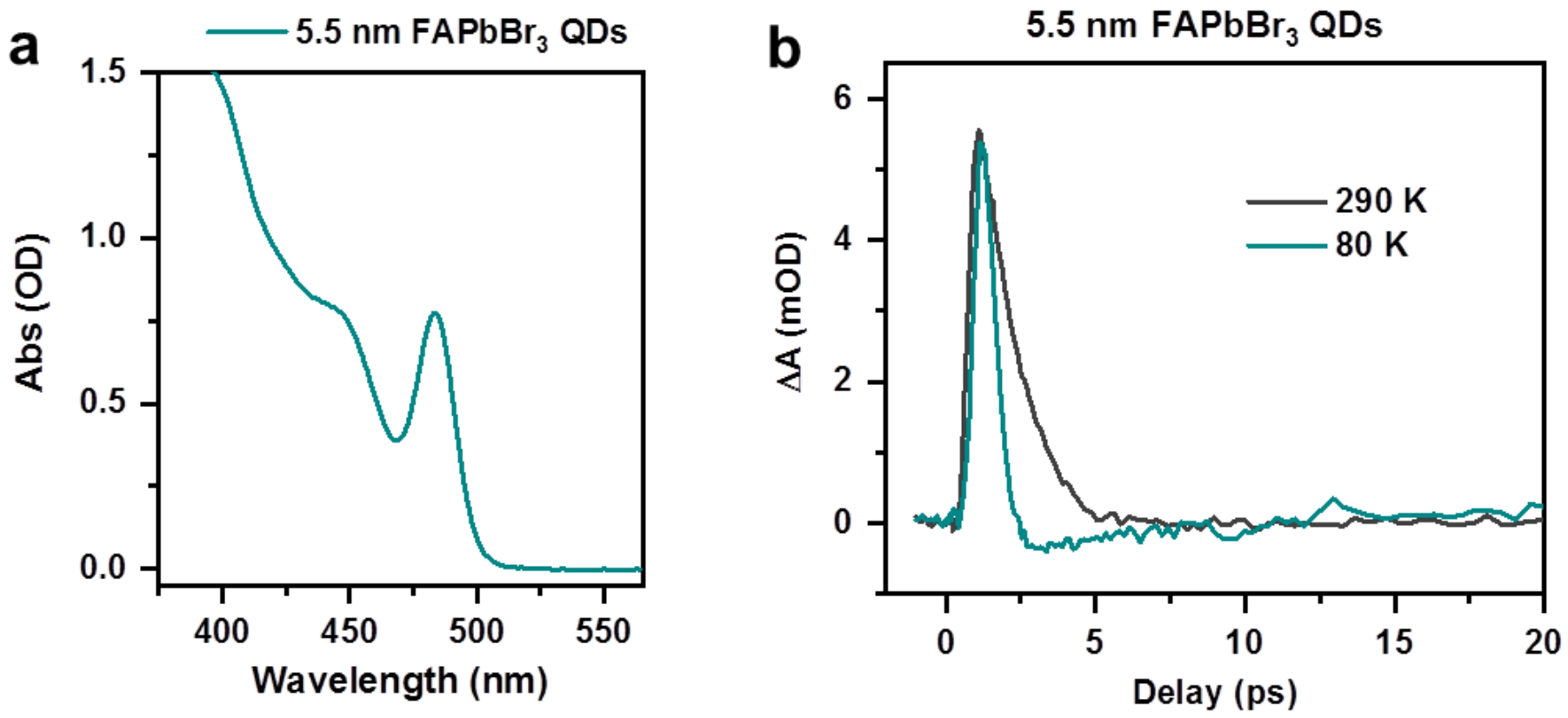

Supplementary Figure 41. Quantum beats in $FAPbBr_3$ QDs.

(a) Absorption spectrum of ~5.5 nm $FAPbBr_3$ QDs. (b) Comparison quantum beating kinetics measured at 80 K and 290 K.

## Use 4.9 nm at 80 K to evaluate different fitting models

**a**

$$S(t) = A_0\left[F\cdot e^{\frac{-t}{T_{dec}}}\cdot e^{-\left(\frac{t}{T_{\delta,1}}\right)^2}\cdot\cos\left(\frac{2\pi}{T_{FSS}}t+\varphi\right)+(1-F)\cdot e^{\frac{-t}{T_{dec}}}\cdot e^{-\left(\frac{t}{T_{\delta,2}}\right)^2}\right]$$

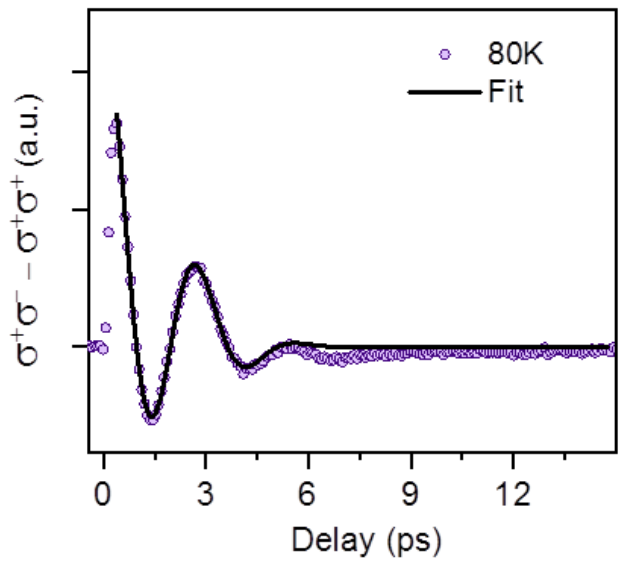

| | |
|---|---|
| $F$ | 0.53 ± 0.03 |
| $T_{dec}$ (ps) | 5.52*10^16 ± 4.35*10^6 |
| $T_{\delta,1}$ (ps) | 2.87 ± 0.07 |
| $T_{\delta,2}$ (ps) | 1.60 ± 0.12 |
| $T_{FSS}$ (ps) | 3.00 ± 0.05 |
| φ | 0.28 ± 0.08 |

**b**

$$S(t) = A_0\left[F\cdot e^{\frac{-t}{T_{dec}}}\cdot e^{-\left(\frac{t}{T_{\delta,1}}\right)^2}\cdot\cos\left(\frac{2\pi}{T_{FSS}}t+\varphi\right)+(1-F)\cdot e^{-\left(\frac{t}{T_{\delta,2}}\right)^2}\right]$$

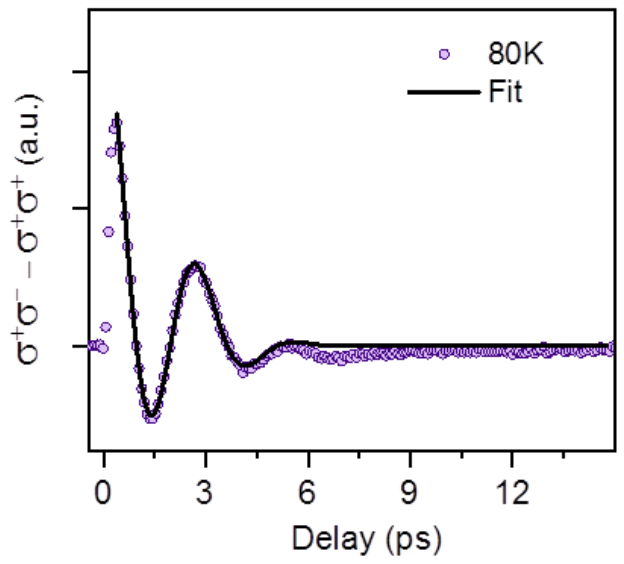

| | |
|---|---|
| $F$ | 0.53 ± 0.03 |
| $T_{dec}$ (ps) | 3.74*10^23 ± 8.03*10^6 |
| $T_{\delta,1}$ (ps) | 2.87 ± 0.07 |
| $T_{\delta,2}$ (ps) | 1.60 ± 0.12 |
| $T_{FSS}$ (ps) | 3.01 ± 0.05 |
| φ | 0.31 ± 0.10 |

**c**

$$S(t) = A_0\left[F\cdot e^{\frac{-t}{T_{dec,1}}}\cdot e^{-\left(\frac{t}{T_{\delta}}\right)^2}\cdot\cos\left(\frac{2\pi}{T_{FSS}}t+\varphi\right)+(1-F)\cdot e^{\frac{-t}{T_{dec,2}}}\right]$$

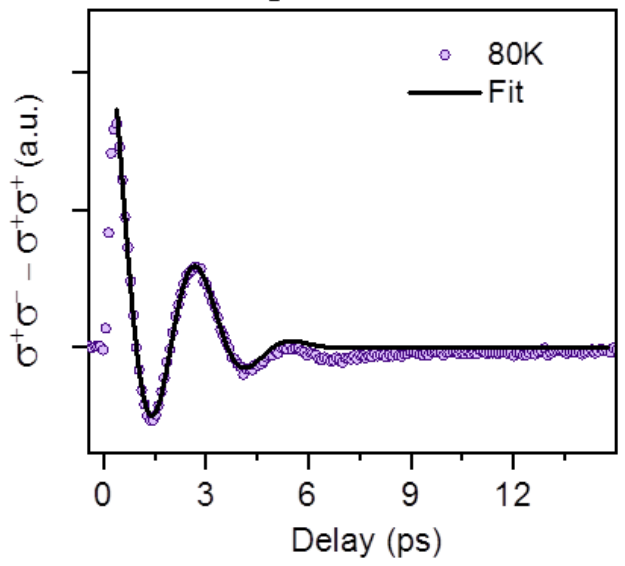

| | |
|---|---|
| $F$ | 0.38 ± 0.02 |
| $T_{dec,1}$ (ps) | 3.13*10^19 ± 1.07*10^7 |
| $T_{dec,2}$ (ps) | 0.93 ± 0.06 |
| $T_{\delta}$ (ps) | 3.08 ± 0.08 |
| $T_{FSS}$ (ps) | 2.92 ± 0.04 |
| φ | 0.19 ± 0.07 |

**d**

$$S(t) = A_0\left[F\cdot e^{\frac{-t}{T_{dec,1}}}\cdot\cos\left(\frac{2\pi}{T_{FSS}}t+\varphi\right)+(1-F)\cdot e^{\frac{-t}{T_{dec,2}}}\right]$$

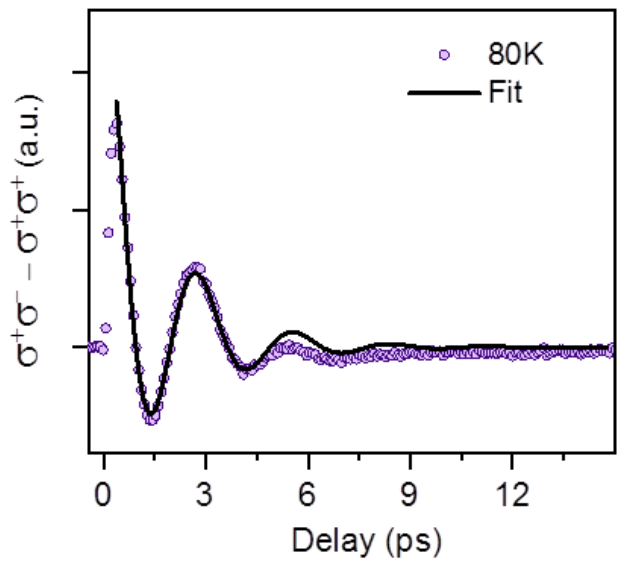

| | |
|---|---|
| $F$ | 0.60 ± 0.03 |
| $T_{dec,1}$ (ps) | 1.94 ± 0.08 |
| $T_{dec,2}$ (ps) | 1.22 ± 0.11 |
| $T_{FSS}$ (ps) | 2.82 ± 0.03 |
| φ | -0.02 ± 0.06 |

**e**

$$S(t) = A_0\left[F\cdot e^{-\left(\frac{t}{T_{\delta,1}}\right)^2}\cdot\cos\left(\frac{2\pi}{T_{FSS}}t+\varphi\right)+(1-F)\cdot e^{-\left(\frac{t}{T_{\delta,2}}\right)^2}\right]$$

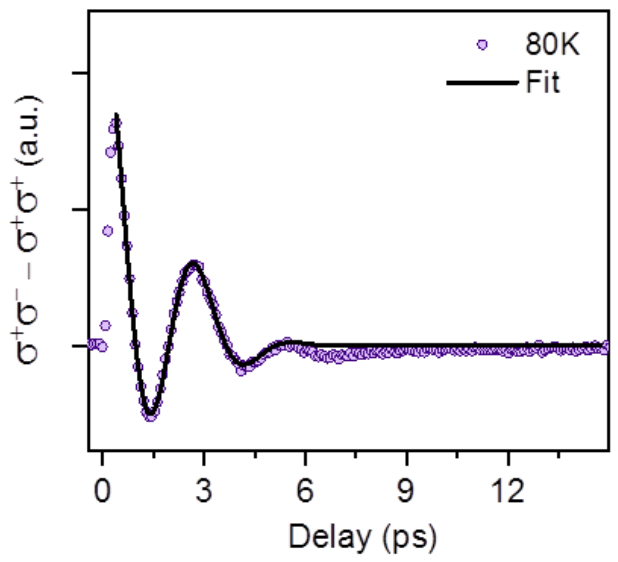

| | |
|---|---|
| $F$ | 0.55 ± 0.02 |
| $T_{\delta,1}$ (ps) | 2.83 ± 0.06 |
| $T_{\delta,2}$ (ps) | 1.69 ± 0.10 |
| $T_{FSS}$ (ps) | 3.00 ± 0.05 |
| φ | 0.28 ± 0.08 |

**f**

$$S(t) = A_0\left[F\cdot e^{\frac{-t}{T_{dec}}}\cdot\cos\left(\frac{2\pi}{T_{FSS}}t+\varphi\right)+(1-F)\cdot e^{-\left(\frac{t}{T_{\delta}}\right)^2}\right]$$

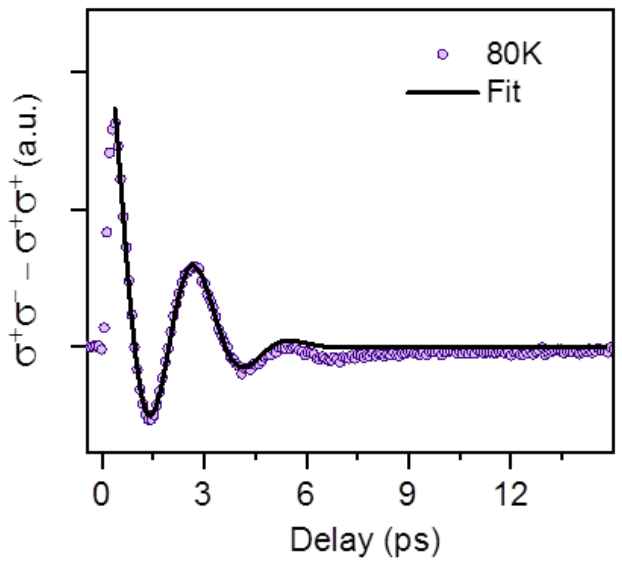

| | |
|---|---|
| $F$ | 0.78 ± 0.01 |
| $T_{dec}$ (ps) | 1.77 ± 0.06 |
| $T_{\delta}$ (ps) | 2.40 ± 0.14 |
| $T_{FSS}$ (ps) | 2.78 ± 0.03 |
| φ | -0.16 ± 0.05 |

**g**

$$S(t) = A_0\left[F\cdot e^{-\left(\frac{t}{T_{\delta}}\right)^2}\cdot\cos\left(\frac{2\pi}{T_{FSS}}t+\varphi\right)+(1-F)\cdot e^{\frac{-t}{T_{dec}}}\right]$$

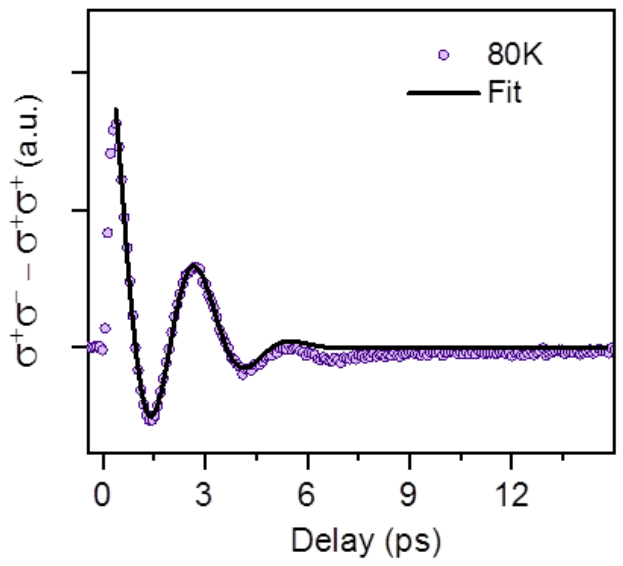

| | |
|---|---|
| $F$ | 0.38 ± 0.02 |
| $T_{dec}$ (ps) | 0.93 ± 0.06 |
| $T_{\delta}$ (ps) | 3.08 ± 0.08 |
| $T_{FSS}$ (ps) | 2.92± 0.04 |
| φ | 0.19 ± 0.07 |

Supplementary Figure 42. Comparison of different empirical fitting models of quantum beats.

## Supplementary Text 7. Discussion on empirical fitting models

In panel a of Supplementary Fig. 42, we consider that the decoherence time $T_{dec}$ should be the same for beating and non-beating components, whereas the Gaussian terms are different for them (reflected as different $T_{\delta,1}$ and $T_{\delta,2}$). Note here for simplicity we use a Gaussian function to represent the energy distribution as an approximation, since the energy distributions are expected to be non-Gaussian. The fit works quite well, but likely because the model is over-parameterized, $T_{dec}$ is unrealistically long. Likewise in panel b and c where we use a simplified pure Gaussian or exponential decay for the non-beating component, the fits are still over-parameterized, again giving unrealistically long $T_{dec}$.

In light of the over-parameterization issue, in panels d-g, we further simplify the model by considering that since the beats are damped so fast, on the timescale of a few ps, mathematically it might be hard to differentiate exponential and Gaussian decays. They work equally well in terms of giving realistic time constants. However, so far we focus only on the time constants in the fitting parameters. In order to differentiate between these models and choose the most suitable one, we pay attention to also *the fitting parameter F*, which represents the fraction of the beating component in the total signal. As explained above, the beating component mainly results from QDs with their c-axis aligned to the laser k-vector whereas the non-beating component from QDs with their c-axis perpendicular to laser. *Assuming that the three bright excitons have equal oscillator strength, we should expect F to be near 1/3. Under this criterion, only the model in panel g seems to be reasonable.*

We further consider why the model in panel g is the most physically meaningful. In this empirical model, the beating part is damped by Gaussian decay whereas the non-beating part is damped by exponential decay. According to Supplementary Fig. 36b, the shape variation leads to a highly-asymmetric energy distribution for the beating part, but the dominant component of this DOS is much narrower than the energy distribution induced by size variation (Supplementary Fig. 35b). So the broadening should be dominated by the latter, which is empirically represented by a Gaussian-like damping in our fitting equation. By contrast, the non-beating part has very broad energy distribution in Supplementary Fig. 36d due to shape variation, much broader than that induced by size variation in (Supplementary Fig. 35b), so the shape broadening is the dominant one. This distribution clearly deviates from a Gaussian distribution. Although we do not have its analytic form, the damping kinetics induced by it is already reproduced by our theoretic model (Supplementary Fig. 38e). Such a decay profile can be phenomenologically fitted by a simple exponential decay in our empirical fitting equation.

At last, we would like to point out that *regardless of the model used, the FSS beating period* $T_{FSS}$ *remains virtually the same, and is consistent with FFT result. So the major conclusions in the paper will not be affected at all by the specific empirical model used.*

## SM References: